\documentclass[12pt]{article}
\usepackage{amsfonts}
\usepackage{amsfonts}
\usepackage[dvips]{graphicx}
\usepackage{latexsym,amsmath,amssymb,stmaryrd}
\usepackage{cite}
\usepackage{array}
\usepackage{subfigure}
\usepackage{multirow}
\usepackage{epsfig}
\usepackage[dvips]{color}
\usepackage{pstricks}
\usepackage{pst-node}
\usepackage{pst-plot}
\usepackage{dsfont}
\usepackage{amsthm}
\usepackage{graphics}
\usepackage{fancyhdr}
\usepackage{hhline} 
\usepackage{longtable}
\usepackage{feynmp-eps}
\usepackage{booktabs}
\usepackage{calc}
\usepackage{rotating}

\allowdisplaybreaks[1] 

\newcommand{\subfigspace}{\phantom{**}} 

\newcommand{\redchi}[1]{\textcolor{red}{\boldsymbol{\chi(#1)}}}
\newcommand{\redperm}[1]{\textcolor{red}{\boldsymbol{\{#1\}}}}
\definecolor{light-gray}{gray}{0.5}

\fancypagestyle{plain}{
\fancyhf{}
\newcommand{\fpage}{\iffloatpage{}{\thepage}}
\fancyfoot[C]{\fpage}

}

\newcommand{\col}{~,}
\newcommand{\pnt}{~.}

\newcommand{\pfour}[4]{{}\{#1,#2,#3,#4\}{}}
\newcommand{\pthree}[3]{{}\{#1,#2,#3\}{}}
\newcommand{\ptwo}[2]{{}\{#1,#2\}{}}
\newcommand{\pone}[1]{{}\{#1\}{}}
\newcommand{\pid}{{}\{ \}{}}

\newlength{\neglength}
\newlength{\diameter}

\newcommand{\svertex}[2]{%
\fmfiequ{#1}{point length(#2)/2 of #2}
}

\newcommand{\plainwrap}[4]{%
\fmfipath{pi[]}
\fmfiset{pi1}{vloc(__#1) ..controls (-0.175w,ypart(vloc(__#1))) and (-0.175w,-0.15w) .. (xpart(vloc(__#2)),-0.15w)}
\fmfiset{pi2}{(xpart(vloc(__#2)),-0.15w) ..(xpart(vloc(__#3)),-0.15w)}

\fmfiset{pi3}{(xpart(vloc(__#3)),-0.15w) ..controls (1.175w,-0.15w) and (1.175w,ypart(vloc(__#4))) .. vloc(__#4)}
\fmfi{plain}{pi1 ..pi2 ..pi3}
}
\newcommand{\wigglywrap}[4]{%
\fmfipath{pi[]}
\fmfiset{pi1}{#1 ..controls (-0.175w,ypart(#1)) and (-0.175w,-0.15w) .. (xpart(vloc(__#2)),-0.15w)}
\fmfiset{pi2}{(xpart(vloc(__#2)),-0.15w) ..(xpart(vloc(__#3)),-0.15w)}

\fmfiset{pi3}{(xpart(vloc(__#3)),-0.15w) ..controls (1.175w,-0.15w) and (1.175w,ypart(#4)) .. #4}
\fmfi{wiggly}{pi1 ..pi2 ..pi3}
}

\newcommand{\Wdqut}{%
\fmftop{v1}
\fmfbottom{v5}
\fmfforce{(0w,h)}{v1}
\fmfforce{(0w,0)}{v5}
\fmffixed{(0.25w,0)}{v1,v2}
\fmffixed{(0.25w,0)}{v2,v3}
\fmffixed{(0.25w,0)}{v3,v4}
\fmffixed{(0.25w,0)}{v4,v9}
\fmffixed{(0.25w,0)}{v5,v6}
\fmffixed{(0.25w,0)}{v6,v7}
\fmffixed{(0.25w,0)}{v7,v8}
\fmffixed{(0.25w,0)}{v8,v10}
\fmffixed{(0,whatever)}{vb2,vc2}
\fmffixed{(0,whatever)}{vb3,vc3}
\fmffixed{(0,whatever)}{vb5,vc5}
\fmffixed{(0,whatever)}{vb7,vc7}
\fmf{plain,tension=0.25,left=0.25}{v5,vc2}
\fmf{plain,tension=0.25,right=0.25}{v6,vc2}
\fmf{phantom,tension=0.25,right=0.25}{v1,vb2}
\fmf{phantom,tension=0.25,left=0.25}{v2,vb2}
\fmf{phantom,tension=0.5}{vc2,vb2}
\fmf{plain,tension=0.25,right=0.25}{v2,vc3}
\fmf{plain,tension=0.25,left=0.25}{v3,vc3}
\fmf{phantom,tension=0.25,right=0.25}{v6,vb3}
\fmf{phantom,tension=0.25,left=0.25}{v7,vb3}
\fmf{phantom,tension=0.5}{vc3,vb3}
\fmf{plain,tension=0.25,left=0.25}{v7,vc5}
\fmf{plain,tension=0.25,right=0.25}{v8,vc5}
\fmf{phantom,tension=0.25,right=0.25}{v3,vb5}
\fmf{phantom,tension=0.25,left=0.25}{v4,vb5}
\fmf{phantom,tension=0.5}{vc5,vb5}
\fmf{plain,tension=0.25,right=0.25}{v4,vc7}
\fmf{plain,tension=0.25,left=0.25}{v9,vc7}
\fmf{phantom,tension=0.25,right=0.25}{v8,vb7}
\fmf{phantom,tension=0.25,left=0.25}{v10,vb7}
\fmf{phantom,tension=0.5}{vc7,vb7}
\fmffreeze
\fmffixed{(0,whatever)}{vc1,vc2}
\fmffixed{(0,whatever)}{vc3,vc4}
\fmffixed{(0,whatever)}{vc5,vc6}
\fmffixed{(0,whatever)}{vc7,vc8}
\fmffixed{(whatever,0)}{vc1,vc6}
\fmffixed{(whatever,0)}{vc4,vc8}
\fmf{plain,tension=0.25,right=0.25}{v1,vc1}
\fmf{plain,tension=0.5}{vc1,vc2}
\fmf{plain,tension=0.5}{vc3,vc4}
\fmf{plain,tension=0.5}{vc1,vc4}
\fmf{plain,tension=0.5}{vc5,vc6}
\fmf{plain,tension=0.5}{vc4,vc6}
\fmf{plain,tension=0.25,right=0.25}{v10,vc8}
\fmf{plain,tension=0.5}{vc7,vc8}
\fmf{plain,tension=0.5}{vc6,vc8}
\fmf{plain,tension=0.5,right=0,width=1mm}{v5,v10}
\fmfposition
\fmfipath{p[]}
\fmfiset{p0}{vpath(__v5,__vc2)}
\fmfiset{p1}{vpath(__vc1,__vc2)}
\fmfiset{p2}{vpath(__v1,__vc1)}
\fmfiset{p3}{vpath(__v10,__vc8)}
\fmfiset{p4}{vpath(__vc8,__vc7)}
\fmfiset{p5}{vpath(__v9,__vc7)}
}

\newcommand{\Wtduq}{%
\fmftop{v1}
\fmfbottom{v5}
\fmfforce{(0w,h)}{v1}
\fmfforce{(0w,0)}{v5}
\fmffixed{(0.25w,0)}{v1,v2}
\fmffixed{(0.25w,0)}{v2,v3}
\fmffixed{(0.25w,0)}{v3,v4}
\fmffixed{(0.25w,0)}{v4,v9}
\fmffixed{(0.25w,0)}{v5,v6}
\fmffixed{(0.25w,0)}{v6,v7}
\fmffixed{(0.25w,0)}{v7,v8}
\fmffixed{(0.25w,0)}{v8,v10}
\fmffixed{(0,whatever)}{vb1,vc2}
\fmffixed{(0,whatever)}{vc3,vb4}
\fmffixed{(0,whatever)}{vb7,vc8}
\fmf{plain,tension=0.25,left=0.25}{v5,vc2}
\fmf{plain,tension=0.25,right=0.25}{v6,vc2}
\fmf{phantom,tension=0.25,right=0.25}{v1,vb1}
\fmf{phantom,tension=0.25,left=0.25}{v2,vb1}
\fmf{phantom,tension=0.5}{vc2,vb1}
\fmf{plain,tension=0.25,left=0.25}{v8,vc8}
\fmf{plain,tension=0.25,right=0.25}{v10,vc8}
\fmf{phantom,tension=0.25,right=0.25}{v4,vb7}
\fmf{phantom,tension=0.25,left=0.25}{v9,vb7}
\fmf{phantom,tension=0.5}{vb7,vc8}
\fmf{plain,tension=0.25,right=0.25}{v3,vc3}
\fmf{plain,tension=0.25,left=0.25}{v4,vc3}
\fmf{phantom,tension=0.25,left=0.25}{v7,vb4}
\fmf{phantom,tension=0.25,right=0.25}{v8,vb4}
\fmf{phantom,tension=0.2}{vc3,vb4}
\fmffreeze
\fmffixed{(0,whatever)}{vc1,vc2}
\fmffixed{(0,whatever)}{vc3,vc4}
\fmffixed{(0,whatever)}{vc5,vc6}
\fmffixed{(0,whatever)}{vc7,vc8}
\fmffixed{(whatever,0)}{vc7,vc5}
\fmf{plain,tension=0.5,right=0.25}{v1,vc1}
\fmf{plain,tension=0.15,right=0.25}{v2,vc5}
\fmf{plain,tension=0.5,left=0.25}{v9,vc7}
\fmf{plain,tension=0.25}{vc3,vc4}
\fmf{plain,tension=0.5}{vc1,vc2}
\fmf{plain,tension=0.5}{vc6,vc1}
\fmf{plain,tension=0.5}{vc7,vc4}
\fmf{plain,tension=0.5}{vc4,vc5}
\fmf{plain,tension=0.5}{vc5,vc6}
\fmf{plain,tension=0.5,left=0.25}{vc6,v7}
\fmf{plain,tension=1}{vc7,vc8}
\fmf{plain,tension=0.5,right=0,width=1mm}{v5,v10}
\fmfposition
\fmfipath{p[]}
\fmfiset{p0}{vpath(__v5,__vc2)}
\fmfiset{p1}{vpath(__vc1,__vc2)}
\fmfiset{p2}{vpath(__v1,__vc1)}
\fmfiset{p3}{vpath(__v10,__vc8)}
\fmfiset{p4}{vpath(__vc8,__vc7)}
\fmfiset{p5}{vpath(__v9,__vc7)}
}

\newcommand{\Wudtq}{%
\fmftop{v1}
\fmfbottom{v5}
\fmfforce{(0w,h)}{v1}
\fmfforce{(0w,0)}{v5}
\fmffixed{(0.25w,0)}{v1,v2}
\fmffixed{(0.25w,0)}{v2,v3}
\fmffixed{(0.25w,0)}{v3,v4}
\fmffixed{(0.25w,0)}{v4,v9}
\fmffixed{(0.25w,0)}{v5,v6}
\fmffixed{(0.25w,0)}{v6,v7}
\fmffixed{(0.25w,0)}{v7,v8}
\fmffixed{(0.25w,0)}{v8,v10}
\fmffixed{(0,whatever)}{vc1,vc2}
\fmffixed{(0,whatever)}{vc3,vc4}
\fmffixed{(0,whatever)}{vc5,vc6}
\fmffixed{(0,whatever)}{vc7,vc8}
\fmf{plain,tension=0.25,right=0.25}{v1,vc1}
\fmf{plain,tension=0.25,left=0.25}{v2,vc1}
\fmf{plain,left=0.25}{v5,vc2}
\fmf{plain,tension=1,left=0.25}{v3,vc3}
\fmf{plain,tension=1,left=0.25}{v4,vc5}
\fmf{plain,tension=1,left=0.25}{v9,vc7}
\fmf{plain,left=0.25}{v7,vc6}
\fmf{plain,tension=0.25,left=0.25}{v8,vc8}
\fmf{plain,tension=0.25,right=0.25}{v10,vc8}
\fmf{plain,left=0.25}{v6,vc4}
\fmf{plain,tension=0.5}{vc1,vc2}
\fmf{plain,tension=0.5}{vc2,vc3}
\fmf{plain,tension=0.5}{vc3,vc4}
\fmf{plain,tension=0.5}{vc4,vc5}
\fmf{plain,tension=0.5}{vc5,vc6}
\fmf{plain,tension=0.5}{vc6,vc7}
\fmf{plain,tension=0.5}{vc7,vc8}
\fmf{plain,tension=0.5,right=0,width=1mm}{v5,v10}
\fmfposition
\fmfipath{p[]}
\fmfiset{p0}{vpath(__v5,__vc2)}
\fmfiset{p1}{vpath(__vc1,__vc2)}
\fmfiset{p2}{vpath(__v1,__vc1)}
\fmfiset{p3}{vpath(__v10,__vc8)}
\fmfiset{p4}{vpath(__vc8,__vc7)}
\fmfiset{p5}{vpath(__v9,__vc7)}
}

\newcommand{\Wuqtd}{%
\fmftop{v1}
\fmfbottom{v5}
\fmfforce{(0w,h)}{v1}
\fmfforce{(0w,0)}{v5}
\fmffixed{(0.25w,0)}{v1,v2}
\fmffixed{(0.25w,0)}{v2,v3}
\fmffixed{(0.25w,0)}{v3,v4}
\fmffixed{(0.25w,0)}{v4,v9}
\fmffixed{(0.25w,0)}{v5,v6}
\fmffixed{(0.25w,0)}{v6,v7}
\fmffixed{(0.25w,0)}{v7,v8}
\fmffixed{(0.25w,0)}{v8,v10}
\fmffixed{(0,whatever)}{vb1,vc1}
\fmffixed{(0,whatever)}{vb2,vc2}
\fmffixed{(0,whatever)}{vb7,vc7}
\fmf{plain,tension=0.25,right=0.25}{v1,vc1}
\fmf{plain,tension=0.25,left=0.25}{v2,vc1}
\fmf{phantom,tension=0.25,right=0.25}{v5,vb1}
\fmf{phantom,tension=0.25,left=0.25}{v6,vb1}
\fmf{phantom,tension=0.5}{vc1,vb1}
\fmf{plain,tension=0.25,left=0.25}{v6,vc2}
\fmf{plain,tension=0.25,right=0.25}{v7,vc2}
\fmf{phantom,tension=0.25,right=0.25}{v2,vb2}
\fmf{phantom,tension=0.25,left=0.25}{v3,vb2}
\fmf{phantom,tension=0.5}{vc2,vb2}
\fmf{plain,tension=0.25,right=0.25}{v4,vc7}
\fmf{plain,tension=0.25,left=0.25}{v9,vc7}
\fmf{phantom,tension=0.25,right=0.25}{v8,vb7}
\fmf{phantom,tension=0.25,left=0.25}{v10,vb7}
\fmf{phantom,tension=0.5}{vc7,vb7}
\fmffreeze
\fmffixed{(0,whatever)}{vc1,vc3}
\fmffixed{(0,whatever)}{vc2,vc4}
\fmffixed{(0,whatever)}{vc5,vc6}
\fmffixed{(0,whatever)}{vc7,vc8}
\fmffixed{(whatever,0)}{vc3,vc5}
\fmf{plain,tension=1}{vc1,vc3}
\fmf{plain,tension=0.25,left=0.25}{v5,vc3}
\fmf{plain,tension=0.5}{vc3,vc4}
\fmf{plain,tension=0.5}{vc2,vc4}
\fmf{plain,tension=0.5}{vc4,vc5}
\fmf{plain,tension=0.25,left=0.25}{vc5,v8}
\fmf{plain,tension=1}{vc5,vc6}
\fmf{plain,tension=1,right=0.25}{v3,vc6}
\fmf{plain,tension=0.5,right=0.25}{v10,vc8}
\fmf{plain,tension=0.5}{vc6,vc8}
\fmf{plain,tension=0.5}{vc7,vc8}
\fmf{plain,tension=0.5,right=0,width=1mm}{v5,v10}
\fmfposition
\fmfipath{p[]}
\fmfiset{p0}{vpath(__v5,__vc3)}
\fmfiset{p1}{vpath(__vc3,__vc1)}
\fmfiset{p2}{vpath(__v1,__vc1)}
\fmfiset{p3}{vpath(__v10,__vc8)}
\fmfiset{p4}{vpath(__vc8,__vc7)}
\fmfiset{p5}{vpath(__v9,__vc7)}
}

\newcommand{\Wutd}{%
\fmftop{v1}
\fmfbottom{v5}
\fmfforce{(0,h)}{v1}
\fmfforce{(0,0)}{v5}
\fmffixed{(0.25w,0)}{v1,v2}
\fmffixed{(0.25w,0)}{v2,v3}
\fmffixed{(0.25w,0)}{v3,v4}
\fmffixed{(0.25w,0)}{v4,v9}
\fmffixed{(0.25w,0)}{v5,v6}
\fmffixed{(0.25w,0)}{v6,v7}
\fmffixed{(0.25w,0)}{v7,v8}
\fmffixed{(0.25w,0)}{v8,v10}
\fmffixed{(0,whatever)}{vc1,vc3}
\fmffixed{(0,whatever)}{vc2,vc4}
\fmffixed{(0,whatever)}{vc5,vc6}
\fmf{plain,tension=1,right=0.25}{v1,vc1}
\fmf{plain,tension=1,left=0.25}{v2,vc1}
\fmf{plain,tension=1,right=0.25}{v3,vc2}
\fmf{plain,tension=1,left=0.25}{v4,vc2}
\fmf{plain,tension=1,left=0.125}{v5,vc3}
\fmf{plain,tension=0.25,left=0.25}{v6,vc6}
\fmf{plain,tension=0.25,right=0.25}{v7,vc6}
\fmf{plain,tension=1,right=0.125}{v8,vc4}
  \fmf{plain,tension=4}{vc1,vc3}
  \fmf{plain,tension=4}{vc2,vc4}
  \fmf{plain,tension=0.5}{vc3,vc5}
  \fmf{plain,tension=0.5}{vc4,vc5}
  \fmf{plain,tension=1}{vc5,vc6}
\fmf{plain}{v9,v10}
\fmf{plain,tension=0.5,right=0,width=1mm}{v5,v10}
\fmfposition
\fmfipath{p[]}
\fmfiset{p0}{vpath(__v5,__vc3)}
\fmfiset{p1}{vpath(__vc1,__vc3)}
\fmfiset{p2}{vpath(__v1,__vc1)}
\fmfiset{p3}{vpath(__v8,__vc4)}
\fmfiset{p4}{vpath(__vc4,__vc2)}
\fmfiset{p5}{vpath(__v4,__vc2)}
\fmfiset{p6}{vpath(__v9,__v10)}
}

\newcommand{\Wdut}{%
\fmftop{v1}
\fmfbottom{v5}
\fmfforce{(0,h)}{v1}
\fmfforce{(0,0)}{v5}
\fmffixed{(0.25w,0)}{v1,v2}
\fmffixed{(0.25w,0)}{v2,v3}
\fmffixed{(0.25w,0)}{v3,v4}
\fmffixed{(0.25w,0)}{v4,v9}
\fmffixed{(0.25w,0)}{v5,v6}
\fmffixed{(0.25w,0)}{v6,v7}
\fmffixed{(0.25w,0)}{v7,v8}
\fmffixed{(0.25w,0)}{v8,v10}
\fmffixed{(0,whatever)}{vc1,vc3}
\fmffixed{(0,whatever)}{vc2,vc4}
\fmffixed{(0,whatever)}{vc5,vc6}
\fmf{plain,tension=1,left=0.25}{v5,vc1}
\fmf{plain,tension=1,right=0.25}{v6,vc1}
\fmf{plain,tension=1,left=0.25}{v7,vc2}
\fmf{plain,tension=1,right=0.25}{v8,vc2}
\fmf{plain,tension=1,right=0.125}{v1,vc3}
\fmf{plain,tension=0.25,right=0.25}{v2,vc6}
\fmf{plain,tension=0.25,left=0.25}{v3,vc6}
\fmf{plain,tension=1,left=0.125}{v4,vc4}
  \fmf{plain,tension=4}{vc1,vc3}
  \fmf{plain,tension=4}{vc2,vc4}
  \fmf{plain,tension=0.5}{vc3,vc5}
  \fmf{plain,tension=0.5}{vc4,vc5}
  \fmf{plain,tension=1}{vc5,vc6}
\fmf{plain}{v9,v10}
\fmf{plain,tension=0.5,left=0,width=1mm}{v5,v10}
\fmfposition
\fmfipath{p[]}
\fmfiset{p0}{vpath(__v5,__vc1)}
\fmfiset{p1}{vpath(__vc1,__vc3)}
\fmfiset{p2}{vpath(__v1,__vc3)}
\fmfiset{p3}{vpath(__v8,__vc2)}
\fmfiset{p4}{vpath(__vc2,__vc4)}
\fmfiset{p5}{vpath(__v4,__vc4)}
\fmfiset{p6}{vpath(__v9,__v10)}
}

\newcommand{\Wudt}{%
\fmftop{v1}
\fmfbottom{v5}
\fmfforce{(0w,h)}{v1}
\fmfforce{(0w,0)}{v5}
\fmffixed{(0.25w,0)}{v1,v2}
\fmffixed{(0.25w,0)}{v2,v3}
\fmffixed{(0.25w,0)}{v3,v4}
\fmffixed{(0.25w,0)}{v4,v9}
\fmffixed{(0.25w,0)}{v5,v6}
\fmffixed{(0.25w,0)}{v6,v7}
\fmffixed{(0.25w,0)}{v7,v8}
\fmffixed{(0.25w,0)}{v8,v10}
\fmffixed{(0,whatever)}{vc1,vc2}
\fmffixed{(0,whatever)}{vc3,vc4}
\fmffixed{(0,whatever)}{vc5,vc6}
\fmf{plain,tension=0.25,right=0.25}{v1,vc1}
\fmf{plain,tension=0.25,left=0.25}{v2,vc1}
\fmf{plain,left=0.25}{v5,vc2}
\fmf{plain,tension=1,left=0.25}{v3,vc3}
\fmf{plain,tension=1,left=0.25}{v4,vc5}
\fmf{plain,left=0.25}{v6,vc4}
\fmf{plain,tension=0.25,left=0.25}{v7,vc6}
\fmf{plain,tension=0.25,right=0.25}{v8,vc6}
  \fmf{plain,tension=0.5}{vc1,vc2}
  \fmf{plain,tension=0.5}{vc2,vc3}
  \fmf{plain,tension=0.5}{vc3,vc4}
  \fmf{plain,tension=0.5}{vc4,vc5}
  \fmf{plain,tension=0.5}{vc5,vc6}
\fmf{plain}{v9,v10}
\fmf{plain,tension=0.5,right=0,width=1mm}{v5,v10}
\fmfposition
\fmfipath{p[]}
\fmfiset{p0}{vpath(__v5,__vc2)}
\fmfiset{p1}{vpath(__vc2,__vc1)}
\fmfiset{p2}{vpath(__v1,__vc1)}
\fmfiset{p3}{vpath(__v8,__vc6)}
\fmfiset{p4}{vpath(__vc6,__vc5)}
\fmfiset{p5}{vpath(__v4,__vc5)}
\fmfiset{p6}{vpath(__v9,__v10)}
}

\newcommand{\Wduq}{%
\fmftop{v1}
\fmfbottom{v5}
\fmfforce{(0w,h)}{v1}
\fmfforce{(0w,0)}{v5}
\fmffixed{(0.25w,0)}{v1,v2}
\fmffixed{(0.25w,0)}{v2,v3}
\fmffixed{(0.25w,0)}{v3,v4}
\fmffixed{(0.25w,0)}{v4,v9}
\fmffixed{(0.25w,0)}{v5,v6}
\fmffixed{(0.25w,0)}{v6,v7}
\fmffixed{(0.25w,0)}{v7,v8}
\fmffixed{(0.25w,0)}{v8,v10}
\fmffixed{(0,whatever)}{vc1,vc2}
\fmffixed{(0,whatever)}{vc3,vc4}
\fmffixed{(whatever,0)}{vc1,vc6}
\fmffixed{(whatever,0)}{vc2,vc4}
\fmf{plain,tension=0.5,right=0.25}{v2,vc1}
\fmf{plain,tension=0.5,left=0.25}{v3,vc1}
\fmf{plain,right=0.25}{v1,vc3}
\fmf{plain,tension=0.5,left=0.25}{v5,vc4}
\fmf{plain,tension=0.5,right=0.25}{v6,vc4}
\fmf{plain,right=0.25}{v7,vc2}
\fmf{plain,tension=0.5,left=0.25}{v8,vc5}
\fmf{plain,tension=0.5,right=0.25}{v10,vc5}
\fmf{plain,tension=0.5,right=0.25}{v4,vc6}
\fmf{plain,tension=0.5,left=0.25}{v9,vc6}
\fmf{plain}{vc5,vc6}
\fmf{plain,tension=1}{vc1,vc2}
\fmf{plain,tension=0.5}{vc2,vc3}
\fmf{plain,tension=1}{vc3,vc4}
\fmf{plain,tension=0.5,right=0,width=1mm}{v5,v10}
\fmfposition
\fmfipath{p[]}
\fmfiset{p0}{vpath(__v5,__vc4)}
\fmfiset{p1}{vpath(__vc4,__vc3)}
\fmfiset{p2}{vpath(__v1,__vc3)}
\fmfiset{p3}{vpath(__v7,__vc2)}
\fmfiset{p4}{vpath(__vc2,__vc1)}
\fmfiset{p5}{vpath(__v3,__vc1)}
\fmfiset{p6}{vpath(__v8,__vc5)}
\fmfiset{p7}{vpath(__vc5,__vc6)}
\fmfiset{p8}{vpath(__v4,__vc6)}
\fmfiset{p9}{vpath(__v10,__vc5)}
\fmfiset{p10}{vpath(__v9,__vc6)}
}

\newcommand{\Wdu}{%
\fmftop{v1}
\fmfbottom{v5}
\fmfforce{(0w,h)}{v1}
\fmfforce{(0w,0)}{v5}
\fmffixed{(0.25w,0)}{v1,v2}
\fmffixed{(0.25w,0)}{v2,v3}
\fmffixed{(0.25w,0)}{v3,v4}
\fmffixed{(0.25w,0)}{v4,v9}
\fmffixed{(0.25w,0)}{v5,v6}
\fmffixed{(0.25w,0)}{v6,v7}
\fmffixed{(0.25w,0)}{v7,v8}
\fmffixed{(0.25w,0)}{v8,v10}
\fmffixed{(0,whatever)}{vc1,vc2}
\fmffixed{(0,whatever)}{vc3,vc4}
\fmf{plain,tension=0.5,right=0.25}{v2,vc1}
\fmf{plain,tension=0.5,left=0.25}{v3,vc1}
\fmf{plain,right=0.25}{v1,vc3}
\fmf{plain,tension=0.5,left=0.25}{v5,vc4}
\fmf{plain,tension=0.5,right=0.25}{v6,vc4}
\fmf{plain,right=0.25}{v7,vc2}
\fmf{plain}{v8,v4}
  \fmf{plain,tension=1}{vc1,vc2}
  \fmf{plain,tension=0.5}{vc2,vc3}
  \fmf{plain,tension=1}{vc3,vc4}
\fmf{plain}{v9,v10}
\fmf{plain,tension=0.5,right=0,width=1mm}{v5,v10}
\fmfposition
\fmfipath{p[]}
\fmfiset{p0}{vpath(__v5,__vc4)}
\fmfiset{p1}{vpath(__vc4,__vc3)}
\fmfiset{p2}{vpath(__v1,__vc3)}
\fmfiset{p3}{vpath(__v7,__vc2)}
\fmfiset{p4}{vpath(__vc2,__vc1)}
\fmfiset{p5}{vpath(__v3,__vc1)}
\fmfiset{p6}{vpath(__v4,__v8)}
\fmfiset{p7}{vpath(__v9,__v10)}
}

\newcommand{\Wuq}{%
\fmftop{v1}
\fmfbottom{v5}
\fmfforce{(0w,h)}{v1}
\fmfforce{(0w,0)}{v5}
\fmffixed{(0.25w,0)}{v1,v2}
\fmffixed{(0.25w,0)}{v2,v3}
\fmffixed{(0.25w,0)}{v3,v4}
\fmffixed{(0.25w,0)}{v4,v9}
\fmffixed{(0.25w,0)}{v5,v6}
\fmffixed{(0.25w,0)}{v6,v7}
\fmffixed{(0.25w,0)}{v7,v8}
\fmffixed{(0.25w,0)}{v8,v10}
\fmffixed{(0,whatever)}{vc1,vc2}
\fmffixed{(0,whatever)}{vc3,vc4}
\fmf{plain,tension=0.5,right=0.25}{v1,vc1}
\fmf{plain,tension=0.5,left=0.25}{v2,vc1}
\fmf{plain,tension=0.5,left=0.25}{v5,vc2}
\fmf{plain,tension=0.5,right=0.25}{v6,vc2}
\fmf{plain,tension=0.5,left=0.25}{v8,vc4}
\fmf{plain,tension=0.5,right=0.25}{v10,vc4}
\fmf{plain,tension=0.5,right=0.25}{v4,vc3}
\fmf{plain,tension=0.5,left=0.25}{v9,vc3}
\fmf{plain,tension=1}{vc1,vc2}
\fmf{plain,tension=1}{vc3,vc4}
\fmf{plain,tension=1}{v3,v7}
\fmf{plain,tension=0.5,right=0,width=1mm}{v5,v10}
\fmfposition
\fmfipath{p[]}
\fmfiset{p0}{vpath(__v5,__vc2)}
\fmfiset{p1}{vpath(__vc2,__vc1)}
\fmfiset{p2}{vpath(__v1,__vc1)}
\fmfiset{p3}{vpath(__v6,__vc2)}
\fmfiset{p4}{vpath(__v2,__vc1)}
\fmfiset{p5}{vpath(__v3,__v7)}
\fmfiset{p6}{vpath(__v8,__vc4)}
\fmfiset{p7}{vpath(__vc4,__vc3)}
\fmfiset{p8}{vpath(__v4,__vc3)}
\fmfiset{p9}{vpath(__v10,__vc4)}
\fmfiset{p10}{vpath(__v9,__vc3)}
}

\newcommand{\Wu}{%
\fmftop{v1}
\fmfbottom{v5}
\fmfforce{(0w,h)}{v1}
\fmfforce{(0w,0)}{v5}
\fmffixed{(0.25w,0)}{v1,v2}
\fmffixed{(0.25w,0)}{v2,v3}
\fmffixed{(0.25w,0)}{v3,v4}
\fmffixed{(0.25w,0)}{v4,v9}
\fmffixed{(0.25w,0)}{v5,v6}
\fmffixed{(0.25w,0)}{v6,v7}
\fmffixed{(0.25w,0)}{v7,v8}
\fmffixed{(0.25w,0)}{v8,v10}
\fmf{plain,tension=0.5,right=0.25}{v2,vc1}
\fmf{plain,tension=0.5,left=0.25}{v3,vc1}
  \fmf{plain}{vc1,vc2}
\fmf{plain,tension=0.5,left=0.125}{vc3,vc2}
\fmf{plain,tension=0.5,left=0.25}{v6,vc2}
\fmf{plain,tension=0.5,right=0.25}{v7,vc2}
\fmf{plain}{v1,v5}
\fmf{plain}{v4,v8}
\fmf{plain}{v9,v10}
\fmf{plain,tension=0.5,right=0,width=1mm}{v5,v10}
\fmfposition
\fmfipath{p[]}
\fmfiset{p0}{vpath(__v1,__v5)}
\fmfiset{p1}{vpath(__v6,__vc2)}
\fmfiset{p2}{vpath(__vc2,__vc1)}
\fmfiset{p3}{vpath(__v2,__vc1)}
\fmfiset{p4}{vpath(__v7,__vc2)}
\fmfiset{p5}{vpath(__v3,__vc1)}
\fmfiset{p6}{vpath(__v4,__v8)}
\fmfiset{p7}{vpath(__v9,__v10)}
}

\DeclareMathOperator{\R}{R}

\DeclareMathOperator{\perm}{P}

\numberwithin{equation}{section}
\newlength{\eqoff}
\newlength{\eqofftwo}

\newlength{\unit}
\psset{xunit=\unit,yunit=\unit,runit=\unit}
\newlength{\linew}
\psset{linewidth=\linew}
\begin{document}
\begin{fmffile}{fullgraphs}

\newlength{\mwidth}
\settowidth{\mwidth}{\widthof{$_m$}}
\newlength{\lwidth}
\settowidth{\lwidth}{\widthof{$_l$}}
\newlength{\mldiff}
\setlength{\mldiff}{\mwidth-\lwidth}
\newlength{\fwidth}
\settowidth{\fwidth}{\widthof{$_f$}}
\newlength{\ewidth}
\settowidth{\ewidth}{\widthof{$_e$}}
\newlength{\fediff}
\setlength{\fediff}{\fwidth-\ewidth}

\fmfcmd{
wiggly_len := 2mm;
vardef wiggly expr p_arg =
 save wpp,len;
 numeric wpp,alen;
 wpp = ceiling (pixlen (p_arg, 10) / wiggly_len);
 len=length p_arg;
 for k=0 upto wpp - 1:
  point arctime k/(wpp-1)*arclength(p_arg) of p_arg of p_arg
    {direction arctime k/(wpp-1)*arclength(p_arg) of p_arg of p_arg rotated wiggly_slope} ..
  point  arctime (k+.5)/(wpp-1)*arclength(p_arg) of p_arg of p_arg
 {direction arctime (k+.5)/(wpp-1)*arclength(p_arg) of p_arg of p_arg rotated - wiggly_slope} ..
 endfor
 if cycle p_arg: cycle else: point infinity of p_arg fi
enddef;
}
\fmfcmd{%
marksize=2mm;
def draw_mark(expr p,a) =
  begingroup
    save t,tip,dma,dmb; pair tip,dma,dmb;
    t=arctime a of p;
    tip =marksize*unitvector direction t of p;
    dma =marksize*unitvector direction t of p rotated -45;
    dmb =marksize*unitvector direction t of p rotated 45;
    linejoin:=beveled;
    draw (-.5dma.. .5tip-- -.5dmb) shifted point t of p;
  endgroup
enddef;
style_def derplain expr p =
    save amid;
    amid=.5*arclength p;
    draw_mark(p, amid);
    draw p;
enddef;
def draw_point(expr p,a) =
  begingroup
    save t,tip,dma,dmb,dmo; pair tip,dma,dmb,dmo;
    t=arctime a of p;
    tip =marksize*unitvector direction t of p;
    dma =marksize*unitvector direction t of p rotated -45;
    dmb =marksize*unitvector direction t of p rotated 45;
    dmo =marksize*unitvector direction t of p rotated 90;
    linejoin:=beveled;
    draw (-.5dma.. .5tip-- -.5dmb) shifted point t of p withcolor 0white;
    draw (-.5dmo.. .5dmo) shifted point t of p;
  endgroup
enddef;
style_def derplainpt expr p =
    save amid;
    amid=.5*arclength p;
    draw_point(p, amid);
    draw p;
enddef;
style_def dblderplain expr p =
    save amidm;
    save amidp;
    amidm=.5*arclength p-0.75mm;
    amidp=.5*arclength p+0.75mm;
    draw_mark(p, amidm);
    draw_point(p, amidp);
    draw p;
enddef;
}

\begin{titlepage}
\begin{flushright}
IFUM-945-FT \\
\end{flushright}
\mbox{ }
\vspace{7ex}

\Large
\begin {center}     
{\bf
Five-loop anomalous dimension at critical wrapping order 
in  ${\cal{N}}=4$ SYM}
\end {center}

\renewcommand{\thefootnote}{\fnsymbol{footnote}}

\large
\vspace{1cm}
\centerline{F.\ Fiamberti ${}^{a,b}$, A.\ Santambrogio ${}^b$, 
C.\ Sieg ${}^c$
\footnote[1]{\noindent \tt
francesco.fiamberti@mi.infn.it \\
\hspace*{6.3mm}alberto.santambrogio@mi.infn.it \\ 
\hspace*{6.3mm}csieg@nbi.dk}}
\vspace{4ex}
\normalsize
\begin{center}
\emph{$^a$  Dipartimento di Fisica, Universit\`a degli Studi di Milano, \\
Via Celoria 16, 20133 Milano, Italy}\\
\vspace{0.2cm}
\emph{$^b$ INFN--Sezione di Milano,\\
Via Celoria 16, 20133 Milano, Italy}\\
\vspace{0.2cm}
\emph{$^c$ The Niels Bohr International Academy,\\
The Niels Bohr Institute,\\
Blegdamsvej 17, DK-2100 Copenhagen, Denmark\\
}
\end{center}
\vspace{0.5cm}
\rm
\abstract
\normalsize 
\noindent We compute the anomalous dimension of a length-five operator
at five-loop order in the $SU(2)$ sector of ${\cal{N}}=4$ SYM theory
in the planar limit. This is critical wrapping order at five
loops. The result is obtained perturbatively by means of ${\cal N}=1$
superspace techniques. Our result from perturbation theory confirms
explicitly the formula conjectured in arXiv:0901.4864 for the
five-loop anomalous dimension of twist-three operators. We also
explicitly obtain the same result by employing the recently proposed
Y-system.
\vspace{1cm}
\vfill
\end{titlepage} 

\section{Introduction}

Finding the anomalous dimension of short operators in ${\cal N}=4$ SYM
theory is a subject that received great attention in the recent
past. The computation of the anomalous dimension of long operators is
now an easily accessible task, thanks to the asymptotic Bethe ansatz
techniques~\cite{Minahan:2002ve,Beisert:2003tq,Beisert:2004ry,Beisert:2005fw}. For
short operators, however, a well established approach allowing us to
similarly compute their anomalous dimensions does not exist yet,
because of the so-called wrapping
effects~\cite{Beisert:2004hm,Sieg:2005kd}.

In the last year several approaches on how to manage this problem have
been proposed. One of them is based on a generalization of L\"uscher
formula~\cite{Janik:2007wt,Bajnok:2008bm,Hatsuda:2008na,Bajnok:2008qj,Beccaria:2009eq,Bajnok:2009vm},
while others are based on the
Y-system~\cite{Gromov:2009tv,Gromov:2009bc,Hegedus:2009ky,Gromov:2009zb}
and on the thermodynamic Bethe
ansatz~\cite{Ambjorn:2005wa,Arutyunov:2009zu,Bombardelli:2009ns,Arutyunov:2009ur}.

It is clearly desirable that the anomalous dimensions obtained from
these approaches are tested by means of explicit perturbative standard
field theory calculations. This has been done in a few cases. In
particular, the anomalous dimension of the Konishi operator has been
computed perturbatively up to four
loops~\cite{us,uslong,Velizhanin:2008jd} and there is agreement with
the value obtained with the proposals of~\cite{Bajnok:2008bm} and
of~\cite{Gromov:2009tv}. Moreover, the anomalous dimensions of
twist-two operators at four loops were predicted from L\"uscher
formula approach~\cite{Bajnok:2008qj} and then confirmed from
perturbative field theory~\cite{Velizhanin:2008pc}.

More computations are needed at the perturbative level in order to
check the various proposed approaches. In this paper we perform a new
test: we compute the anomalous dimension of the operator
$\mathrm{tr}(\phi Z\phi ZZ-\phi\phi ZZZ)$ at five-loop order by means
of ${\cal N}=1$ superspace techniques. This is a length-five operator,
so we make a computation at critical wrapping order.\footnote{The
  five-loop anomalous dimension of Konishi operator, computed recently
  in~\cite{Bajnok:2009vm}, would require a calculation one step beyond
  critical wrapping order.} This operator belongs to a supermultiplet
which has a representative also in the $SL(2)$
sector~\cite{Beisert:2005fw}, namely the length-three operator with
two derivatives. The anomalous dimension of this twist-three operator
has been predicted in~\cite{Beccaria:2009eq} from a general formula
conjectured on the basis of the maximal transcendentality
principle~\cite{Kotikov:2002ab}. Our result matches the prediction
of~\cite{Beccaria:2009eq}, thus providing a confirmation from
perturbative field theory of the general formulas proposed there.

To perform our five-loop perturbative calculation we make use of the
same strategy adopted at four loops in our papers~\cite{us,uslong}. We
first construct the complete five-loop asymptotic dilatation operator,
from which we extract the contribution of all non-wrapping graphs by
subtracting the range six contributions. In this way we avoid the
computation of all the graphs with interaction range from one to
five. The explicit Feynman graph computation is then reduced to
the consideration of only wrapping diagrams. We compute them by making use of
${\cal N}=1$ superspace techniques and by taking advantage of all the
cancellations between supergaphs discussed
in~\cite{uslong,Fiamberti:2008sn}. The integrals are then computed
with the Gegenbauer polynomial x-space
technique~\cite{Chetyrkin:1980pr,Kotikov:1995cw,uslong}.

We then also explicitly compare our result with the one obtained by
applying the Y-system proposed in~\cite{Gromov:2009tv,Gromov:2009bc},
and we find agreement.

The paper is organized as follows. In section 2 we build the complete
five-loop asymptotic dilatation operator in the $SU(2)$
sector. Section 3 contains the main result of the paper, namely the
computation of the anomalous dimension of the operator
$\mathrm{tr}(\phi Z\phi ZZ-\phi\phi ZZZ)$ at five loops. The strategy
adopted for this calculation is a straightforward generalization of
the one used at four loops in our previous papers~\cite{us,uslong}. We
then compute the same five-loop anomalous dimension in section 4 by
using the approach introduced in~\cite{Gromov:2009tv,Gromov:2009bc}
and based on the Y-system technique. We then conclude with some final
remarks in section 5. Details about the Feynman diagrams and loop
integrals can be found in the appendices.

\section{The dilatation operator at five loops}

In this section we want to calculate the complete five-loop dilatation
operator in the $SU(2)$ sector. Such operator was already computed
in~\cite{Beisert:2004ry} but only in the case without the dressing
phase~\cite{Beisert:2006ez,Beisert:2007hz,Arutyunov:2004vx,Hernandez:2006tk,Beisert:2006ib}, which according to~\cite{Beisert:2006ez} should have two
relevant components at five loops, $\beta_{2,3,3}$ and
$\beta_{2,3,4}$. We thus have to repeat the steps
of~\cite{Beisert:2004ry} in order to restore the dependence on the
components of the dressing phase. Moreover, we need to find the most
general expression for the dilatation operator by taking similarity
transformations into account. We follow the same strategy adopted
in~\cite{Beisert:2004ry}, and we rely on the assumption of
integrability.

Consider the perturbative expansion in terms of the rescaled `t Hooft
coupling constant
\begin{equation}\label{lambdadef}
\lambda=\frac{g^2N}{(4\pi)^2}
\pnt
\end{equation}
The asymptotic dilatation operator expands as
\begin{equation}
D(\lambda)=Q_2=L+\sum_{k=1}^\infty\lambda^k\,D_k \pnt
\end{equation}
We also assume that a conserved charge $Q_3$ exists, with expansion
\begin{equation}
Q_3(\lambda)=\sum_{k=1}^\infty\lambda^k\,Q_3^{(k)} \pnt
\end{equation}
The dilatation operator and the charge $Q_3$ can be written in a basis
of operators built using permutations
\begin{equation}\label{permstrucdef}
\pthree{a_1}{\dots}{a_n}=\sum_{r=0}^{L-1}\perm_{a_1+r\;a_1+r+1}\cdots
\perm_{a_n+r\;a_n+r+1}
\end{equation}
with range
\begin{equation}\label{nneighbourint}
\R=2+\max_{a_1\dots a_n}-\min_{a_1\dots a_n}\pnt
\end{equation}
Some rules valid in the asymptotic case for the manipulation of these structures can be found in~\cite{Beisert:2005wv}.

The operators $D_k$ and $Q_3^{(k)}$ contain structures with range up
to $k+1$ and $k+2$ respectively.
We write $D_5$ and $Q_3^{(5)}$ as linear combinations of the relevant
basis operators with unknown coefficients, which are fixed by
requiring the following constraints~\cite{Beisert:2004ry}:
\begin{itemize}
\item $D$ and $Q_3$ respectively have even and odd parity
\item $D$ is symmetric, $Q_3$ is antisymmetric,
\item $D$ and $Q_3$ have the right BMN scaling on single-impurity states,
\item $D$ and $Q_3$ are perturbatively commuting up to order $\lambda^6$, i.e.\
\begin{equation}
\sum_{k=1}^5[D_k,Q_3^{(6-k)}]=0 \col
\end{equation}
\item the spectrum of $D$ agrees with the asymptotic Bethe equations
  up to five loops.
\end{itemize}
The components $D_1$ to $D_4$ and thus the dilatation operator up to
four loops are known. The same holds for the charge components
$Q_3^{(1)}$, $Q_3^{(2)}$ and $Q_3^{(3)}$~\cite{Beisert:2004ry,Beisert:2007hz}. 
In our case we also need the expression for $Q_3^{(4)}$, which is not
present in the literature. So we applied the described procedure also
at four loops. This yields $Q_3^{(4)}$, which depends on a single
undetermined coefficient. The result is shown in
Table~\ref{Q3_4}. Simultaneously the procedure also reproduces the
known expression for $D_4$ as a check.

\begin{table}
\footnotesize
\begin{equation*}
\begin{aligned}
Q_3^{\,(4)} &= {}-{}2i \,(373+2\beta_{2,3,3}+\epsilon _{3 a})\,(\{1,2\}-\{2,1\}) \\
&\phantom{{}={}} +2i\,(180+\beta _{2,3,3}+2\epsilon _{3 a})\,(\{1,2,3\}-\{3,2,1\}) \\
&\phantom{{}={}} +i\,(40+3\beta _{2,3,3})\,(\{1,2,4\}-\{1,4,3\}+\{1,3,4\}-\{2,1,4\}) \\
&\phantom{{}={}} +2i\,(\{1,2,5\}-\{1,5,4\}+\{1,4,5\}-\{2,1,5\}) \\
&\phantom{{}={}} -2i\,(57+\epsilon _{3 a})\,(\{1,2,3,4\}-\{4,3,2,1\}) \\
&\phantom{{}={}} +i\,(23-2\beta _{2,3,3}-\epsilon _{3 a})\,(\{1,2,4,3\}-\{1,4,3,2\}+\{2,1,3,4\}-\{3,2,1,4\}) \\
&\phantom{{}={}} +4i\,(8-\beta _{2,3,3})\,(\{1,3,2,4\}-\{2,1,4,3\}) \\
&\phantom{{}={}} -4i\,(\{1,2,3,5\}-\{1,5,4,3\}+\{1,2,4,5\}-\{2,1,5,4\}+\{1,3,4,5\}-\{3,2,1,5\}) \\
&\phantom{{}={}} +i\,(1+\beta _{2,3,3}+\epsilon _{3 a})\,(\{1,3,2,4,3\}-\{2,1,4,3,2\}+\{2,1,3,2,4\}-\{3,2,1,4,3\}) \\
&\phantom{{}={}} +i \,(1+\epsilon _{3 a})\,(\{1,2,3,5,4\}-\{1,5,4,3,2\}+\{2,1,3,4,5\}-\{4,3,2,1,5\}) \\
&\phantom{{}={}} -i\,(7+\epsilon _{3 a})\,(\{1,2,4,3,5\}-\{2,1,5,4,3\}+\{1,3,2,4,5\}-\{3,2,1,5,4\}) \\
&\phantom{{}={}} +2i\,(4+\epsilon _{3 a})\,(\{1,4,3,2,5\}-\{2,1,3,5,4\}) \\
&\phantom{{}={}} +14i\,(\{1,2,3,4,5\}-\{5,4,3,2,1\}) \\
\end{aligned}
\end{equation*}
\normalsize
\caption{Four-loop component of the first higher conserved charge
  $Q_3$ before the application of similarity transformations.}
\label{Q3_4}
\end{table}

After applying the procedure at five loops, some of the coefficients
in $D_5$ remain undetermined. They are related to similarity
transformations and do not enter the spectrum of the asymptotic
dilatation operator~\cite{Beisert:2007hz}. 
Other such coefficients can be found by applying the most general
similarity transformation 
\begin{equation}
\label{similarity}
D\rightarrow D'=e^{-i\chi}D e^{i\chi} \col
\end{equation}
where the generating function $\chi$ can be expanded perturbatively as
\begin{equation}
\chi=\sum_{k=0}^{\infty}\lambda^k \chi_k \pnt
\end{equation}
The components of $D'$ are given explicitly by
\begin{equation}
\label{sim}
\begin{aligned}
D_0'&= D_0 \col \\
D_1'&= D_1 \col \\
D_2'&= D_2 \col \\
D_3'&= D_3 + i\,[D_1,\chi_2] +i\,[D_2,\chi_1] \col \\
D_4'&= D_4 + i\,[D_1,\chi_3] + i\,[D_2,\chi_2] +i\,[D_3,\chi_1]+\frac{1}{2}\big[\chi_1,[D_1,\chi_2]+[D_2,\chi_1]\big] \col \\
D_5'&= D_5 + i\,[D_1,\chi_4] + i\,[D_2,\chi_3] + i\,[D_3,\chi_2] + i\,[D_4,\chi_1] \\ 
&\phantom{{}={} D_5} + \frac{1}{2}\big[\chi_1,[D_3,\chi_1]+[D_2,\chi_2]+[D_1,\chi_3]\big]+\frac{1}{2}\big[\chi_2,[D_1,\chi_2]+[D_2,\chi_1]\big] \\
&\phantom{{}={} D_5} -\frac{i}{6}\Big[\chi_1,\big[\chi_1,[D_1,\chi_2]+[D_2,\chi_1]\big]\Big]
\pnt
\end{aligned}
\end{equation}
To determine $\chi$, we require for consistency that each $\chi_k$ is
writable in terms of operators with range up to $k+1$, so that the
dilatation operator will maintain its range after the transformation.
Therefore, $\chi_0$ is irrelevant being proportional to the identity
$\{\}$, while 
\begin{equation}
\label{chisim1}
\chi_1=\tilde{\epsilon}_1\{1\} \pnt
\end{equation}
To explicitly preserve the invariance of the dilatation operator under
parity, we demand that $\chi$ is parity-invariant.
As far as Hermiticity is concerned, it is possible to choose a $\chi$
which is Hermitian for real values of the coefficients multiplying the
permutation operators, so that the transformed Hamiltonian is still
Hermitian. In general, however, the explicitly computation in a given
renormalization scheme will produce complex values for the similarity
coefficients $\tilde \epsilon_x$.
This can be seen already at four loops: if we consider the most
general Hermitian form for $\chi_2$ and $\chi_3$ as in
~\cite{Beisert:2007hz}
\begin{equation}
\label{chisim23}
\begin{aligned}
\chi_2 &= \tilde{\epsilon}_{2a}(\{1,2\}+\{2,1\}) + \tilde{\epsilon}_{2b}\{1\}\col \\
\chi_3 &= i\,\tilde{\epsilon}_{3a}(\{2,1,3\}-\{1,3,2\})+\tilde{\epsilon}_{3b}(\{1,2,3\}+\{3,2,1\})\\
&\phantom{{}={}}+\tilde{\epsilon}_{3c}\{1,3\}+\tilde{\epsilon}_{3d}(\{1,2\}+\{2,1\}) + \tilde{\epsilon}_{3e}\{1\}\col
\end{aligned}
\end{equation}
the transformed dilatation operator will be in turn Hermitian only for
real values of the $\tilde{\epsilon}_x$ coefficients, but a Feynman
diagram computation in a generic scheme will typically produce complex
values for them~\cite{uslong,Beisert:2007hz}.
The same happens at five loops.
Anyway, the non-Hermiticity of the dilatation operator after a
similarity transformation depends on the renormalization scheme and
does not constitute a problem as explained
in~\cite{Beisert:2007hz}. \\
For $\chi_4$, we look for the most general parity-invariant, Hermitian
operator built using operators with range up to five:
\begin{equation}
\chi_4=\sum_{\alpha\in\{a,b,\ldots,n\}}\tilde{\epsilon}_{4\alpha}\chi_{4\alpha} \col
\end{equation}
we thereby use linear combinations of the permutation structures which
are eigenstates with eigenvalue one under parity transformation and
Hermitian conjugation.
The transformation $\chi_4$ then preserves parity and Hermiticity if
all its coefficients $\tilde{\epsilon}_{4\alpha}$ are real, and
imaginary parts of $\tilde{\epsilon}_{4\alpha}$ are responsible for breaking the
 Hermiticity.
The combinations $\chi_{4\alpha}$ are given in Table~\ref{chi4}. \begin{table}
\begin{equation*}
\begin{aligned}
\label{chisim4}
\chi_{4a}&=\pfour{1}{2}{3}{4}+\pfour{4}{3}{2}{1} \\
\chi_{4b}&=i(\pfour{1}{2}{4}{3}+\pfour{1}{4}{3}{2}-\pfour{2}{1}{3}{4}-\pfour{3}{2}{1}{4}) \\
\chi_{4c}&=\pfour{1}{2}{4}{3}+\pfour{1}{4}{3}{2}+\pfour{2}{1}{3}{4}+\pfour{3}{2}{1}{4} \\
\chi_{4d}&=\pfour{1}{3}{2}{4}+\pfour{2}{1}{4}{3} \\
\chi_{4e}&=\pfour{2}{1}{3}{2} \\
\chi_{4f}&=i(\pthree{1}{2}{4}+\pthree{1}{4}{3}-\pthree{1}{3}{4}-\pthree{2}{1}{4}) \\
\chi_{4g}&=\pthree{1}{2}{4}+\pthree{1}{4}{3}+\pthree{1}{3}{4}+\pthree{2}{1}{4} \\
\end{aligned}
\quad
\begin{aligned}
\chi_{4h}&=i(\pthree{1}{3}{2}-\pthree{2}{1}{3}) \\
\chi_{4i}&=\pthree{1}{3}{2}+\pthree{2}{1}{3}\\
\chi_{4j}&=\pthree{1}{2}{3}+\pthree{3}{2}{1} \\
\chi_{4k}&=\ptwo{1}{2}+\ptwo{2}{1} \\
\chi_{4l}&=\ptwo{1}{4} \\
\chi_{4m}&=\ptwo{1}{3} \\
\chi_{4n}&=\pone{1} \\
\end{aligned}
\end{equation*}
\caption{Components of the generating function}
\label{chi4}
\end{table}
Using~\eqref{sim} we find the most general expression for the
five-loop asymptotic dilatation operator comprehensive of the dressing
phase and similarity coefficients, which is shown in
Table~\ref{D5-perm}. The $\epsilon_x$ coefficients are redefinitions
of the original $\tilde{\epsilon}_x$ ones which simplify the final
result.

\begin{table}
\vspace{-1cm}
\footnotesize
\newcommand{\eqspace}{\phantom{D_5={}}}
\begin{equation*}
\begin{aligned}
D_5&={}
+{}4(1479+14\beta_{2,3,3}-\beta_{2,3,4})\redperm{}
-4(2902+42\beta_{2,3,3}-3\beta_{2,3,4}+32\epsilon_{4h})\redperm{1}\\
&\phantom{{}={}}
-4(-816-32\epsilon_{4h}-13\beta_{2,3,3}+\beta_{2,3,4}+16\epsilon_{4b})(\redperm{1,2}+\redperm{2,1})\\
&\phantom{{}={}}
+2(512+48\beta_{2,3,3}-3\beta_{2,3,4}+4\epsilon_{2a}^2+4\epsilon_{3a}+64\epsilon_{4f}+32\epsilon_{4h})\redperm{1,3}\\
&\phantom{{}={}}
+8(20+\beta_{2,3,3}+2\epsilon_{2a}^2+2\epsilon_{3a}-16\epsilon _{4b}-16\epsilon_{4f})\redperm{1,4}
+4\redperm{1,5}\\
&\phantom{{}={}}
-4(326+16\epsilon_{4h}-\beta_{2,3,3}+2\epsilon_{2a}^2+2\epsilon_{3a}-32\epsilon_{4b})(\redperm{1,2,3}+\redperm{3,2,1})\\
&\phantom{{}={}}
+4(94-15\beta_{2,3,3}+\beta_{2,3,4}+16\epsilon_{4b}-16\epsilon_{4h}+16i\epsilon_{4k})\redperm{1,3,2}\\
&\phantom{{}={}}
+4(94-15\beta_{2,3,3}+\beta_{2,3,4}+16\epsilon_{4b}-16\epsilon_{4h}-16i\epsilon_{4k})\redperm{2,1,3}\\
&\phantom{{}={}}
-8(12+\beta_{2,3,3}+\epsilon _{2a}^2+\epsilon_{3a}-8\epsilon_{4b}+4i\epsilon_{4m})(\redperm{1,3,4}+\redperm{2,1,4})\\
&\phantom{{}={}}
-8(12+\beta_{2,3,3}+\epsilon_{2a}^2+\epsilon _{3a}-8\epsilon_{4b}-4i\epsilon_{4m})(\redperm{1,2,4}+\redperm{1,4,3})\\
&\phantom{{}={}}
-4(1-8i\epsilon_{4l})(\redperm{1,2,5}+\redperm{1,5,4})\\
&\phantom{{}={}}
-4(1+8i\epsilon_{4l})(\redperm{1,4,5}+\redperm{2,1,5})
-8\redperm{1,3,5}\\
&\phantom{{}={}}
-2(40-12\beta_{2,3,3}+\beta_{2,3,4}+4\epsilon_{2a}^2+4\epsilon_{3a}-32\epsilon_{4h})\redperm{2,1,3,2}\\
&\phantom{{}={}}
+8(35+\epsilon_{2a}^2+\epsilon_{3a}-8\epsilon_{4b})(\redperm{1,2,3,4}+\redperm{4,3,2,1})\\
&\phantom{{}={}}
+8(-21+2\beta_{2,3,3}+8\epsilon_{4f}-8\epsilon_{4h})(\redperm{1,3,2,4}+\redperm{2,1,4,3})\\
&\phantom{{}={}}
+4(\epsilon_{2a}^2+\epsilon_{3a}-8(3\epsilon_{4b}+\epsilon_{4f}-\epsilon_{4h}+i\epsilon_{4j}))(\redperm{2,1,3,4}+\redperm{3,2,1,4})\\
&\phantom{{}={}}
+4(\epsilon_{2a}^2+\epsilon_{3a}-8(3\epsilon_{4b}+\epsilon_{4f}-\epsilon_{4h}-i\epsilon_{4j}))(\redperm{1,2,4,3}+\redperm{1,4,3,2})\\
&\phantom{{}={}}
+32(1-2 \epsilon_{4f})(\redperm{1,2,4,5}+\redperm{2,1,5,4})\\
&\phantom{{}={}}
-8(3-8\epsilon_{4f}-8i\epsilon_{4g})\redperm{1,2,5,4}
-8(3-8\epsilon_{4f}+8i\epsilon_{4g})\redperm{2,1,4,5}\\
&\phantom{{}={}}
+2(-2i\epsilon_{2a}-i\epsilon_{3c}+16\epsilon_{4f}+16i\epsilon_{4g})(\redperm{1,2,3,5}+\redperm{1,5,4,3})\\
&\phantom{{}={}}
+2(2i\epsilon_{2a}+i\epsilon_{3c}+16\epsilon_{4f}-16i\epsilon _{4g})(\redperm{1,3,4,5}+\redperm{3,2,1,5})\\
&\phantom{{}={}}
+2(4-2i\epsilon_{2a}-i\epsilon_{3c}-16\epsilon_{4f}-16i\epsilon_{4g})(\redperm{1,4,3,5}+\redperm{2,1,3,5})\\
&\phantom{{}={}}
+2(4+2i\epsilon_{2a}+i\epsilon_{3c}-16\epsilon_{4f}+16i\epsilon_{4g})(\redperm{1,3,2,5}+\redperm{1,3,5,4})\\
&\phantom{{}={}}
+2(10-\beta_{2,3,3}+16\epsilon_{4b}+16i\epsilon_{4e})(\redperm{1,3,2,4,3}+\redperm{2,1,4,3,2})\\
&\phantom{{}={}}
+2(10-\beta_{2,3,3}+16\epsilon_{4b}-16i\epsilon_{4e})(\redperm{2,1,3,2,4}+\redperm{3,2,1,4,3})\\
&\phantom{{}={}}
+4(4+\epsilon _{2a}^2+\epsilon_{3a}-16i\epsilon_{4d})\redperm{2,1,4,3,5}\\
&\phantom{{}={}}
+4(4+\epsilon_{2a}^2+\epsilon_{3a}+16i\epsilon_{4d})\redperm{1,3,2,5,4}\\
&\phantom{{}={}}
+4(2+\epsilon_{2a}^2+2i\epsilon_{2a}+\epsilon_{3a}-i\epsilon_{3b}-8\epsilon_{4b}-8i\epsilon_{4c}+8i\epsilon_{4d})(\redperm{1,2,4,3,5}+\redperm{2,1,5,4,3})\\
&\phantom{{}={}}
+4(2+\epsilon_{2a}^2-2i\epsilon_{2a}+\epsilon_{3a}+i\epsilon_{3b}-8\epsilon_{4b}+8i \epsilon_{4c}-8i\epsilon_{4d})(\redperm{1,3,2,4,5}+\redperm{3,2,1,5,4})\\
&\phantom{{}={}}
+4(2-\epsilon_{2a}^2-\epsilon_{3a}+16\epsilon_{4b}+16i\epsilon_{4c})\redperm{1,2,5,4,3}\\
&\phantom{{}={}}
+4(2-\epsilon_{2a}^2-\epsilon_{3a}+16\epsilon_{4b}-16i\epsilon_{4c})\redperm{3,2,1,4,5}\\
&\phantom{{}={}}
-4(16\epsilon _{4b}+7)(\redperm{1,4,3,2,5}+\redperm{2,1,3,5,4})\\
&\phantom{{}={}}
-4(\epsilon_{2a}^2+\epsilon_{3a}-8i\epsilon_{4a}-8\epsilon_{4b})(\redperm{1,2,3,5,4}+\redperm{1,5,4,3,2})\\
&\phantom{{}={}}
-4(\epsilon_{2a}^2+\epsilon_{3a}+8i\epsilon_{4a}-8\epsilon_{4b})(\redperm{2,1,3,4,5}+\redperm{4,3,2,1,5})\\
&\phantom{{}={}}
-28(\redperm{1,2,3,4,5}+\redperm{5,4,3,2,1})
\end{aligned}
\end{equation*}
\normalsize
\caption{Asymptotic five-loop dilatation operator in the permutation basis}
\label{D5-perm}
\end{table}

\section{Computation of the anomalous dimension at five loops}

In this section we are going to describe the computation of the
anomalous dimension of the operator $\mathrm{tr}(\phi Z\phi
ZZ-\phi\phi ZZZ)$ at five loops. As already mentioned, the strategy
which we adopt here is the same as the one used in our previous
papers~\cite{us,uslong}. We first compute the contribution from the
diagrams with range from one to five by taking advantage of the
asymptotic five-loop dilatation operator $D_5$ computed in the
previous section. Indeed, we exploit the fact that these diagrams are
relevant also in the asymptotic case, and therefore all the
information on them is encoded in $D_5$. However, in addition to these
diagrams the dilatation operator also gets contributions from
range-six diagrams, which must be subtracted. The second step will be
the explicit calculation of the wrapping contributions by means of
${\cal N}=1$ superspace techniques~\cite{Gates:1983nr}.

\subsection{Subtraction of range-six diagrams from the asymptotic
  dilatation operator}

Since our procedure is a straightforward generalization of the one
used in~\cite{us,uslong}, we do not repeat here all the steps, and we
refer the reader to those papers. First of all, we need to write the
five-loop Hamiltonian in terms of chiral structures, which are
directly related to the chiral structures of the underlying Feynman
supergraphs (see~\cite{uslong} for a discussion of these
functions). The transformation between the chiral structure basis and
the one made of permutation operators is easily performed by means of
the rules shown in Table~\ref{chistruc}.

\begin{table}
\begin{equation*}\begin{aligned}
\chi(a,b,c,d,e)&=
-\pid+5\pone{1}-\ptwo{a}{b}-
\ptwo{a}{c}-\ptwo{a}{d}-\ptwo{a}{e}-\ptwo{b}{c}-\ptwo{b}{d}-\ptwo{b}{e} \\
&\phantom{{}={}}
-\ptwo{c}{d}-\ptwo{c}{e}-\ptwo{d}{e}+\pthree{a}{b}{c}+
\pthree{a}{b}{d}+\pthree{a}{b}{e}+\pthree{a}{c}{d} \\
&\phantom{{}={}}
+\pthree{a}{c}{e}+
\pthree{a}{d}{e}+\pthree{b}{c}{d}+\pthree{b}{c}{e}+\pthree{b}{d}{e}+
\pthree{c}{d}{e} \\
&\phantom{{}={}}
-\pfour{a}{b}{c}{d}-\pfour{a}{b}{c}{e}-\pfour{a}{b}{d}{e}-
\pfour{a}{c}{d}{e}-\pfour{b}{c}{d}{e} \\
&\phantom{{}={}}
+\{a,b,c,d,e\}\col \\
\chi(a,b,c,d)&=\pid-4\pone1
+\ptwo ab+\ptwo{a}{c}+\ptwo ad+\ptwo bc+\ptwo bd+\ptwo cd\\
&\phantom{{}={}}
-\pthree abc-\pthree abd-\pthree acd-\pthree bcd
+\pfour abcd\col\\
\chi(a,b,c)&=-\pid+3\pone1
-\ptwo ab-\ptwo ac-\ptwo bc+\pthree abc\col\\
\chi(a,b)&=\pid-2\pone1+\ptwo ab\col\\
\chi(1)&=-\pid+\pone1\col\\
\chi() &=\pid
\pnt
\end{aligned}
\end{equation*}
\begin{equation*}
\begin{aligned}
\{a,b,c,d,e\} &=
\chi (a,b,c,d,e)+\chi (b,c,d,e)+\chi (a,c,d,e)+
\chi (a,b,d,e)+\chi (a,b,c,e) \\
&\phantom{{}={}}
+\chi (a,b,c,d)+\chi (c,d,e)+
\chi (b,d,e)+\chi (b,c,e)+\chi (b,c,d)+\chi (a,d,e) \\
&\phantom{{}={}}
+\chi (a,c,e)+\chi (a,c,d)+\chi (a,b,e)+\chi (a,b,d)+
\chi (a,b,c)+\chi (d,e) \\
&\phantom{{}={}}
+\chi (c,e)+\chi (c,d)+\chi (b,e)+
\chi (b,d)+\chi (b,c)+\chi (a,e)+\chi (a,d) \\
&\phantom{{}={}}
+\chi (a,c)+
\chi (a,b)+5\chi (1)+\chi ()\col \\
\pfour abcd &= \chi(a,b,c,d) + \chi(a,b,c) + \chi(a,b,d) + \chi(a,c,d)
+ \chi(b,c,d)\\
&\phantom{{}={}}
+ \chi(a,b) + \chi(a,c) + \chi(a,d) + \chi(b,c) + \chi(b,d) + \chi(c,d)\\
&\phantom{{}={}}
+4 \chi(1) + \chi()\col\\
\pthree abc &= \chi(a,b,c) + \chi(a,b) + \chi(a,c) + \chi(b,c) + 3\chi(1)
+\chi()\col\\
\ptwo ab &= \chi(a,b) + 2\chi(1) + \chi()\col\\
\pone1 &= \chi(1) + \chi()\col\\
\pid &= \chi()\pnt
\end{aligned}
\end{equation*}
\caption{Rules for the conversion between permutation operators and
  chiral structures}
\label{chistruc}
\end{table}
The dilatation operator written in terms of chiral structures is given
in Table~\ref{D5}. As discussed in the previous section, the
$\epsilon_x$ coefficients in Table~\ref{D5} are related to similarity
transformations. Note that even if these coefficients (which are
scheme-dependent) do not enter the asymptotic spectrum, some of them
may (and actually do) enter the spectrum of short operators once the
subtraction of range-six contributions is performed. 
From computations at three and four loops we found in our scheme~\cite{uslong}
\begin{equation}
\epsilon_{3a}=-4\col\quad \epsilon_{3b}=-\frac{4}{3}i\col\quad \epsilon_{3c}=\frac{4}{3}i\col\quad \epsilon_{2a}=-\frac{i}{2} \col
\end{equation}
while all the other $\epsilon_x$ which are relevant here can be computed from range-six diagrams, as explained in appendix~\ref{app:range6}.

\begin{table}[p!]
\vspace{-1cm}
\footnotesize
\begin{equation*}
\begin{aligned}
D_5&={}
-1960\redchi{1}
+1568(\redchi{1,2}+\redchi{2,1})
+16(\epsilon_{2a}^2+\epsilon_{3a}+8\epsilon_{4f}-8\epsilon_{4h}+\beta_{2,3,3}-40)\redchi{1,3}\\
&\phantom{{}={}}
+16(\epsilon_{2a}^2+\epsilon_{3a}-16\epsilon_{4b}-8\epsilon_{4f}+4)\redchi{1,4}
-4\redchi{1,5}
-784(\redchi{1,2,3}+\redchi{3,2,1})\\
&\phantom{{}={}}
+2(64-8\beta_{2,3,3}+\beta_{2,3,4}
+8\epsilon_{2a}^2+4i\epsilon_{2a}+8\epsilon_{3a}+2i\epsilon_{3c}\\
&\phantom{{}={}+2(}
-32(\epsilon_{4b}+\epsilon_{4h})+32i(\epsilon_{4a}+\epsilon_{4c}+\epsilon_{4d}+\epsilon_{4e}+\epsilon_{4g}+\epsilon_{4j}+\epsilon_{4k}))\redchi{1,3,2}\\
&\phantom{{}={}}
+2(64-8\beta_{2,3,3}+\beta_{2,3,4}
+8\epsilon_{2a}^2-4i\epsilon_{2a}+8\epsilon_{3a}-2i\epsilon_{3c}\\
&\phantom{{}={}+2(}
-32(\epsilon_{4b}+\epsilon_{4h})-32i(\epsilon_{4a}+\epsilon_{4c}+\epsilon_{4d}+\epsilon_{4e}+\epsilon_{4g}+\epsilon_{4j}+\epsilon_{4k}))\redchi{2,1,3}\\
&\phantom{{}={}}
+2(30+\beta_{2,3,3}
+4\epsilon_{2a}^2-4i\epsilon_{2a}+4\epsilon_{3a}+4i\epsilon_{3b}+2i\epsilon_{3c}\\
&\phantom{{}={}+2(}
-48\epsilon_{4b}-16i(2\epsilon_{4a}+2\epsilon_{4c}+2\epsilon_{4d}+\epsilon_{4e}+2\epsilon_{4g}+2\epsilon_{4j}+\epsilon_{4m}))(\redchi{1,3,4}+\redchi{2,1,4})\\
&\phantom{{}={}}
+2(30+\beta_{2,3,3}
+4\epsilon_{2a}^2+4i\epsilon_{2a}+4\epsilon_{3a}-4i\epsilon_{3b}-2i\epsilon_{3c}\\
&\phantom{{}={}+2(}
-48\epsilon_{4b}+16i(2\epsilon_{4a}+2\epsilon_{4c}+2\epsilon_{4d}+\epsilon_{4e}+2\epsilon_{4g}+2\epsilon_{4j}+\epsilon_{4m}))(\redchi{1,2,4}+\redchi{1,4,3})\\
&\phantom{{}={}}
-4(1-8i(2\epsilon_{4a}+2\epsilon_{4c}+2\epsilon_{4d}+4\epsilon_{4g}+\epsilon_{4l}))(\redchi{1,2,5}+\redchi{1,5,4})\\
&\phantom{{}={}}
-4(1+8i(2\epsilon_{4a}+2\epsilon_{4c}+2\epsilon_{4d}+4\epsilon_{4g}+\epsilon_{4l}))(\redchi{1,4,5}+\redchi{2,1,5})\\
&\phantom{{}={}}
-8\redchi{1,3,5}\\
&\phantom{{}={}}
+2(8\beta_{2,3,3}-\beta_{2,3,4}-4\epsilon_{2a}^2-4\epsilon_{3a}+64\epsilon_{4b}+32\epsilon_{4h})\redchi{2,1,3,2}\\
&\phantom{{}={}}
+224(\redchi{1,2,3,4}+\redchi{4,3,2,1})\\
&\phantom{{}={}}
-4(20-3\beta_{2,3,3}-4\epsilon_{2a}^2-4\epsilon_{3a}-16\epsilon_{4f}+16\epsilon_{4h})(\redchi{1,3,2,4}+\redchi{2,1,4,3})\\
&\phantom{{}={}}
+2(4-\beta_{2,3,3}-4i\epsilon_{2a}+2i\epsilon_{3b}\\
&\phantom{{}={}+2(}
-32\epsilon_{4b}-16\epsilon_{4f}+16\epsilon_{4h}-16i(\epsilon_{4a}+\epsilon_{4c}+\epsilon_{4d}+\epsilon_{4e}+\epsilon_{4j}))(\redchi{2,1,3,4}+\redchi{3,2,1,4})\\
&\phantom{{}={}}
+2(4-\beta_{2,3,3}+4i\epsilon_{2a}-2i\epsilon_{3b}\\
&\phantom{{}={}+2(}
-32\epsilon_{4b}-16\epsilon_{4f}+16\epsilon_{4h}+16i(\epsilon_{4a}+\epsilon_{4c}+\epsilon_{4d}+\epsilon_{4e}+\epsilon_{4j}))(\redchi{1,2,4,3}+\redchi{1,4,3,2})\\
&\phantom{{}={}}
-8(1-\epsilon_{2a}^2-\epsilon_{3a}+16\epsilon_{4b}+8\epsilon_{4f})(\redchi{1,2,4,5}+\redchi{2,1,5,4})\\
&\phantom{{}={}}
-8(\epsilon_{2a}^2+\epsilon_{3a}
-16\epsilon_{4b}-8\epsilon_{4f}-8i(\epsilon_{4a}+\epsilon_{4c}+\epsilon_{4d}+\epsilon_{4g}))\redchi{1,2,5,4}\\
&\phantom{{}={}}
-8(\epsilon_{2a}^2+\epsilon_{3a}
-16\epsilon_{4b}-8\epsilon_{4f}+8i(\epsilon_{4a}+\epsilon_{4c}+\epsilon_{4d}+\epsilon_{4g}))\redchi{2,1,4,5}\\
&\phantom{{}={}}
-2(6
+2\epsilon_{2a}^2-2i\epsilon_{2a}+2\epsilon_{3a}+2i\epsilon_{3b}+i\epsilon_{3c}\\
&\phantom{{}={}-2(}
-32\epsilon_{4b}-16\epsilon_{4f}
-16i(\epsilon_{4a}+\epsilon_{4c}+\epsilon_{4d}+\epsilon_{4g}))(\redchi{1,2,3,5}+\redchi{1,5,4,3})\\
&\phantom{{}={}}
-2(6
+2\epsilon_{2a}^2+2i\epsilon_{2a}+2\epsilon_{3a}-2i\epsilon_{3b}-i\epsilon_{3c}\\
&\phantom{{}={}-2(}
-32\epsilon_{4b}-16\epsilon_{4f}
+16i(\epsilon_{4a}+\epsilon_{4c}+\epsilon_{4d}+\epsilon_{4g}))(\redchi{1,3,4,5}+\redchi{3,2,1,5})\\
&\phantom{{}={}}
+2(2
+2\epsilon_{2a}^2+2i\epsilon_{2a}+2\epsilon_{3a}-2i\epsilon_{3b}-i\epsilon_{3c}\\
&\phantom{{}={}+2(}
-32\epsilon_{4b}-16\epsilon_{4f}
-16i(\epsilon_{4a}+\epsilon_{4c}+\epsilon_{4d}+\epsilon_{4g}))(\redchi{1,4,3,5}+\redchi{2,1,3,5})\\
&\phantom{{}={}}
+2(2
+2\epsilon_{2a}^2-2i\epsilon_{2a}+2\epsilon_{3a}+2i\epsilon_{3b}+i\epsilon_{3c}\\
&\phantom{{}={}+2(}
-32\epsilon_{4b}-16\epsilon_{4f}
+16i(\epsilon_{4a}+\epsilon_{4c}+\epsilon_{4d}+\epsilon_{4g}))(\redchi{1,3,2,5}+\redchi{1,3,5,4})\\
&\phantom{{}={}}
+2(10-\beta_{2,3,3}+16\epsilon_{4b}+16i\epsilon_{4e})(\redchi{1,3,2,4,3}+\redchi{2,1,4,3,2})\\
&\phantom{{}={}}
+2(10-\beta_{2,3,3}+16\epsilon_{4b}-16i\epsilon_{4e})(\redchi{2,1,3,2,4}+\redchi{3,2,1,4,3})\\
&\phantom{{}={}}
+4(4+\epsilon_{2a}^2+\epsilon_{3a}-16i\epsilon_{4d})\redchi{2,1,4,3,5}\\
&\phantom{{}={}}
+4(4+\epsilon_{2a}^2+\epsilon_{3a}+16i\epsilon_{4d})\redchi{1,3,2,5,4}\\
&\phantom{{}={}}
+4(2+\epsilon_{2a}^2+2i\epsilon_{2a}+\epsilon_{3a}-i\epsilon_{3b}-8\epsilon_{4b}-8i(\epsilon_{4c}-\epsilon_{4d}))(\redchi{1,2,4,3,5}+\redchi{2,1,5,4,3})\\
&\phantom{{}={}}
+4(2+\epsilon_{2a}^2-2i\epsilon_{2a}+\epsilon_{3a}+i\epsilon_{3b}-8\epsilon_{4b}+8i(\epsilon_{4c}-\epsilon_{4d}))(\redchi{1,3,2,4,5}+\redchi{3,2,1,5,4})\\
&\phantom{{}={}}
+4(2-\epsilon_{2a}^2-\epsilon_{3a}+16\epsilon_{4b}+16i\epsilon_{4c})\redchi{1,2,5,4,3}\\
&\phantom{{}={}}
+4(2-\epsilon_{2a}^2-\epsilon_{3a}+16\epsilon_{4b}-16i\epsilon_{4c})\redchi{3,2,1,4,5}\\
&\phantom{{}={}}
-4(7+16\epsilon _{4b})(\redchi{1,4,3,2,5}+\redchi{2,1,3,5,4})\\
&\phantom{{}={}}
-4(\epsilon _{2a}^2+\epsilon_{3a}-8i\epsilon_{4a}-8\epsilon_{4b})(\redchi{1,2,3,5,4}+\redchi{1,5,4,3,2})\\
&\phantom{{}={}}
-4(\epsilon_{2a}^2+\epsilon_{3a}+8i\epsilon_{4a}-8\epsilon_{4b})(\redchi{2,1,3,4,5}+\redchi{4,3,2,1,5})\\
&\phantom{{}={}}
-28(\redchi{1,2,3,4,5}+\redchi{5,4,3,2,1})
\end{aligned}
\end{equation*}
\normalsize
\caption{Asymptotic five-loop dilatation operator in terms of chiral
  structures}
\label{D5}
\end{table}

Now we can subtract from $D_5$ the contributions of range-six
diagrams. As in the four-loop case, these can be divided into two
classes:
\begin{itemize}
\item the diagrams whose chiral structure has range six can be
  subtracted simply by erasing the corresponding structures from
  $D_5$,
\item the diagrams with a chiral structure of range not greater than
  five, which become range-six because of vector interactions, sum up
  to zero thanks to the general argument described in~\cite{uslong}.
\end{itemize}
So we obtain the five-loop subtracted dilatation operator, shown in
Table~\ref{D5-sub}, simply by removing all the range-six chiral
structures from the full asymptotic operator of Table~\ref{D5}.
\begin{table}[t]
\newcommand{\eqsubspace}{\phantom{D_5^{\mathrm{sub}}={}}}
\footnotesize
\begin{equation*}
\begin{aligned}
D_5^{\mathrm{sub}}&={}
-1960\redchi{1}
+1568(\redchi{1,2}+\redchi{2,1})
+16(\epsilon_{2a}^2+\epsilon_{3a}+8\epsilon_{4f}-8\epsilon_{4h}+\beta_{2,3,3}-40)\redchi{1,3}\\
&\phantom{{}={}}
+16(\epsilon_{2a}^2+\epsilon_{3a}-16\epsilon_{4b}-8\epsilon_{4f}+4)\redchi{1,4}
-784(\redchi{1,2,3}+\redchi{3,2,1})\\
&\phantom{{}={}}
+2(64-8\beta_{2,3,3}+\beta_{2,3,4}
+8\epsilon_{2a}^2+4i\epsilon_{2a}+8\epsilon_{3a}+2i\epsilon_{3c}\\
&\phantom{{}={}+2(}
-32(\epsilon_{4b}+\epsilon_{4h})+32i(\epsilon_{4a}+\epsilon_{4c}+\epsilon_{4d}+\epsilon_{4e}+\epsilon_{4g}+\epsilon_{4j}+\epsilon_{4k}))\redchi{1,3,2}\\
&\phantom{{}={}}
+2(64-8\beta_{2,3,3}+\beta_{2,3,4}
+8\epsilon_{2a}^2-4i\epsilon_{2a}+8\epsilon_{3a}-2i\epsilon_{3c}\\
&\phantom{{}={}+2(}
-32(\epsilon_{4b}+\epsilon_{4h})-32i(\epsilon_{4a}+\epsilon_{4c}+\epsilon_{4d}+\epsilon_{4e}+\epsilon_{4g}+\epsilon_{4j}+\epsilon_{4k}))\redchi{2,1,3}\\
&\phantom{{}={}}
+2(30+\beta_{2,3,3}
+4\epsilon_{2a}^2-4i\epsilon_{2a}+4\epsilon_{3a}+4i\epsilon_{3b}+2i\epsilon_{3c}\\
&\phantom{{}={}+2(}
-48\epsilon_{4b}-16i(2\epsilon_{4a}+2\epsilon_{4c}+2\epsilon_{4d}+\epsilon_{4e}+2\epsilon_{4g}+2\epsilon_{4j}+\epsilon_{4m}))(\redchi{1,3,4}+\redchi{2,1,4})\\
&\phantom{{}={}}
+2(30+\beta_{2,3,3}
+4\epsilon_{2a}^2+4i\epsilon_{2a}+4\epsilon_{3a}-4i\epsilon_{3b}-2i\epsilon_{3c}\\
&\phantom{{}={}+2(}
-48\epsilon_{4b}+16i(2\epsilon_{4a}+2\epsilon_{4c}+2\epsilon_{4d}+\epsilon_{4e}+2\epsilon_{4g}+2\epsilon_{4j}+\epsilon_{4m}))(\redchi{1,2,4}+\redchi{1,4,3})\\
&\phantom{{}={}}
+2(8\beta_{2,3,3}-\beta_{2,3,4}-4\epsilon_{2a}^2-4\epsilon_{3a}+64\epsilon_{4b}+32\epsilon_{4h})\redchi{2,1,3,2}\\
&\phantom{{}={}}
+224(\redchi{1,2,3,4}+\redchi{4,3,2,1})\\
&\phantom{{}={}}
-4(20-3\beta_{2,3,3}-4\epsilon_{2a}^2-4\epsilon_{3a}-16\epsilon_{4f}+16\epsilon_{4h})(\redchi{1,3,2,4}+\redchi{2,1,4,3})\\
&\phantom{{}={}}
+2(4-\beta_{2,3,3}-4i\epsilon_{2a}+2i\epsilon_{3b}\\
&\phantom{{}={}+2(}
-32\epsilon_{4b}-16\epsilon_{4f}+16\epsilon_{4h}-16i(\epsilon_{4a}+\epsilon_{4c}+\epsilon_{4d}+\epsilon_{4e}+\epsilon_{4j}))(\redchi{2,1,3,4}+\redchi{3,2,1,4})\\
&\phantom{{}={}}
+2(4-\beta_{2,3,3}+4i\epsilon_{2a}-2i\epsilon_{3b}\\
&\phantom{{}={}+2(}
-32\epsilon_{4b}-16\epsilon_{4f}+16\epsilon_{4h}+16i(\epsilon_{4a}+\epsilon_{4c}+\epsilon_{4d}+\epsilon_{4e}+\epsilon_{4j}))(\redchi{1,2,4,3}+\redchi{1,4,3,2})\\
&\phantom{{}={}}
+2(10-\beta_{2,3,3}+16\epsilon_{4b}+16i\epsilon_{4e})(\redchi{1,3,2,4,3}+\redchi{2,1,4,3,2})\\
&\phantom{{}={}}
+2(10-\beta_{2,3,3}+16\epsilon_{4b}-16i\epsilon_{4e})(\redchi{2,1,3,2,4}+\redchi{3,2,1,4,3})\\
\end{aligned}
\end{equation*}
\normalsize
\caption{Subtracted five-loop dilatation operator}
\label{D5-sub}
\end{table}

Now we can apply this operator to the two length-five states of the
$SU(2)$ sector, which mix under renormalization:
\begin{equation}
\label{Opbasis}
\mathcal{O}_1=\mathrm{tr}(\phi Z\phi ZZ)\col\qquad\mathcal{O}_2=\mathrm{tr}(\phi\phi ZZZ) \pnt
\end{equation}
All the short-range chiral structures have an action on these states
which is proportional to the mixing matrix 
\begin{equation}
\label{mixingM}
M=
\left(
\begin{array}{cc}
1 & -1 \\
-1 & 1
\end{array}
\right) \pnt
\end{equation}
After the range-six subtraction, the matrix expression of the
subtracted dilatation operator on the length-five subsector still
depends on a subset of the $\epsilon_x$ coefficients (more precisely,
it depends on $\epsilon_{4b}$, $\epsilon_{4f}$). Using the values
given in appendix~\ref{app:range6}, we find
\begin{equation}
D_5^{\mathrm{sub}}\rightarrow 2(1665+104\beta_{2,3,3}-8\beta_{2,3,4})M \pnt
\end{equation}
The $\beta_{2,3,3}$ component of the dressing phase is known to be
equal to $4\zeta(3)$~\cite{Beisert:2007hz}, while for $\beta_{2,3,4}$
we use the value $-40\zeta(5)$ conjectured in~\cite{Beisert:2006ez}.
So we have
\begin{equation}
\label{D5-sub-mat}
D_5^{\mathrm{sub}}\rightarrow 2(1665+416\zeta(3)+320\zeta(5))M \pnt
\end{equation}

\subsection{Wrapping diagrams}

We now consider the contribution from wrapping diagrams. First of all,
we must list all the possible wrapping graphs and classify them
according to their chiral structures. Then, thanks to the general
argument described in~\cite{uslong}, several pairs of diagrams can be
immediately cancelled. For the remaining diagrams, we apply the
standard D-algebra procedure and obtain momentum integrals which can
be computed using the
GPXT~\cite{Chetyrkin:1980pr,Kotikov:1995cw,uslong}. The relevant
diagrams after pair cancellations are listed in
appendix~\ref{app:wrapping}.

There are three completely chiral wrapping structures, shown in
figure~\ref{wrap-chiral}.
By identifying the sixth and the first lines of the diagrams, these
structures can be written as $\chi(2,1,3,4,5)$, $\chi(1,2,3,4,5)$ and
$\chi(1,3,2,5,4)$, respectively.
All the other wrapping diagrams can be obtained from range-five graphs
by adding a wrapping vector interaction, and so have chiral structures
of range not greater than five.

A minimal set of \emph{independent} chiral structures, with up to four
loops and range less than or equal to five, can be chosen as
$\chi(2,4,1,3)$, $\chi(3,2,1,4)$, $\chi(1,2,3,4)$, $\chi(1,4,3,2)$,
$\chi(1,3,2)$, $\chi(2,1,3)$, $\chi(1,2,3)$, $\chi(2,1,4)$,
$\chi(2,1)$, $\chi(1,4)$ and $\chi(1)$. All the other structures with
the required range, which appear in the expansion of the subtracted
dilatation operator, are either a reflection or simply a different way
of writing one element of the minimal set. In particular, when acting
on a length-five state, $\chi(1,2,4)$ and $\chi(1,3)$ are equivalent
to $\chi(2,1,4)$ and $\chi(1,4)$, respectively. The wrapping diagrams
for all the independent structures, together with the results of
D-algebra and color factors for the length-five states, are shown in
Figs.~\ref{wrap-2413}-\ref{wrap-14}. For non-symmetric structures, the
corresponding reflection is indicated.

We now collect all the contributions to find the leading wrapping
correction to the dilatation operator:
\begin{equation}
\begin{aligned}
\label{D5w}
D_5^\text{w}
=-10\Big(&{}
-{}\Big(\frac{1}{6}-\frac{4}{5}\zeta(3)\Big) (\redchi{2,1,3,4,5}+\redchi{4,5,3,2,1}) \\
&
+\Big(\frac{14}{5}-4 \zeta(5)\Big) (\redchi{1,2,3,4,5}+\redchi{5,4,3,2,1}) \\
&
-\Big(\frac{19}{30}-\frac{4}{5} \zeta(3)\Big) (\redchi{1,3,2,5,4}+\redchi{5,3,4,1,2}) \\
&
-\Big(\frac{1}{3}+\frac{12}{5} \zeta(3)-4 \zeta(5)\Big) (\redchi{2,4,1,3}+\redchi{1,3,2,4}) \\
&
+\Big(\frac{1}{3}-\frac{12}{5} \zeta(3)+4 \zeta(5)\Big) (\redchi{3,2,1,4}+\redchi{2,1,3,4}) \\
&
+(8 \zeta(5)-14 \zeta(7)) (\redchi{1,2,3,4}+\redchi{4,3,2,1}) \\
&
-\Big(\frac{2}{5}+\frac{12}{5} \zeta(3)-4 \zeta(5)\Big) (\redchi{1,4,3,2}+\redchi{1,2,4,3}) \\
&
+\Big(\frac{13}{10}+\frac{8}{5} \zeta(3)\Big) \redchi{1,3,2}
+\Big(\frac{19}{10}+\frac{8}{5} \zeta(3)\Big) \redchi{2,1,3} \\ 
&
+\Big(\frac{18}{5}+\frac{44}{5} \zeta(3)-12 \zeta(5)\Big) (\redchi{2,1,4}+\redchi{1,3,4}) \\
&
-(8 \zeta(5)-14 \zeta(7)) (\redchi{2,1}+\redchi{1,2}) \\
&
-\Big(\frac{18}{5}+8 \zeta(3)+8 \zeta(5)-28 \zeta(7)\Big) \redchi{1,4}
+8 \zeta(5) \redchi{1}
\Big)
\pnt
\end{aligned}
\end{equation}
In the basis~\eqref{Opbasis} this expression reads
\begin{equation}
D_5^{\mathrm{w}}\rightarrow 2(1-128\zeta(3)+640\zeta(5)-560\zeta(7))M \pnt
\end{equation}
The correct five-loop dilatation operator for the length-five states
is obtained by adding this wrapping contribution to the subtracted
operator of Table~\ref{D5-sub}. The matrix form of the result is
\begin{equation}
D_4^\text{sub}+D_4^\text{w}\rightarrow 4(833+144\zeta(3)+480\zeta(5)-280\zeta(7))M
\pnt
\end{equation}
Since the asymptotic dilatation operator in the $SU(2)$ subsector is
proportional to the mixing matrix~\eqref{mixingM} also at all the
lower loop orders, the five-loop part of the anomalous dimension of
the non-protected eigenstate is simply the non-zero eigenvalue of
$D_5^\text{sub}+D_5^\text{w}$:
\begin{equation}
\gamma_5=6664+1152\zeta(3)+3840\zeta(5)-2240\zeta(7) \pnt
\end{equation}
Including the lower orders, the anomalous dimension up to five loops thus reads
\begin{equation}
\gamma=8\lambda-24\lambda^2+136\lambda^3-8[115+16\zeta(3)]\lambda^4
+[6664+1152\zeta(3)+3840\zeta(5)-2240\zeta(7)]\lambda^5 \pnt
\end{equation}
This result coincides with the one presented in~\cite{Beccaria:2009eq}
for the anomalous dimension of the twist-three operator in the $SL(2)$
sector. This last operator and our operator $\mathrm{tr}(\phi Z\phi
ZZ-\phi\phi ZZZ)$ belong indeed to the same
supermultiplet.\footnote{All two-impurity states belong to
  supermultiplets which have representatives both in $SU(2)$ and
  $SL(2)$~\cite{Beisert:2005fw}. In particular, the two-impurity,
  length-$L$ subsector of $SU(2)$ is mapped onto the two-impurity,
  length-$(L-2)$ subsector of $SL(2)$. So in our case we have to
  consider $L=3$.} Our result gives then a direct field theoretical
confirmation of the conjectures and assumptions made
in~\cite{Beccaria:2009eq}.

\section{Computation with the Y-system}

In this section we want to compare the result of our direct,
field-theoretical computation with the value of the anomalous
dimension which is obtained using the Y-system
technique~\cite{Gromov:2009tv,Gromov:2009bc,Bombardelli:2009ns} when
applied to the twist-three operator of the $SL(2)$ sector.

To compute the leading wrapping correction to the anomalous dimension,
we extend to five loops the explicit computation
of~\cite{Gromov:2009tv} for the four-loop case. In order to restore
the required dependence of $Y_{a,0}^*(u)$ on $L$ and on the Bethe root
$u_{4,1}$, we need to repeat all the steps of that computation. 

We start from the general expression for $Y_{a,0}^*$~\cite{Gromov:2009tv}:
\begin{equation}
\label{Ya0-general}
Y_{a,0}^*(u)=\Big(\frac{x^{[-a]}}{x^{[+a]}}\Big)^L\frac{\phi^{[-a]}}{\phi^{[+a]}}T_{a,-1}^L T_{a,1}^R \col
\end{equation}
where $f^{[\pm a]}(u)=f(u\pm i a/2)$, $x$ is a function of $u$ defined
by $u/\sqrt{\lambda}=x+1/x$, $\phi$ is a fixed function of $u$ whose
expression is given in~\cite{Gromov:2009tv} and $T_{a,-1}^L$ and
$T_{a,1}^R$ are the transfer matrix eigenvalues of anti-symmetric
irreducible representations of the $SU(2\vert 2)_L$ and $SU(2\vert
2)_R$ subgroups of the full $SU(2,2\vert 4)$ symmetry.\\
Since for a state in the $SL(2)$ sector the only non-zero Bethe roots
are those of type $u_{4,j}$, at leading order the term involving the
function $\phi$ simplifies. In the notation of~\cite{Gromov:2009tv} it
becomes
\begin{equation}
\frac{\phi^{[-a]}}{\phi^{[+a]}}=\frac{B^{[+a](+)}R^{[-a](-)}}{B^{[-a](-)}R^{[+a](+)}} \pnt
\end{equation}
The contribution from the first two factors of~\eqref{Ya0-general} hence yields
\begin{equation}
\Big(\frac{x^{[-a]}}{x^{[+a]}}\Big)^L\frac{\phi^{[-a]}}{\phi^{[+a]}}\rightarrow9\frac{2^{2L}\lambda^{L}(4u_{4,1}^2+1)^2}{(a^2+4u^2)^L\tilde{y}_{-a}(u)} \col
\end{equation}
where
\begin{equation}
\tilde{y}_a(u)=9[((1-a)^2+4u^2)^2+8u_{4,1}^2((1-a)^2-4u^2+2u_{4,1}^2)] \pnt
\end{equation}
The action of the two $SU(2\vert 2)$ subgroups of the full symmetry
group is the same on the $SL(2)$ sector and therefore $T_{a,-1}^L$ and
$T_{a,1}^R$ can be computed from the same generating functional
\begin{equation}
\mathcal{W}=\Big[1-\frac{B^{+(+)}R^{-(+)}}{B^{+(-)}R^{-(-)}}D\Big]\Big[1-\frac{R^{-(+)}}{R^{-(-)}}D\Big]^{-2}[1-D]\col\qquad D=e^{-i\partial_u} \col
\end{equation}
using
\begin{equation}
\mathcal{W}^{-1}=\sum_{a=0}^\infty(-1)^a T_{a,1}^{[1-a]}D^a \pnt
\end{equation}
In this way we obtain the contribution from the third factor of~\eqref{Ya0-general} as
\begin{equation}
T_{a,-1}^LT_{a,1}^R\rightarrow2^{10}\lambda^2\frac{[12a(u^2-u_{4,1}^2)+3a(a^2-1)]^2}{(4u_{4,1}^2+1)^2(a^2+4u^2)^2\tilde{y}_a(u)} \pnt
\end{equation}
Putting all the factors together, we find
\begin{equation}
Y_{a,0}^*(u)=9\lambda^{L+2}2^{10+2L}\frac{[12a(u^2-u_{4,1}^2)+3a(a^2-1)]^2}{(a^2+4u^2)^{L+2}\tilde{y}_a(u)\tilde{y}_{-a}(u)} \pnt
\end{equation}
Here, the root $u_{4,1}$ is the solution of the two-impurity Bethe
equations for the $SL(2)$ sector, which at order $\lambda^0$ read
\begin{equation}
\Big(\frac{u_{4,1}+i/2}{u_{4,1}-i/2}\Big)^{L+1}=1 \pnt
\end{equation}
For $L=3$ we get $u_{4,1}=1/2$.\\
Using the expression for $Y_{a,0}^*(u)$, the leading wrapping
contribution to the anomalous dimension can be found with the help of
the formula~\cite{Gromov:2009tv}
\begin{equation}
\delta\gamma_{L+2}=\sum_{a=1}^\infty\int_{-\infty}^\infty\frac{\mathrm{d}u}{2\pi i}\frac{\partial\epsilon_a^*}{\partial u}\mathrm{log}[1+Y_{a,0}^*(u)] \pnt
\end{equation}
At the lowest order in $\lambda$, we find
\begin{equation}
\frac{\partial\epsilon_a^*}{\partial u}=-2i \pnt
\end{equation}
Moreover, since $Y_{a,0}^*(u)\sim\lambda^{L+2}$, we can approximate
$\log[1+Y_{a,0}^*(u)]= Y_{a,0}^*(u)+o(\lambda^{L+2})$, so that at
leading order we have
\begin{equation}
\delta\gamma_{L+2}=-\frac{1}{\pi}\sum_{a=1}^\infty\int_{-\infty}^\infty Y_{a,0}^*(u) \pnt
\end{equation}
The integral can now be computed by using the residue method, closing
the integration path at infinity in the upper half of the complex
plane. In the end, the following result is found at five loops
\begin{equation}
\delta\gamma_5=-\lambda^5[128+512 \zeta(3)-2560\zeta(5)+2240\zeta(7)] \pnt
\end{equation}
This is the correction which must be added to the asymptotic five-loop
anomalous dimension computed using the Bethe equations: 
\begin{equation}
\gamma_5^{\mathrm{as}}=\lambda^5[6792+1664\zeta(3)+1280\zeta(5)] \pnt
\end{equation}
The five-loop contribution to the anomalous dimension of the
two-impurity, length-five operator is
\begin{equation}
\gamma_5=\gamma_5^{\mathrm{as}}+\delta\gamma_5=\lambda^5[6664+1152\zeta(3)+3840\zeta(5)-2240\zeta(7)] \col
\end{equation}
which agrees with our result.

\section{Concluding remarks}

In this paper we have computed perturbatively the planar anomalous dimension
of a length-five operator at five loops by extending the procedure of our
paper~\cite{uslong}. Moreover, we have explicitly
shown that our result agrees with the one obtained by applying the
Y-system proposed in~\cite{Gromov:2009tv,Gromov:2009bc}. 

The anomalous dimension computed here was predicted
in~\cite{Beccaria:2009eq} from a general formula for the five-loop
anomalous dimension of twist-three operators. This formula was
conjectured on the basis of the maximal transcendentality
principle~\cite{Kotikov:2002ab}.

With our calculation we have then given a new test of these existing
proposals on how to compute the anomalous dimension of short
operators.

It would be important to test the recently obtained five-loop
anomalous dimension of the Konishi
operator~\cite{Bajnok:2009vm}. However, this would require a
calculation beyond critical wrapping order. The complexity of the calculation 
in this case increases dramatically, even in the context of the ${\cal N}=1$
superspace techniques used here. In order to make this calculation manageable it
would be necessary to find new cancellation patterns beyond the ones
discovered in~\cite{uslong,Fiamberti:2008sn}.

\section*{Acknowledgements}
\noindent This work has been supported in part by INFN and by the
Italian MIUR-PRIN contract 20075ATT78. The work of C.S.\ has 
been supported by the European Marie Curie Research and Training Network 
ENRAGE (MRTN-CT-2004-005616).

\clearpage
\newpage

\appendix
\renewcommand{\thefigure}{A.\arabic{figure}}
\setcounter{figure}{0}
\renewcommand{\thetable}{A.\arabic{table}}
\setcounter{table}{0}

\section{Range-six diagrams}
\label{app:range6}
\begin{figure}[!h]
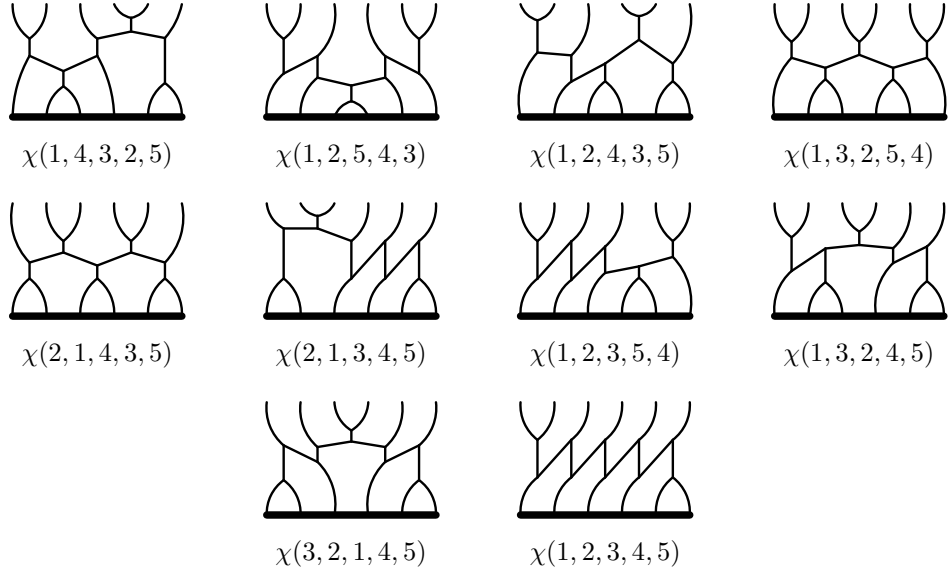

\centering
\unitlength=0.75mm
\settoheight{\eqoff}{$\times$}%
\setlength{\eqoff}{0.5\eqoff}%
\addtolength{\eqoff}{-12.5\unitlength}%
\settoheight{\eqofftwo}{$\times$}%
\setlength{\eqofftwo}{0.5\eqofftwo}%
\addtolength{\eqofftwo}{-7.5\unitlength}%
\subfigure[$\chi(1,4,3,2,5)$]{
\raisebox{\eqoff}{%
\fmfframe(3,1)(1,4){%
\begin{fmfchar*}(30,20)
\fmftop{v1}
\fmfbottom{v7}
\fmfforce{(0w,h)}{v1}
\fmfforce{(0w,0)}{v7}
\fmffixed{(0.2w,0)}{v1,v2}
\fmffixed{(0.2w,0)}{v2,v3}
\fmffixed{(0.2w,0)}{v3,v4}
\fmffixed{(0.2w,0)}{v4,v5}
\fmffixed{(0.2w,0)}{v5,v6}
\fmffixed{(0.2w,0)}{v7,v8}
\fmffixed{(0.2w,0)}{v8,v9}
\fmffixed{(0.2w,0)}{v9,v10}
\fmffixed{(0.2w,0)}{v10,v11}
\fmffixed{(0.2w,0)}{v11,v12}
\fmffixed{(0,whatever)}{vc1,vc2}
\fmffixed{(0,whatever)}{vc3,vc4}
\fmffixed{(0,whatever)}{vc5,vc6}
\fmffixed{(0,whatever)}{vc7,vc8}
\fmffixed{(0,whatever)}{vc9,vc10}
\fmf{phantom,tension=1,right=0.25}{v1,vc1}
\fmf{phantom,tension=1,left=0.25}{v2,vc1}
\fmf{phantom,tension=1,right=0.25}{v3,vc5}
\fmf{phantom,tension=1,left=0.25}{v4,vc5}
\fmf{phantom,tension=1,left=0.25}{v7,vc2}
\fmf{phantom,tension=0.25,left=0.25}{v8,vc4}
\fmf{phantom,tension=0.25,right=0.25}{v9,vc4}
\fmf{phantom,tension=1,right=0.25}{v10,vc6}
\fmf{phantom,tension=1,right=0.25}{v5,vc9}
\fmf{phantom,tension=1,left=0.25}{v6,vc9}
\fmf{phantom,tension=1,left=0.25}{v11,vc10}
\fmf{phantom,tension=1,right=0.25}{v12,vc10}
  \fmf{phantom,tension=4}{vc1,vc2}
  \fmf{phantom,tension=4}{vc5,vc6}
  \fmf{phantom,tension=0.5}{vc2,vc3}
  \fmf{phantom,tension=0.5}{vc3,vc6}
  \fmf{phantom,tension=1}{vc3,vc4}
  \fmf{phantom,tension=1.5}{vc9,vc10}
\fmffreeze
\fmf{plain,tension=0.25,right=0.25}{v1,vc1}
\fmf{plain,tension=0.25,left=0.25}{v2,vc1}
\fmf{plain,tension=1,left=0.125}{v7,vc2}
\fmf{plain,tension=0.25,right=0.25}{v3,vc5}
\fmf{plain,tension=1,right=0.125}{v10,vc6}
\fmf{plain,tension=0.25,left=0.25}{v8,vc4}
\fmf{plain,tension=0.25,right=0.25}{v9,vc4}
\fmf{plain,tension=0.25,right=0.25}{v4,vc7}
\fmf{plain,tension=0.25,left=0.25}{v5,vc7}
\fmf{plain,tension=0.25,left=0.125}{v6,vc9}
\fmf{plain,tension=0.25,left=0.25}{v11,vc10}
\fmf{plain,tension=0.25,right=0.25}{v12,vc10}
  \fmf{plain,tension=0.5}{vc1,vc2}
  \fmf{plain,tension=0.5}{vc2,vc3}
  \fmf{plain,tension=0.5}{vc3,vc4}
  \fmf{plain,tension=0.5}{vc3,vc6}
  \fmf{plain,tension=0.5}{vc5,vc6}
  \fmf{plain,tension=0.5}{vc5,vc8}
  \fmf{plain,tension=0.5}{vc7,vc8}
  \fmf{plain,tension=0.5}{vc8,vc9}
  \fmf{plain,tension=0.5}{vc9,vc10}
\fmffreeze
\fmfposition
\fmf{plain,tension=1,left=0,width=1mm}{v7,v12}
\fmffreeze
\end{fmfchar*}}}
}
\subfigspace
\subfigure[$\chi(1,2,5,4,3)$]{
\raisebox{\eqoff}{%
\fmfframe(3,1)(1,4){%
\begin{fmfchar*}(30,20)
\fmftop{v1}
\fmfbottom{v7}
\fmfforce{(0w,h)}{v1}
\fmfforce{(0w,0)}{v7}
\fmffixed{(0.2w,0)}{v1,v2}
\fmffixed{(0.2w,0)}{v2,v3}
\fmffixed{(0.2w,0)}{v3,v4}
\fmffixed{(0.2w,0)}{v4,v5}
\fmffixed{(0.2w,0)}{v5,v6}
\fmffixed{(0.2w,0)}{v7,v8}
\fmffixed{(0.2w,0)}{v8,v9}
\fmffixed{(0.2w,0)}{v9,v10}
\fmffixed{(0.2w,0)}{v10,v11}
\fmffixed{(0.2w,0)}{v11,v12}
\fmffixed{(0,whatever)}{vc1,vc2}
\fmffixed{(0,whatever)}{vc3,vc4}
\fmffixed{(0,whatever)}{vc5,vc6}
\fmffixed{(0,whatever)}{vc7,vc8}
\fmffixed{(0,whatever)}{vc9,vc10}
\fmf{plain,tension=0.5,right=0.25}{v1,vc1}
\fmf{plain,tension=0.5,left=0.25}{v2,vc1}
\fmf{plain,tension=1,left=0.25}{v7,vc2}
\fmf{plain,tension=1,left=0.25}{v3,vc3}
\fmf{plain,tension=1,left=0.25}{v8,vc4}
\fmf{plain,tension=0.25,left=0.25}{v9,vc6}
\fmf{plain,tension=0.25,right=0.25}{v10,vc6}
\fmf{plain,tension=1,right=0.25}{v4,vc7}
\fmf{plain,tension=1,right=0.25}{v11,vc8}
\fmf{plain,tension=1,right=0.25}{v12,vc10}
\fmf{plain,tension=0.5,right=0.25}{v5,vc9}
\fmf{plain,tension=0.5,left=0.25}{v6,vc9}
  \fmf{plain,tension=1}{vc1,vc2}
  \fmf{plain,tension=0.5}{vc2,vc3}
  \fmf{plain,tension=2}{vc3,vc4}
  \fmf{plain,tension=0.5}{vc4,vc5}
  \fmf{plain,tension=0.5}{vc5,vc6}
  \fmf{plain,tension=0.5}{vc5,vc8}
  \fmf{plain,tension=2}{vc7,vc8}
  \fmf{plain,tension=0.5}{vc7,vc10}
  \fmf{plain,tension=1}{vc9,vc10}
\fmffreeze
\fmfposition
\fmf{plain,tension=1,left=0,width=1mm}{v7,v12}
\fmffreeze
\end{fmfchar*}}}
}
\subfigspace
\subfigure[$\chi(1,2,4,3,5)$]{
\raisebox{\eqoff}{%
\fmfframe(3,1)(1,4){%
\begin{fmfchar*}(30,20)
\fmftop{v1}
\fmfbottom{v7}
\fmfforce{(0w,h)}{v1}
\fmfforce{(0w,0)}{v7}
\fmffixed{(0.2w,0)}{v1,v2}
\fmffixed{(0.2w,0)}{v2,v3}
\fmffixed{(0.2w,0)}{v3,v4}
\fmffixed{(0.2w,0)}{v4,v5}
\fmffixed{(0.2w,0)}{v5,v6}
\fmffixed{(0.2w,0)}{v7,v8}
\fmffixed{(0.2w,0)}{v8,v9}
\fmffixed{(0.2w,0)}{v9,v10}
\fmffixed{(0.2w,0)}{v10,v11}
\fmffixed{(0.2w,0)}{v11,v12}
\fmffixed{(0,whatever)}{vc1,vc2}
\fmffixed{(0,whatever)}{vc5,vc6}
\fmffixed{(0,whatever)}{vc7,vc8}
\fmffixed{(0,whatever)}{vc9,vc10}
\fmf{plain,tension=1,right=0.25}{v1,vc1}
\fmf{plain,tension=1,left=0.25}{v2,vc1}
\fmf{phantom,tension=0.25,right=0.25}{v8,vc2}
\fmf{phantom,tension=0.25,left=0.25}{v7,vc2}
\fmf{plain,tension=0.25,right=0.25}{v4,vc7}
\fmf{plain,tension=0.25,left=0.25}{v5,vc7}
\fmf{plain,tension=0.25,right=0.25}{v10,vc6}
\fmf{plain,tension=0.25,left=0.25}{v9,vc6}
\fmf{plain,tension=0.25,right=0.25}{v12,vc10}
\fmf{plain,tension=0.25,left=0.25}{v11,vc10}
\fmf{plain,tension=0.25,left=0.25}{v6,vc9}
\fmf{phantom,tension=0.25,right=0.25}{v3,vc5}
\fmf{phantom,tension=1}{vc1,vc2}
\fmf{plain,tension=1}{vc5,vc6}
\fmf{plain,tension=0.25}{vc7,vc8}
\fmf{plain,tension=0.125}{vc5,vc8}
\fmf{plain,tension=0.125}{vc8,vc9}
\fmf{plain,tension=1}{vc9,vc10}
\fmffreeze
\fmf{plain,tension=2}{vc1,vc2}
\fmf{plain,tension=0.5}{vc3,vc4}
\fmf{plain,tension=0.125}{vc2,vc3}
\fmf{plain,tension=0.25}{vc4,vc5}
\fmf{plain,tension=0.25,left=0.25}{v7,vc2}
\fmf{plain,tension=0.5,left=0.25}{v8,vc4}
\fmf{plain,tension=0.25,left=0.25}{v3,vc3}
\fmfposition
\fmf{plain,tension=1,left=0,width=1mm}{v7,v12}
\fmffreeze
\end{fmfchar*}}}
}
\subfigspace
\subfigure[$\chi(1,3,2,5,4)$]{
\raisebox{\eqoff}{%
\fmfframe(3,1)(1,4){%
\begin{fmfchar*}(30,20)
\fmftop{v1}
\fmfbottom{v7}
\fmfforce{(0w,h)}{v1}
\fmfforce{(0w,0)}{v7}
\fmffixed{(0.2w,0)}{v1,v2}
\fmffixed{(0.2w,0)}{v2,v3}
\fmffixed{(0.2w,0)}{v3,v4}
\fmffixed{(0.2w,0)}{v4,v5}
\fmffixed{(0.2w,0)}{v5,v6}
\fmffixed{(0.2w,0)}{v7,v8}
\fmffixed{(0.2w,0)}{v8,v9}
\fmffixed{(0.2w,0)}{v9,v10}
\fmffixed{(0.2w,0)}{v10,v11}
\fmffixed{(0.2w,0)}{v11,v12}
\fmf{plain,tension=1,right=0.25}{v1,vc1}
\fmf{plain,tension=1,left=0.25}{v2,vc1}
\fmf{phantom,tension=1,right=0.25}{v8,va2}
\fmf{phantom,tension=1,left=0.25}{v7,va2}
\fmf{phantom,tension=2}{vc1,va2}
\fmf{phantom,tension=1,right=0.25}{v2,va3}
\fmf{phantom,tension=1,left=0.25}{v3,va3}
\fmf{plain,tension=1,right=0.25}{v9,vc4}
\fmf{plain,tension=1,left=0.25}{v8,vc4}
\fmf{phantom,tension=2}{va3,vc4}
\fmf{plain,tension=1,right=0.25}{v3,vc5}
\fmf{plain,tension=1,left=0.25}{v4,vc5}
\fmf{phantom,tension=1,right=0.25}{v10,va6}
\fmf{phantom,tension=1,left=0.25}{v9,va6}
\fmf{phantom,tension=2}{vc5,va6}
\fmf{phantom,tension=1,right=0.25}{v4,va7}
\fmf{phantom,tension=1,left=0.25}{v5,va7}
\fmf{plain,tension=1,right=0.25}{v11,vc8}
\fmf{plain,tension=1,left=0.25}{v10,vc8}
\fmf{phantom,tension=2}{va7,vc8}
\fmf{plain,tension=1,right=0.25}{v5,vc9}
\fmf{plain,tension=1,left=0.25}{v6,vc9}
\fmf{phantom,tension=1,right=0.25}{v12,va10}
\fmf{phantom,tension=1,left=0.25}{v11,va10}
\fmf{phantom,tension=2}{vc9,va10}
\fmffreeze
\fmf{plain,tension=1}{vc1,vc2}
\fmf{plain,tension=0.25}{vc5,vc6}
\fmf{plain,tension=1}{vc9,vc10}
\fmf{plain,tension=0.25}{vc3,vc4}
\fmf{plain,tension=0.25}{vc7,vc8}
\fmf{plain,tension=0.125}{vc2,vc3}
\fmf{plain,tension=0.125}{vc3,vc6}
\fmf{plain,tension=0.125}{vc6,vc7}
\fmf{plain,tension=0.125}{vc7,vc10}
\fmf{plain,tension=0.25,left=0.25}{v7,vc2}
\fmf{plain,tension=0.25,right=0.25}{v12,vc10}
\fmffreeze
\fmfposition
\fmf{plain,tension=1,left=0,width=1mm}{v7,v12}
\fmffreeze
\end{fmfchar*}}}
}
\\
\subfigure[$\chi(2,1,4,3,5)$]{
\raisebox{\eqoff}{%
\fmfframe(3,1)(1,4){%
\begin{fmfchar*}(30,20)
\fmftop{v7}
\fmfbottom{v1}
\fmfforce{(0w,h)}{v7}
\fmfforce{(0w,0)}{v1}
\fmffixed{(0.2w,0)}{v1,v2}
\fmffixed{(0.2w,0)}{v2,v3}
\fmffixed{(0.2w,0)}{v3,v4}
\fmffixed{(0.2w,0)}{v4,v5}
\fmffixed{(0.2w,0)}{v5,v6}
\fmffixed{(0.2w,0)}{v7,v8}
\fmffixed{(0.2w,0)}{v8,v9}
\fmffixed{(0.2w,0)}{v9,v10}
\fmffixed{(0.2w,0)}{v10,v11}
\fmffixed{(0.2w,0)}{v11,v12}
\fmf{plain,tension=1,left=0.25}{v1,vc1}
\fmf{plain,tension=1,right=0.25}{v2,vc1}
\fmf{phantom,tension=1,left=0.25}{v8,va2}
\fmf{phantom,tension=1,right=0.25}{v7,va2}
\fmf{phantom,tension=2}{vc1,va2}
\fmf{phantom,tension=1,left=0.25}{v2,va3}
\fmf{phantom,tension=1,right=0.25}{v3,va3}
\fmf{plain,tension=1,left=0.25}{v9,vc4}
\fmf{plain,tension=1,right=0.25}{v8,vc4}
\fmf{phantom,tension=2}{va3,vc4}
\fmf{plain,tension=1,left=0.25}{v3,vc5}
\fmf{plain,tension=1,right=0.25}{v4,vc5}
\fmf{phantom,tension=1,left=0.25}{v10,va6}
\fmf{phantom,tension=1,right=0.25}{v9,va6}
\fmf{phantom,tension=2}{vc5,va6}
\fmf{phantom,tension=1,left=0.25}{v4,va7}
\fmf{phantom,tension=1,right=0.25}{v5,va7}
\fmf{plain,tension=1,left=0.25}{v11,vc8}
\fmf{plain,tension=1,right=0.25}{v10,vc8}
\fmf{phantom,tension=2}{va7,vc8}
\fmf{plain,tension=1,left=0.25}{v5,vc9}
\fmf{plain,tension=1,right=0.25}{v6,vc9}
\fmf{phantom,tension=1,left=0.25}{v12,va10}
\fmf{phantom,tension=1,right=0.25}{v11,va10}
\fmf{phantom,tension=2}{vc9,va10}
\fmffreeze
\fmf{plain,tension=1}{vc1,vc2}
\fmf{plain,tension=0.25}{vc5,vc6}
\fmf{plain,tension=1}{vc9,vc10}
\fmf{plain,tension=0.25}{vc3,vc4}
\fmf{plain,tension=0.25}{vc7,vc8}
\fmf{plain,tension=0.125}{vc2,vc3}
\fmf{plain,tension=0.125}{vc3,vc6}
\fmf{plain,tension=0.125}{vc6,vc7}
\fmf{plain,tension=0.125}{vc7,vc10}
\fmf{plain,tension=0.25,right=0.25}{v7,vc2}
\fmf{plain,tension=0.25,left=0.25}{v12,vc10}
\fmffreeze
\fmfposition
\fmf{plain,tension=1,right=0,width=1mm}{v1,v6}
\fmffreeze
\end{fmfchar*}}}
}
\subfigspace
\subfigure[$\chi(2,1,3,4,5)$]{
\raisebox{\eqoff}{%
\fmfframe(3,1)(1,4){%
\begin{fmfchar*}(30,20)
\fmftop{v1}
\fmfbottom{v7}
\fmfforce{(0w,h)}{v1}
\fmfforce{(0w,0)}{v7}
\fmffixed{(0.2w,0)}{v1,v2}
\fmffixed{(0.2w,0)}{v2,v3}
\fmffixed{(0.2w,0)}{v3,v4}
\fmffixed{(0.2w,0)}{v4,v5}
\fmffixed{(0.2w,0)}{v5,v6}
\fmffixed{(0.2w,0)}{v7,v8}
\fmffixed{(0.2w,0)}{v8,v9}
\fmffixed{(0.2w,0)}{v9,v10}
\fmffixed{(0.2w,0)}{v10,v11}
\fmffixed{(0.2w,0)}{v11,v12}
\fmf{phantom,tension=0.25,right=0.25}{v3,vc5}
\fmf{plain,tension=0.25,left=0.25}{v4,vc5}
\fmf{plain,left=0.25}{v9,vc6}
\fmf{plain,tension=1,left=0.25}{v5,vc7}
\fmf{plain,tension=1,left=0.25}{v6,vc9}
\fmf{plain,left=0.25}{v10,vc8}
\fmf{plain,tension=0.25,left=0.25}{v11,vc10}
\fmf{plain,tension=0.25,right=0.25}{v12,vc10}
\fmf{plain,tension=0.5}{vc5,vc6}
\fmf{plain,tension=0.5}{vc6,vc7}
\fmf{plain,tension=0.5}{vc7,vc8}
\fmf{plain,tension=0.5}{vc8,vc9}
\fmf{plain,tension=0.5}{vc9,vc10}
\fmf{plain,tension=0.25,right=0.25}{v8,vc2}
\fmf{plain,tension=0.25,left=0.25}{v7,vc2}
\fmf{phantom,tension=0.25,right=0.25}{v1,va1}
\fmf{phantom,tension=0.25,left=0.25}{v2,va1}
\fmf{phantom,tension=0.5}{va1,vc2}
\fmffreeze
\fmf{plain,tension=0.25,right=0.25}{v2,vc3}
\fmf{plain,tension=0.25,left=0.25}{v3,vc3}
\fmf{plain,tension=0.5}{vc1,vc4}
\fmf{plain,tension=0.5}{vc3,vc4}
\fmf{plain,tension=0.5}{vc4,vc5}
\fmf{plain,tension=0.5}{vc1,vc2}
\fmf{plain,tension=1,right=0.25}{v1,vc1}
\fmffreeze
\fmfposition
\fmf{plain,tension=1,right=0,width=1mm}{v7,v12}
\fmffreeze
\end{fmfchar*}}}
}
\subfigspace
\subfigure[$\chi(1,2,3,5,4)$]{
\raisebox{\eqoff}{%
\fmfframe(3,1)(1,4){%
\begin{fmfchar*}(30,20)
\fmftop{v1}
\fmfbottom{v7}
\fmfforce{(0w,h)}{v1}
\fmfforce{(0w,0)}{v7}
\fmffixed{(0.2w,0)}{v1,v2}
\fmffixed{(0.2w,0)}{v2,v3}
\fmffixed{(0.2w,0)}{v3,v4}
\fmffixed{(0.2w,0)}{v4,v5}
\fmffixed{(0.2w,0)}{v5,v6}
\fmffixed{(0.2w,0)}{v7,v8}
\fmffixed{(0.2w,0)}{v8,v9}
\fmffixed{(0.2w,0)}{v9,v10}
\fmffixed{(0.2w,0)}{v10,v11}
\fmffixed{(0.2w,0)}{v11,v12}
\fmffixed{(0,whatever)}{vc1,vc2}
\fmffixed{(0,whatever)}{vc3,vc4}
\fmffixed{(0,whatever)}{vc5,vc6}
\fmffixed{(0,whatever)}{va7,vc8}
\fmf{plain,tension=0.25,right=0.25}{v1,vc1}
\fmf{plain,tension=0.25,left=0.25}{v2,vc1}
\fmf{plain,left=0.25}{v7,vc2}
\fmf{plain,tension=1,left=0.25}{v3,vc3}
\fmf{plain,tension=1,left=0.25}{v4,vc5}
\fmf{phantom,tension=1,left=0.25}{v5,va7}
\fmf{plain,left=0.25}{v9,vc6}
\fmf{plain,tension=0.25,left=0.25}{v10,vc8}
\fmf{plain,tension=0.25,right=0.25}{v11,vc8}
\fmf{plain,left=0.25}{v8,vc4}
\fmf{plain,tension=0.5}{vc1,vc2}
\fmf{plain,tension=0.5}{vc2,vc3}
\fmf{plain,tension=0.5}{vc3,vc4}
\fmf{plain,tension=0.5}{vc4,vc5}
\fmf{plain,tension=1}{vc5,vc6}
\fmf{phantom,tension=0.5}{vc6,va7}
\fmf{phantom,tension=0.5}{va7,vc8}
\fmf{plain,tension=0.25,right=0.25}{v5,vc9}
\fmf{plain,tension=0.25,left=0.25}{v6,vc9}
\fmf{phantom,tension=0.25,right=0.25}{v12,va10}
\fmf{phantom,tension=0.25,left=0.25}{v11,va10}
\fmf{phantom,tension=0.5}{vc9,va10}
\fmffreeze
\fmf{plain,tension=0.5}{vc6,vc7}
\fmf{plain,tension=0.125}{vc7,vc8}
\fmf{plain,tension=0.5}{vc7,vc10}
\fmf{plain,tension=4}{vc9,vc10}
\fmf{plain,right=0.25}{v12,vc10}
\fmffreeze
\fmfposition
\fmf{plain,tension=1,right=0,width=1mm}{v7,v12}
\fmffreeze
\end{fmfchar*}}}
}
\subfigspace
\subfigure[$\chi(1,3,2,4,5)$]{
\raisebox{\eqoff}{%
\fmfframe(3,1)(1,4){%
\begin{fmfchar*}(30,20)
\fmftop{v6}
\fmfbottom{v12}
\fmfforce{(0w,h)}{v6}
\fmfforce{(0w,0)}{v12}
\fmffixed{(0.2w,0)}{v6,v5}
\fmffixed{(0.2w,0)}{v5,v4}
\fmffixed{(0.2w,0)}{v4,v3}
\fmffixed{(0.2w,0)}{v3,v2}
\fmffixed{(0.2w,0)}{v2,v1}
\fmffixed{(0.2w,0)}{v12,v11}
\fmffixed{(0.2w,0)}{v11,v10}
\fmffixed{(0.2w,0)}{v10,v9}
\fmffixed{(0.2w,0)}{v9,v8}
\fmffixed{(0.2w,0)}{v8,v7}
\fmf{plain,tension=0.5,left=0.25}{v2,vc3}
\fmf{phantom,tension=0.5,right=0.25}{v3,vc3}
\fmf{plain,left=0.25}{v1,vc1}
\fmf{plain,tension=0.5,right=0.25}{v7,vc2}
\fmf{plain,tension=0.5,left=0.25}{v8,vc2}
\fmf{plain,left=0.25}{v9,vc4}
\fmf{plain,tension=1}{vc1,vc2}
\fmf{plain,tension=0.5}{vc1,vc4}
\fmf{plain,tension=3}{vc3,vc4}
\fmf{plain,tension=0.5,left=0.25}{v5,vc9}
\fmf{plain,tension=0.5,right=0.25}{v6,vc9}
\fmf{phantom,left=0.25}{v4,vc7}
\fmf{plain,tension=0.5,right=0.25}{v10,vc8}
\fmf{plain,tension=0.5,left=0.25}{v11,vc8}
\fmf{plain,left=0.25}{v12,vc10}
\fmf{plain,tension=1}{vc7,vc8}
\fmf{plain,tension=0.5}{vc7,vc10}
\fmf{plain,tension=1}{vc9,vc10}
\fmf{plain,tension=0.5,left=0.25}{v3,vc5}
\fmf{plain,tension=0.5,right=0.25}{v4,vc5}
\fmf{phantom,tension=0.5,left=0.25}{v10,va2}
\fmf{phantom,tension=0.5,right=0.25}{v9,va2}
\fmf{phantom,tension=0.5}{vc5,va2}
\fmffreeze
\fmf{plain,tension=0.5}{vc3,vc6}
\fmf{plain,tension=0.25}{vc5,vc6}
\fmf{plain,tension=0.5}{vc6,vc7}
\fmffreeze
\fmfposition
\fmf{plain,tension=1,right=0,width=1mm}{v7,v12}
\fmffreeze
\end{fmfchar*}}}
}
\\
\subfigure[$\chi(3,2,1,4,5)$]{
\raisebox{\eqoff}{%
\fmfframe(3,1)(1,4){%
\begin{fmfchar*}(30,20)
\fmftop{v1}
\fmfbottom{v7}
\fmfforce{(0w,h)}{v1}
\fmfforce{(0w,0)}{v7}
\fmffixed{(0.2w,0)}{v1,v2}
\fmffixed{(0.2w,0)}{v2,v3}
\fmffixed{(0.2w,0)}{v3,v4}
\fmffixed{(0.2w,0)}{v4,v5}
\fmffixed{(0.2w,0)}{v5,v6}
\fmffixed{(0.2w,0)}{v7,v8}
\fmffixed{(0.2w,0)}{v8,v9}
\fmffixed{(0.2w,0)}{v9,v10}
\fmffixed{(0.2w,0)}{v10,v11}
\fmffixed{(0.2w,0)}{v11,v12}
\fmf{plain,tension=0.5,right=0.25}{v2,vc3}
\fmf{phantom,tension=0.5,left=0.25}{v3,vc3}
\fmf{plain,right=0.25}{v1,vc1}
\fmf{plain,tension=0.5,left=0.25}{v7,vc2}
\fmf{plain,tension=0.5,right=0.25}{v8,vc2}
\fmf{plain,right=0.25}{v9,vc4}
\fmf{plain,tension=1}{vc1,vc2}
\fmf{plain,tension=0.5}{vc1,vc4}
\fmf{plain,tension=3}{vc3,vc4}
\fmf{phantom,tension=0.5,right=0.25}{v4,vc7}
\fmf{plain,tension=0.5,left=0.25}{v5,vc7}
\fmf{plain,left=0.25}{v6,vc9}
\fmf{plain,tension=0.5,left=0.25}{v11,vc10}
\fmf{plain,tension=0.5,right=0.25}{v12,vc10}
\fmf{plain,left=0.25}{v10,vc8}
\fmf{plain,tension=3}{vc7,vc8}
\fmf{plain,tension=0.5}{vc8,vc9}
\fmf{plain,tension=1}{vc9,vc10}
\fmf{plain,tension=0.5,right=0.25}{v3,vc5}
\fmf{plain,tension=0.5,left=0.25}{v4,vc5}
\fmf{phantom,tension=0.5,right=0.25}{v10,va2}
\fmf{phantom,tension=0.5,left=0.25}{v9,va2}
\fmf{phantom,tension=0.5}{vc5,va2}
\fmffreeze
\fmf{plain,tension=0.5}{vc5,vc6}
\fmf{plain,tension=0.5}{vc3,vc6}
\fmf{plain,tension=0.5}{vc6,vc7}
\fmffreeze
\fmfposition
\fmf{plain,tension=1,right=0,width=1mm}{v7,v12}
\fmffreeze
\end{fmfchar*}}}
}
\subfigspace
\subfigure[$\chi(1,2,3,4,5)$]{
\raisebox{\eqoff}{%
\fmfframe(3,1)(1,4){%
\begin{fmfchar*}(30,20)
\fmftop{v1}
\fmfbottom{v7}
\fmfforce{(0w,h)}{v1}
\fmfforce{(0w,0)}{v7}
\fmffixed{(0.2w,0)}{v1,v2}
\fmffixed{(0.2w,0)}{v2,v3}
\fmffixed{(0.2w,0)}{v3,v4}
\fmffixed{(0.2w,0)}{v4,v5}
\fmffixed{(0.2w,0)}{v5,v6}
\fmffixed{(0.2w,0)}{v7,v8}
\fmffixed{(0.2w,0)}{v8,v9}
\fmffixed{(0.2w,0)}{v9,v10}
\fmffixed{(0.2w,0)}{v10,v11}
\fmffixed{(0.2w,0)}{v11,v12}
\fmf{plain,tension=0.25,right=0.25}{v1,vc1}
\fmf{plain,tension=0.25,left=0.25}{v2,vc1}
\fmf{plain,left=0.25}{v7,vc2}
\fmf{plain,tension=1,left=0.25}{v3,vc3}
\fmf{plain,tension=1,left=0.25}{v4,vc5}
\fmf{plain,tension=1,left=0.25}{v5,vc7}
\fmf{plain,tension=1,left=0.25}{v6,vc9}
\fmf{plain,left=0.25}{v9,vc6}
\fmf{plain,tension=0.25,left=0.25}{v11,vc10}
\fmf{plain,tension=0.25,right=0.25}{v12,vc10}
\fmf{plain,left=0.25}{v10,vc8}
\fmf{plain,left=0.25}{v8,vc4}
\fmf{plain,tension=0.5}{vc1,vc2}
\fmf{plain,tension=0.5}{vc2,vc3}
\fmf{plain,tension=0.5}{vc3,vc4}
\fmf{plain,tension=0.5}{vc4,vc5}
\fmf{plain,tension=0.5}{vc5,vc6}
\fmf{plain,tension=0.5}{vc6,vc7}
\fmf{plain,tension=0.5}{vc7,vc8}
\fmf{plain,tension=0.5}{vc8,vc9}
\fmf{plain,tension=0.5}{vc9,vc10}
\fmffreeze
\fmfposition
\fmf{plain,tension=1,right=0,width=1mm}{v7,v12}
\fmffreeze
\end{fmfchar*}}}
}
\caption{Completely chiral range-six diagrams}
\label{r6-chiral}
\end{figure}

\begin{figure}[!h]
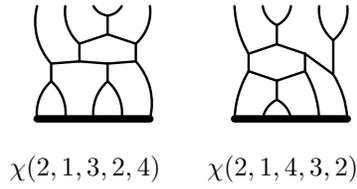

\centering
\unitlength=0.75mm
\settoheight{\eqoff}{$\times$}%
\setlength{\eqoff}{0.5\eqoff}%
\addtolength{\eqoff}{-12.5\unitlength}%
\settoheight{\eqofftwo}{$\times$}%
\setlength{\eqofftwo}{0.5\eqofftwo}%
\addtolength{\eqofftwo}{-7.5\unitlength}%
\subfigure[$\chi(2,1,3,2,4)$]{
\raisebox{\eqoff}{%
\fmfframe(3,1)(1,4){%
\begin{fmfchar*}(20,20)
\fmftop{v1}
\fmfbottom{v5}
\fmfforce{(0w,h)}{v1}
\fmfforce{(0w,0)}{v5}
\fmffixed{(0.25w,0)}{v1,v2}
\fmffixed{(0.25w,0)}{v2,v3}
\fmffixed{(0.25w,0)}{v3,v4}
\fmffixed{(0.25w,0)}{v4,v9}
\fmffixed{(0.25w,0)}{v5,v6}
\fmffixed{(0.25w,0)}{v6,v7}
\fmffixed{(0.25w,0)}{v7,v8}
\fmffixed{(0.25w,0)}{v8,v10}
\fmffixed{(0,whatever)}{vb2,vc2}
\fmffixed{(0,whatever)}{vb3,vc3}
\fmffixed{(0,whatever)}{vb5,vc5}
\fmffixed{(0,whatever)}{vb7,vc7}
\fmf{plain,tension=0.25,left=0.25}{v5,vc2}
\fmf{plain,tension=0.25,right=0.25}{v6,vc2}
\fmf{phantom,tension=0.25,right=0.25}{v1,vb2}
\fmf{phantom,tension=0.25,left=0.25}{v2,vb2}
\fmf{phantom,tension=0.5}{vc2,vb2}
\fmf{plain,tension=0.25,right=0.25}{v2,vc3}
\fmf{phantom,tension=0.25,left=0.25}{v3,vc3}
\fmf{phantom,tension=0.25,right=0.25}{v6,vb3}
\fmf{phantom,tension=0.25,left=0.25}{v7,vb3}
\fmf{phantom,tension=0.5}{vc3,vb3}
\fmf{plain,tension=0.25,left=0.25}{v7,vc5}
\fmf{plain,tension=0.25,right=0.25}{v8,vc5}
\fmf{phantom,tension=0.25,right=0.25}{v3,vb5}
\fmf{phantom,tension=0.25,left=0.25}{v4,vb5}
\fmf{phantom,tension=0.5}{vc5,vb5}
\fmf{phantom,tension=0.25,right=0.25}{v4,vc7}
\fmf{plain,tension=0.25,left=0.25}{v9,vc7}
\fmf{phantom,tension=0.25,right=0.25}{v8,vb7}
\fmf{phantom,tension=0.25,left=0.25}{v10,vb7}
\fmf{phantom,tension=0.5}{vc7,vb7}
\fmffreeze
\fmf{plain,tension=0.25,right=0.25}{v3,vc9}
\fmf{plain,tension=0.25,left=0.25}{v4,vc9}
\fmf{plain,tension=0.25}{vc3,vc10}
\fmf{plain,tension=0.25}{vc7,vc10}
\fmf{plain,tension=0.25}{vc9,vc10}
\fmffixed{(0,whatever)}{vc1,vc2}
\fmffixed{(0,whatever)}{vc3,vc4}
\fmffixed{(0,whatever)}{vc5,vc6}
\fmffixed{(0,whatever)}{vc7,vc8}
\fmffixed{(whatever,0)}{vc1,vc6}
\fmffixed{(whatever,0)}{vc4,vc8}
\fmf{plain,tension=0.25,right=0.25}{v1,vc1}
\fmf{plain,tension=0.5}{vc1,vc2}
\fmf{plain,tension=0.5}{vc3,vc4}
\fmf{plain,tension=0.5}{vc1,vc4}
\fmf{plain,tension=0.5}{vc5,vc6}
\fmf{plain,tension=0.5}{vc4,vc6}
\fmf{plain,tension=0.25,right=0.25}{v10,vc8}
\fmf{plain,tension=0.5}{vc7,vc8}
\fmf{plain,tension=0.5}{vc6,vc8}
\fmf{plain,tension=0.5,right=0,width=1mm}{v5,v10}
\fmfposition
\fmfipath{p[]}
\end{fmfchar*}}}
}
\subfigspace
\subfigure[$\chi(2,1,4,3,2)$]{
\raisebox{\eqoff}{%
\fmfframe(3,1)(1,4){%
\begin{fmfchar*}(20,20)
\fmftop{v1}
\fmfbottom{v5}
\fmfforce{(0w,h)}{v1}
\fmfforce{(0w,0)}{v5}
\fmffixed{(0.25w,0)}{v1,v2}
\fmffixed{(0.25w,0)}{v2,v3}
\fmffixed{(0.25w,0)}{v3,v4}
\fmffixed{(0.25w,0)}{v4,v9}
\fmffixed{(0.25w,0)}{v5,v6}
\fmffixed{(0.25w,0)}{v6,v7}
\fmffixed{(0.25w,0)}{v7,v8}
\fmffixed{(0.25w,0)}{v8,v10}
\fmffixed{(0,whatever)}{vc1,vc5}
\fmffixed{(0,whatever)}{vc2,vc3}
\fmffixed{(0,whatever)}{vc3,vc6}
\fmffixed{(0,whatever)}{vc6,vc7}
\fmffixed{(0,whatever)}{vc4,vc8}
\fmffixed{(0,whatever)}{vc9,vc10}
\fmffixed{(0.5w,0)}{vc1,vc4}
\fmffixed{(0.5w,0)}{vc5,vc8}
\fmf{plain,tension=1,right=0.125}{v1,vc1}
\fmf{plain,tension=0.25,right=0.25}{v2,vc2}
\fmf{plain,tension=0.25,left=0.25}{v3,vc2}
\fmf{phantom,tension=1,left=0.125}{v4,vc4}
\fmf{plain,tension=1,left=0.125}{v5,vc5}
\fmf{plain,tension=0.25,left=0.25}{v6,vc6}
\fmf{plain,tension=0.25,right=0.25}{v7,vc6}
\fmf{plain,tension=1,right=0.125}{v8,vc8}
  \fmf{plain,tension=0.5}{vc1,vc3}
  \fmf{plain,tension=0.5}{vc2,vc3}
  \fmf{plain,tension=0.5}{vc3,vc4}
  \fmf{plain,tension=0.5}{vc5,vc7}
  \fmf{plain,tension=0.5}{vc6,vc7}
  \fmf{plain,tension=0.5}{vc7,vc8}
  \fmf{plain,tension=2}{vc1,vc5}
  \fmf{plain,tension=2}{vc4,vc8}
  \fmf{phantom,tension=2}{vc5,vc4}
\fmffreeze
\fmf{plain,tension=0.5,left=0.25}{v9,vc10}
\fmf{plain,tension=0.5,right=0.25}{v4,vc10}
\fmf{plain,tension=0.5}{vc4,vc9}
\fmf{plain,tension=1}{vc10,vc9}
\fmf{plain,tension=1,right=0.125}{v10,vc9}
\fmfposition
\fmf{plain,tension=1,left=0,width=1mm}{v5,v10}
\fmffreeze
\end{fmfchar*}}}
}
\caption{Completely chiral range-five diagrams}
\label{r5-chiral}
\end{figure}

In this appendix we consider the range-six diagrams relevant for the
computation of the $\epsilon_x$ coefficients, which enter the
subtracted dilatation operator on length-five states. To this end, we
compute all the range-six diagrams which are either completely chiral
(Fig.~\ref{r6-chiral}) or contain a single vector interaction
(Figs.~\ref{r6-1245}-\ref{r6-2145}), and the two completely chiral
range-five graphs (Fig.~\ref{r5-chiral}). 
In all the figures containing lists of diagrams, the symmetry factor
is explicitly shown when it differs from 1.
By comparing the results from these diagrams with the corresponding
coefficients in the five-loop asymptotic dilatation operator, we find
a set of relations for the coefficients. Some of these can be used to
determine the needed $\epsilon_x$ coefficients, while all the other ones
reduce to identities which allow us to perform consistency checks.

Let us define $\mathcal{C}[\chi(\ldots)]$ as the coefficient of the
chiral structure $\chi(\ldots)$ in the asymptotic dilatation operator
$D_5$ of Table~\ref{D5}. $\mathcal{C}[\chi(\ldots)]$ is computed from
the coefficient of the $1/\varepsilon$ pole of the sum of all the
diagrams with structure $\chi(\ldots)$, multiplied by a factor $-10$
according to the definition of the anomalous dimension which, in the 
case of a multiplicatively renormalized operator, is simply given by
\begin{equation}
\gamma(\mathcal{O})=\lim_{\varepsilon\rightarrow0}\left[\varepsilon g\frac{\mathrm{d} }{\mathrm{d} g}\log\mathcal{Z}_{\mathcal{O}}(g,\varepsilon)\right]\col
\label{anomdim}
\end{equation}
where
\begin{equation}
\mathcal{O}_{\mathrm{ren}}=\mathcal{Z}_\mathcal{O}\mathcal{O}_{\mathrm{bare}}
\label{opren}
\end{equation}
and $\varepsilon$ is the dimensional regularization parameter.
The $(g^2N)^5$
factor coming from color and the $1/(4\pi)^{10}$ from the momentum
integrals combine into the $\lambda^5$ coupling \eqref{lambdadef}
which multiplies the five-loop Hamiltonian, and is not shown
explicitly.

The constraints from diagrams are summarized in Table~\ref{r6-res},
where the results are written in terms of the momentum integrals
$J_i$, which are listed in Appendix~\ref{app:integrals}. The partial
contributions coming from the single diagrams with vector interactions
are given in Table~\ref{r6-vector}, where the final results for the
chiral structures already contain all the symmetry factors and
contributions from possible reflected structures.

Using the conditions on the coefficients of the dilatation operator,
several equations relating a subset of the $\epsilon_{4x}$
coefficients can be found:
\begin{equation}
\begin{aligned}
\epsilon_{4a} &=\frac{13}{64}i\col & \epsilon_{4b} &= -\frac{85}{192} \col & \epsilon_{4c} &= \frac{i}{96} \col \\
\epsilon_{4d} &= -\frac{5}{192}i \col & \epsilon_{4e} &= \frac{\pi^4-75}{960}i \col & \epsilon_{4f} &= \frac{35}{96} \col \\
& & \epsilon_{4g} &= -\frac{3}{32}i \pnt
\end{aligned}
\end{equation}
These relations are sufficient to completely determine the action of
the subtracted dilatation operator on the length-five states.

\begin{table}
\begin{center}
\begin{equation*}
\begin{aligned}
& \mathcal{C}[\chi(1,2,3,4,5)]=-10\ J_{1}=-28 \\
& \mathcal{C}[\chi(2,1,4,3,5)]=-10\ J_{2}=-8/3 \\
& \mathcal{C}[\chi(1,3,2,5,4)]=-10\ J_{3}=2/3 \\
& \mathcal{C}[\chi(1,2,5,4,3)]=-10\ J_{4}=-4 \\
& \mathcal{C}[\chi(1,2,4,3,5)]=-10\ J_{5}=5 \\
& \mathcal{C}[\chi(3,2,1,4,5)]=-10\ J_{6}=-8/3 \\
& \mathcal{C}[\chi(1,3,2,4,5)]=-10\ J_{7}=16/3 \\
& \mathcal{C}[\chi(1,2,3,5,4)]=-10\ J_{8}=-11/3 \\
& \mathcal{C}[\chi(2,1,3,4,5)]=-10\ J_{9}=28/3 \\
& \mathcal{C}[\chi(1,4,3,2,5)]=-10\ J_{10}=1/3 \\
& \mathcal{C}[\chi(1,5,4,3)]=-10(-2J_{11})=-46/3 \\
& \mathcal{C}[\chi(1,3,4,5)]=-10(-2J_{12})=-8 \\
& \mathcal{C}[\chi(1,2,4,5)]=-10(-2J_{13})=-26/3 \\
& \mathcal{C}[\chi(1,2,5,4)]=-10(-2J_{14})=-16/3 \\
& \mathcal{C}[\chi(2,1,4,5)]=-10(-2J_{15})=20/3 \\
& \mathcal{C}[\chi(1,4,3,5)]=-10(-2J_{16})=6 \\
& \mathcal{C}[\chi(1,3,2,5)]=-10(-2J_{17})=4/3 \\
& \mathcal{C}[\chi(1,3,2,4,3)]=-10\,J_{18}=25/3-\pi^4/30-8\zeta(3) \\
& \mathcal{C}[\chi(2,1,3,2,4)]=-10\,J_{19}=10/3+\pi^4/30-8\zeta(3) \\
\end{aligned}
\end{equation*}
\end{center}
\caption{Constraints on the coefficients of the dilatation operator}
\label{r6-res}
\end{table}

\begin{figure}[!h]
\centering
\unitlength=0.75mm
\settoheight{\eqoff}{$\times$}%
\setlength{\eqoff}{0.5\eqoff}%
\addtolength{\eqoff}{-12.5\unitlength}%
\settoheight{\eqofftwo}{$\times$}%
\setlength{\eqofftwo}{0.5\eqofftwo}%
\addtolength{\eqofftwo}{-7.5\unitlength}%
\subfigure[$G_{1}$]{
\raisebox{\eqoff}{%
\fmfframe(3,1)(1,4){%
\begin{fmfchar*}(30,20)
\fmftop{v1}
\fmfbottom{v7}
\fmfforce{(0w,h)}{v1}
\fmfforce{(0w,0)}{v7}
\fmffixed{(0.2w,0)}{v1,v2}
\fmffixed{(0.2w,0)}{v2,v3}
\fmffixed{(0.2w,0)}{v3,v4}
\fmffixed{(0.2w,0)}{v4,v5}
\fmffixed{(0.2w,0)}{v5,v6}
\fmffixed{(0.2w,0)}{v7,v8}
\fmffixed{(0.2w,0)}{v8,v9}
\fmffixed{(0.2w,0)}{v9,v10}
\fmffixed{(0.2w,0)}{v10,v11}
\fmffixed{(0.2w,0)}{v11,v12}
\fmf{plain,tension=0.5,right=0.25}{v1,vc1}
\fmf{plain,tension=0.5,left=0.25}{v2,vc1}
\fmf{plain,left=0.25}{v3,vc3}
\fmf{plain,tension=0.5,left=0.25}{v8,vc4}
\fmf{plain,tension=0.5,right=0.25}{v9,vc4}
\fmf{plain,left=0.25}{v7,vc2}
\fmf{plain,tension=1}{vc1,vc2}
\fmf{plain,tension=0.5}{vc2,vc3}
\fmf{plain,tension=1}{vc3,vc4}
\fmf{plain,tension=0.5,right=0.25}{v4,vc7}
\fmf{plain,tension=0.5,left=0.25}{v5,vc7}
\fmf{plain,left=0.25}{v6,vc9}
\fmf{plain,tension=0.5,left=0.25}{v11,vc10}
\fmf{plain,tension=0.5,right=0.25}{v12,vc10}
\fmf{plain,left=0.25}{v10,vc8}
\fmf{plain,tension=1}{vc7,vc8}
\fmf{plain,tension=0.5}{vc8,vc9}
\fmf{plain,tension=1}{vc9,vc10}
\fmffreeze
\fmfposition
\fmf{plain,tension=1,right=0,width=1mm}{v7,v12}
\fmfipath{p[]}
\fmfiset{p1}{vpath(__vc3,__vc4)}
\fmfiset{p2}{vpath(__v10,__vc8)}
\fmfipair{w[]}
\svertex{w1}{p1}
\svertex{w2}{p2}
\fmfi{wiggly}{w1..w2}
\fmffreeze
\end{fmfchar*}}}
}
\subfigure[$G_{2}$]{
\raisebox{\eqoff}{%
\fmfframe(3,1)(1,4){%
\begin{fmfchar*}(30,20)
\fmftop{v1}
\fmfbottom{v7}
\fmfforce{(0w,h)}{v1}
\fmfforce{(0w,0)}{v7}
\fmffixed{(0.2w,0)}{v1,v2}
\fmffixed{(0.2w,0)}{v2,v3}
\fmffixed{(0.2w,0)}{v3,v4}
\fmffixed{(0.2w,0)}{v4,v5}
\fmffixed{(0.2w,0)}{v5,v6}
\fmffixed{(0.2w,0)}{v7,v8}
\fmffixed{(0.2w,0)}{v8,v9}
\fmffixed{(0.2w,0)}{v9,v10}
\fmffixed{(0.2w,0)}{v10,v11}
\fmffixed{(0.2w,0)}{v11,v12}
\fmf{plain,tension=0.5,right=0.25}{v1,vc1}
\fmf{plain,tension=0.5,left=0.25}{v2,vc1}
\fmf{plain,left=0.25}{v3,vc3}
\fmf{plain,tension=0.5,left=0.25}{v8,vc4}
\fmf{plain,tension=0.5,right=0.25}{v9,vc4}
\fmf{plain,left=0.25}{v7,vc2}
\fmf{plain,tension=1}{vc1,vc2}
\fmf{plain,tension=0.5}{vc2,vc3}
\fmf{plain,tension=1}{vc3,vc4}
\fmf{plain,tension=0.5,right=0.25}{v4,vc7}
\fmf{plain,tension=0.5,left=0.25}{v5,vc7}
\fmf{plain,left=0.25}{v6,vc9}
\fmf{plain,tension=0.5,left=0.25}{v11,vc10}
\fmf{plain,tension=0.5,right=0.25}{v12,vc10}
\fmf{plain,left=0.25}{v10,vc8}
\fmf{plain,tension=1}{vc7,vc8}
\fmf{plain,tension=0.5}{vc8,vc9}
\fmf{plain,tension=1}{vc9,vc10}
\fmffreeze
\fmfposition
\fmf{plain,tension=1,right=0,width=1mm}{v7,v12}
\fmfipath{p[]}
\fmfiset{p1}{vpath(__vc3,__vc4)}
\fmfiset{p2}{vpath(__vc7,__vc8)}
\fmfipair{w[]}
\svertex{w1}{p1}
\svertex{w2}{p2}
\fmfi{wiggly}{w1..w2}
\fmffreeze
\end{fmfchar*}}}
}
\subfigure[$G_{3}$]{
\raisebox{\eqoff}{%
\fmfframe(3,1)(1,4){%
\begin{fmfchar*}(30,20)
\fmftop{v1}
\fmfbottom{v7}
\fmfforce{(0w,h)}{v1}
\fmfforce{(0w,0)}{v7}
\fmffixed{(0.2w,0)}{v1,v2}
\fmffixed{(0.2w,0)}{v2,v3}
\fmffixed{(0.2w,0)}{v3,v4}
\fmffixed{(0.2w,0)}{v4,v5}
\fmffixed{(0.2w,0)}{v5,v6}
\fmffixed{(0.2w,0)}{v7,v8}
\fmffixed{(0.2w,0)}{v8,v9}
\fmffixed{(0.2w,0)}{v9,v10}
\fmffixed{(0.2w,0)}{v10,v11}
\fmffixed{(0.2w,0)}{v11,v12}
\fmf{plain,tension=0.5,right=0.25}{v1,vc1}
\fmf{plain,tension=0.5,left=0.25}{v2,vc1}
\fmf{plain,left=0.25}{v3,vc3}
\fmf{plain,tension=0.5,left=0.25}{v8,vc4}
\fmf{plain,tension=0.5,right=0.25}{v9,vc4}
\fmf{plain,left=0.25}{v7,vc2}
\fmf{plain,tension=1}{vc1,vc2}
\fmf{plain,tension=0.5}{vc2,vc3}
\fmf{plain,tension=1}{vc3,vc4}
\fmf{plain,tension=0.5,right=0.25}{v4,vc7}
\fmf{plain,tension=0.5,left=0.25}{v5,vc7}
\fmf{plain,left=0.25}{v6,vc9}
\fmf{plain,tension=0.5,left=0.25}{v11,vc10}
\fmf{plain,tension=0.5,right=0.25}{v12,vc10}
\fmf{plain,left=0.25}{v10,vc8}
\fmf{plain,tension=1}{vc7,vc8}
\fmf{plain,tension=0.5}{vc8,vc9}
\fmf{plain,tension=1}{vc9,vc10}
\fmffreeze
\fmfposition
\fmf{plain,tension=1,right=0,width=1mm}{v7,v12}
\fmfipath{p[]}
\fmfiset{p1}{vpath(__v9,__vc4)}
\fmfiset{p2}{vpath(__v10,__vc8)}
\fmfipair{w[]}
\svertex{w1}{p1}
\svertex{w2}{p2}
\fmfi{wiggly}{w1..w2}
\fmffreeze
\end{fmfchar*}}}
}
\subfigure[$G_{4}$]{
\raisebox{\eqoff}{%
\fmfframe(3,1)(1,4){%
\begin{fmfchar*}(30,20)
\fmftop{v1}
\fmfbottom{v7}
\fmfforce{(0w,h)}{v1}
\fmfforce{(0w,0)}{v7}
\fmffixed{(0.2w,0)}{v1,v2}
\fmffixed{(0.2w,0)}{v2,v3}
\fmffixed{(0.2w,0)}{v3,v4}
\fmffixed{(0.2w,0)}{v4,v5}
\fmffixed{(0.2w,0)}{v5,v6}
\fmffixed{(0.2w,0)}{v7,v8}
\fmffixed{(0.2w,0)}{v8,v9}
\fmffixed{(0.2w,0)}{v9,v10}
\fmffixed{(0.2w,0)}{v10,v11}
\fmffixed{(0.2w,0)}{v11,v12}
\fmf{plain,tension=0.5,right=0.25}{v1,vc1}
\fmf{plain,tension=0.5,left=0.25}{v2,vc1}
\fmf{plain,left=0.25}{v3,vc3}
\fmf{plain,tension=0.5,left=0.25}{v8,vc4}
\fmf{plain,tension=0.5,right=0.25}{v9,vc4}
\fmf{plain,left=0.25}{v7,vc2}
\fmf{plain,tension=1}{vc1,vc2}
\fmf{plain,tension=0.5}{vc2,vc3}
\fmf{plain,tension=1}{vc3,vc4}
\fmf{plain,tension=0.5,right=0.25}{v4,vc7}
\fmf{plain,tension=0.5,left=0.25}{v5,vc7}
\fmf{plain,left=0.25}{v6,vc9}
\fmf{plain,tension=0.5,left=0.25}{v11,vc10}
\fmf{plain,tension=0.5,right=0.25}{v12,vc10}
\fmf{plain,left=0.25}{v10,vc8}
\fmf{plain,tension=1}{vc7,vc8}
\fmf{plain,tension=0.5}{vc8,vc9}
\fmf{plain,tension=1}{vc9,vc10}
\fmffreeze
\fmfposition
\fmf{plain,tension=1,right=0,width=1mm}{v7,v12}
\fmfipath{p[]}
\fmfiset{p1}{vpath(__v9,__vc4)}
\fmfiset{p2}{vpath(__vc7,__vc8)}
\fmfipair{w[]}
\svertex{w1}{p1}
\svertex{w2}{p2}
\fmfi{wiggly}{w1..w2}
\fmffreeze
\end{fmfchar*}}}
}
\caption{Range-six diagrams with structure $\chi(1,2,4,5)$}
\label{r6-1245}
\end{figure}

\begin{figure}[!h]
\centering
\unitlength=0.75mm
\settoheight{\eqoff}{$\times$}%
\setlength{\eqoff}{0.5\eqoff}%
\addtolength{\eqoff}{-12.5\unitlength}%
\settoheight{\eqofftwo}{$\times$}%
\setlength{\eqofftwo}{0.5\eqofftwo}%
\addtolength{\eqofftwo}{-7.5\unitlength}%
\subfigure[$G_{1}$]{
\raisebox{\eqoff}{%
\fmfframe(3,1)(1,4){%
\begin{fmfchar*}(30,20)
\fmftop{v1}
\fmfbottom{v7}
\fmfforce{(0w,h)}{v1}
\fmfforce{(0w,0)}{v7}
\fmffixed{(0.2w,0)}{v1,v2}
\fmffixed{(0.2w,0)}{v2,v3}
\fmffixed{(0.2w,0)}{v3,v4}
\fmffixed{(0.2w,0)}{v4,v5}
\fmffixed{(0.2w,0)}{v5,v6}
\fmffixed{(0.2w,0)}{v7,v8}
\fmffixed{(0.2w,0)}{v8,v9}
\fmffixed{(0.2w,0)}{v9,v10}
\fmffixed{(0.2w,0)}{v10,v11}
\fmffixed{(0.2w,0)}{v11,v12}
\fmf{plain,tension=0.25,right=0.25}{v5,vc9}
\fmf{plain,tension=0.25,left=0.25}{v6,vc9}
\fmf{plain,right=0.25}{v11,vc8}
\fmf{plain,tension=1,right=0.25}{v3,vc5}
\fmf{plain,tension=1,right=0.25}{v4,vc7}
\fmf{plain,right=0.25}{v12,vc10}
\fmf{plain,tension=0.25,left=0.25}{v9,vc6}
\fmf{plain,tension=0.25,right=0.25}{v10,vc6}
\fmf{plain,tension=0.5}{vc5,vc6}
\fmf{plain,tension=0.5}{vc5,vc8}
\fmf{plain,tension=0.5}{vc7,vc8}
\fmf{plain,tension=0.5}{vc7,vc10}
\fmf{plain,tension=0.5}{vc9,vc10}
\fmf{plain,tension=0.25,right=0.25}{v1,vc1}
\fmf{plain,tension=0.25,left=0.25}{v2,vc1}
\fmf{plain,tension=0.25,left=0.25}{v7,vc2}
\fmf{plain,tension=0.25,right=0.25}{v8,vc2}
\fmf{plain,tension=0.5}{vc1,vc2}
\fmffreeze
\fmfposition
\fmf{plain,tension=1,right=0,width=1mm}{v7,v12}
\fmfipath{p[]}
\fmfiset{p1}{vpath(__vc1,__vc2)}
\fmfiset{p2}{vpath(__vc5,__vc6)}
\fmfipair{w[]}
\svertex{w1}{p1}
\svertex{w2}{p2}
\fmfi{wiggly}{w1..w2}
\fmffreeze
\end{fmfchar*}}}
}
\subfigure[$G_{2}$]{
\raisebox{\eqoff}{%
\fmfframe(3,1)(1,4){%
\begin{fmfchar*}(30,20)
\fmftop{v1}
\fmfbottom{v7}
\fmfforce{(0w,h)}{v1}
\fmfforce{(0w,0)}{v7}
\fmffixed{(0.2w,0)}{v1,v2}
\fmffixed{(0.2w,0)}{v2,v3}
\fmffixed{(0.2w,0)}{v3,v4}
\fmffixed{(0.2w,0)}{v4,v5}
\fmffixed{(0.2w,0)}{v5,v6}
\fmffixed{(0.2w,0)}{v7,v8}
\fmffixed{(0.2w,0)}{v8,v9}
\fmffixed{(0.2w,0)}{v9,v10}
\fmffixed{(0.2w,0)}{v10,v11}
\fmffixed{(0.2w,0)}{v11,v12}
\fmf{plain,tension=0.25,right=0.25}{v5,vc9}
\fmf{plain,tension=0.25,left=0.25}{v6,vc9}
\fmf{plain,right=0.25}{v11,vc8}
\fmf{plain,tension=1,right=0.25}{v3,vc5}
\fmf{plain,tension=1,right=0.25}{v4,vc7}
\fmf{plain,right=0.25}{v12,vc10}
\fmf{plain,tension=0.25,left=0.25}{v9,vc6}
\fmf{plain,tension=0.25,right=0.25}{v10,vc6}
\fmf{plain,tension=0.5}{vc5,vc6}
\fmf{plain,tension=0.5}{vc5,vc8}
\fmf{plain,tension=0.5}{vc7,vc8}
\fmf{plain,tension=0.5}{vc7,vc10}
\fmf{plain,tension=0.5}{vc9,vc10}
\fmf{plain,tension=0.25,right=0.25}{v1,vc1}
\fmf{plain,tension=0.25,left=0.25}{v2,vc1}
\fmf{plain,tension=0.25,left=0.25}{v7,vc2}
\fmf{plain,tension=0.25,right=0.25}{v8,vc2}
\fmf{plain,tension=0.5}{vc1,vc2}
\fmffreeze
\fmfposition
\fmf{plain,tension=1,right=0,width=1mm}{v7,v12}
\fmfipath{p[]}
\fmfiset{p1}{vpath(__vc1,__vc2)}
\fmfiset{p2}{vpath(__v9,__vc6)}
\fmfipair{w[]}
\svertex{w1}{p1}
\svertex{w2}{p2}
\fmfi{wiggly}{w1..w2}
\fmffreeze
\end{fmfchar*}}}
}
\subfigure[$G_{3}$]{
\raisebox{\eqoff}{%
\fmfframe(3,1)(1,4){%
\begin{fmfchar*}(30,20)
\fmftop{v1}
\fmfbottom{v7}
\fmfforce{(0w,h)}{v1}
\fmfforce{(0w,0)}{v7}
\fmffixed{(0.2w,0)}{v1,v2}
\fmffixed{(0.2w,0)}{v2,v3}
\fmffixed{(0.2w,0)}{v3,v4}
\fmffixed{(0.2w,0)}{v4,v5}
\fmffixed{(0.2w,0)}{v5,v6}
\fmffixed{(0.2w,0)}{v7,v8}
\fmffixed{(0.2w,0)}{v8,v9}
\fmffixed{(0.2w,0)}{v9,v10}
\fmffixed{(0.2w,0)}{v10,v11}
\fmffixed{(0.2w,0)}{v11,v12}
\fmf{plain,tension=0.25,right=0.25}{v5,vc9}
\fmf{plain,tension=0.25,left=0.25}{v6,vc9}
\fmf{plain,right=0.25}{v11,vc8}
\fmf{plain,tension=1,right=0.25}{v3,vc5}
\fmf{plain,tension=1,right=0.25}{v4,vc7}
\fmf{plain,right=0.25}{v12,vc10}
\fmf{plain,tension=0.25,left=0.25}{v9,vc6}
\fmf{plain,tension=0.25,right=0.25}{v10,vc6}
\fmf{plain,tension=0.5}{vc5,vc6}
\fmf{plain,tension=0.5}{vc5,vc8}
\fmf{plain,tension=0.5}{vc7,vc8}
\fmf{plain,tension=0.5}{vc7,vc10}
\fmf{plain,tension=0.5}{vc9,vc10}
\fmf{plain,tension=0.25,right=0.25}{v1,vc1}
\fmf{plain,tension=0.25,left=0.25}{v2,vc1}
\fmf{plain,tension=0.25,left=0.25}{v7,vc2}
\fmf{plain,tension=0.25,right=0.25}{v8,vc2}
\fmf{plain,tension=0.5}{vc1,vc2}
\fmffreeze
\fmfposition
\fmf{plain,tension=1,right=0,width=1mm}{v7,v12}
\fmfipath{p[]}
\fmfiset{p1}{vpath(__v8,__vc2)}
\fmfiset{p2}{vpath(__vc5,__vc6)}
\fmfipair{w[]}
\svertex{w1}{p1}
\svertex{w2}{p2}
\fmfi{wiggly}{w1..w2}
\fmffreeze
\end{fmfchar*}}}
}
\subfigure[$G_{4}$]{
\raisebox{\eqoff}{%
\fmfframe(3,1)(1,4){%
\begin{fmfchar*}(30,20)
\fmftop{v1}
\fmfbottom{v7}
\fmfforce{(0w,h)}{v1}
\fmfforce{(0w,0)}{v7}
\fmffixed{(0.2w,0)}{v1,v2}
\fmffixed{(0.2w,0)}{v2,v3}
\fmffixed{(0.2w,0)}{v3,v4}
\fmffixed{(0.2w,0)}{v4,v5}
\fmffixed{(0.2w,0)}{v5,v6}
\fmffixed{(0.2w,0)}{v7,v8}
\fmffixed{(0.2w,0)}{v8,v9}
\fmffixed{(0.2w,0)}{v9,v10}
\fmffixed{(0.2w,0)}{v10,v11}
\fmffixed{(0.2w,0)}{v11,v12}
\fmf{plain,tension=0.25,right=0.25}{v5,vc9}
\fmf{plain,tension=0.25,left=0.25}{v6,vc9}
\fmf{plain,right=0.25}{v11,vc8}
\fmf{plain,tension=1,right=0.25}{v3,vc5}
\fmf{plain,tension=1,right=0.25}{v4,vc7}
\fmf{plain,right=0.25}{v12,vc10}
\fmf{plain,tension=0.25,left=0.25}{v9,vc6}
\fmf{plain,tension=0.25,right=0.25}{v10,vc6}
\fmf{plain,tension=0.5}{vc5,vc6}
\fmf{plain,tension=0.5}{vc5,vc8}
\fmf{plain,tension=0.5}{vc7,vc8}
\fmf{plain,tension=0.5}{vc7,vc10}
\fmf{plain,tension=0.5}{vc9,vc10}
\fmf{plain,tension=0.25,right=0.25}{v1,vc1}
\fmf{plain,tension=0.25,left=0.25}{v2,vc1}
\fmf{plain,tension=0.25,left=0.25}{v7,vc2}
\fmf{plain,tension=0.25,right=0.25}{v8,vc2}
\fmf{plain,tension=0.5}{vc1,vc2}
\fmffreeze
\fmfposition
\fmf{plain,tension=1,right=0,width=1mm}{v7,v12}
\fmfipath{p[]}
\fmfiset{p1}{vpath(__v8,__vc2)}
\fmfiset{p2}{vpath(__v9,__vc6)}
\fmfipair{w[]}
\svertex{w1}{p1}
\svertex{w2}{p2}
\fmfi{wiggly}{w1..w2}
\fmffreeze
\end{fmfchar*}}}
}
\caption{Range-six diagrams with structure $\chi(1,5,4,3)$}
\label{r6-1543}
\end{figure}

\begin{figure}[!h]
\centering
\unitlength=0.75mm
\settoheight{\eqoff}{$\times$}%
\setlength{\eqoff}{0.5\eqoff}%
\addtolength{\eqoff}{-12.5\unitlength}%
\settoheight{\eqofftwo}{$\times$}%
\setlength{\eqofftwo}{0.5\eqofftwo}%
\addtolength{\eqofftwo}{-7.5\unitlength}%
\subfigure[$G_{1}$]{
\raisebox{\eqoff}{%
\fmfframe(3,1)(1,4){%
\begin{fmfchar*}(30,20)
\fmftop{v1}
\fmfbottom{v7}
\fmfforce{(0w,h)}{v1}
\fmfforce{(0w,0)}{v7}
\fmffixed{(0.2w,0)}{v1,v2}
\fmffixed{(0.2w,0)}{v2,v3}
\fmffixed{(0.2w,0)}{v3,v4}
\fmffixed{(0.2w,0)}{v4,v5}
\fmffixed{(0.2w,0)}{v5,v6}
\fmffixed{(0.2w,0)}{v7,v8}
\fmffixed{(0.2w,0)}{v8,v9}
\fmffixed{(0.2w,0)}{v9,v10}
\fmffixed{(0.2w,0)}{v10,v11}
\fmffixed{(0.2w,0)}{v11,v12}
\fmf{plain,tension=0.25,right=0.25}{v3,vc5}
\fmf{plain,tension=0.25,left=0.25}{v4,vc5}
\fmf{plain,left=0.25}{v9,vc6}
\fmf{plain,tension=1,left=0.25}{v5,vc7}
\fmf{plain,tension=1,left=0.25}{v6,vc9}
\fmf{plain,left=0.25}{v10,vc8}
\fmf{plain,tension=0.25,left=0.25}{v11,vc10}
\fmf{plain,tension=0.25,right=0.25}{v12,vc10}
\fmf{plain,tension=0.5}{vc5,vc6}
\fmf{plain,tension=0.5}{vc6,vc7}
\fmf{plain,tension=0.5}{vc7,vc8}
\fmf{plain,tension=0.5}{vc8,vc9}
\fmf{plain,tension=0.5}{vc9,vc10}
\fmf{plain,tension=0.25,right=0.25}{v1,vc1}
\fmf{plain,tension=0.25,left=0.25}{v2,vc1}
\fmf{plain,tension=0.25,left=0.25}{v7,vc2}
\fmf{plain,tension=0.25,right=0.25}{v8,vc2}
\fmf{plain,tension=0.5}{vc1,vc2}
\fmffreeze
\fmfposition
\fmf{plain,tension=1,right=0,width=1mm}{v7,v12}
\fmfipath{p[]}
\fmfiset{p1}{vpath(__vc1,__vc2)}
\fmfiset{p2}{vpath(__v9,__vc6)}
\fmfipair{w[]}
\svertex{w1}{p1}
\svertex{w2}{p2}
\fmfi{wiggly}{w1..w2}
\fmffreeze
\end{fmfchar*}}}
}
\subfigure[$G_{2}$]{
\raisebox{\eqoff}{%
\fmfframe(3,1)(1,4){%
\begin{fmfchar*}(30,20)
\fmftop{v1}
\fmfbottom{v7}
\fmfforce{(0w,h)}{v1}
\fmfforce{(0w,0)}{v7}
\fmffixed{(0.2w,0)}{v1,v2}
\fmffixed{(0.2w,0)}{v2,v3}
\fmffixed{(0.2w,0)}{v3,v4}
\fmffixed{(0.2w,0)}{v4,v5}
\fmffixed{(0.2w,0)}{v5,v6}
\fmffixed{(0.2w,0)}{v7,v8}
\fmffixed{(0.2w,0)}{v8,v9}
\fmffixed{(0.2w,0)}{v9,v10}
\fmffixed{(0.2w,0)}{v10,v11}
\fmffixed{(0.2w,0)}{v11,v12}
\fmf{plain,tension=0.25,right=0.25}{v3,vc5}
\fmf{plain,tension=0.25,left=0.25}{v4,vc5}
\fmf{plain,left=0.25}{v9,vc6}
\fmf{plain,tension=1,left=0.25}{v5,vc7}
\fmf{plain,tension=1,left=0.25}{v6,vc9}
\fmf{plain,left=0.25}{v10,vc8}
\fmf{plain,tension=0.25,left=0.25}{v11,vc10}
\fmf{plain,tension=0.25,right=0.25}{v12,vc10}
\fmf{plain,tension=0.5}{vc5,vc6}
\fmf{plain,tension=0.5}{vc6,vc7}
\fmf{plain,tension=0.5}{vc7,vc8}
\fmf{plain,tension=0.5}{vc8,vc9}
\fmf{plain,tension=0.5}{vc9,vc10}
\fmf{plain,tension=0.25,right=0.25}{v1,vc1}
\fmf{plain,tension=0.25,left=0.25}{v2,vc1}
\fmf{plain,tension=0.25,left=0.25}{v7,vc2}
\fmf{plain,tension=0.25,right=0.25}{v8,vc2}
\fmf{plain,tension=0.5}{vc1,vc2}
\fmffreeze
\fmfposition
\fmf{plain,tension=1,right=0,width=1mm}{v7,v12}
\fmfipath{p[]}
\fmfiset{p1}{vpath(__vc1,__vc2)}
\fmfiset{p2}{vpath(__vc5,__vc6)}
\fmfipair{w[]}
\svertex{w1}{p1}
\svertex{w2}{p2}
\fmfi{wiggly}{w1..w2}
\fmffreeze
\end{fmfchar*}}}
}
\subfigure[$G_{3}$]{
\raisebox{\eqoff}{%
\fmfframe(3,1)(1,4){%
\begin{fmfchar*}(30,20)
\fmftop{v1}
\fmfbottom{v7}
\fmfforce{(0w,h)}{v1}
\fmfforce{(0w,0)}{v7}
\fmffixed{(0.2w,0)}{v1,v2}
\fmffixed{(0.2w,0)}{v2,v3}
\fmffixed{(0.2w,0)}{v3,v4}
\fmffixed{(0.2w,0)}{v4,v5}
\fmffixed{(0.2w,0)}{v5,v6}
\fmffixed{(0.2w,0)}{v7,v8}
\fmffixed{(0.2w,0)}{v8,v9}
\fmffixed{(0.2w,0)}{v9,v10}
\fmffixed{(0.2w,0)}{v10,v11}
\fmffixed{(0.2w,0)}{v11,v12}
\fmf{plain,tension=0.25,right=0.25}{v3,vc5}
\fmf{plain,tension=0.25,left=0.25}{v4,vc5}
\fmf{plain,left=0.25}{v9,vc6}
\fmf{plain,tension=1,left=0.25}{v5,vc7}
\fmf{plain,tension=1,left=0.25}{v6,vc9}
\fmf{plain,left=0.25}{v10,vc8}
\fmf{plain,tension=0.25,left=0.25}{v11,vc10}
\fmf{plain,tension=0.25,right=0.25}{v12,vc10}
\fmf{plain,tension=0.5}{vc5,vc6}
\fmf{plain,tension=0.5}{vc6,vc7}
\fmf{plain,tension=0.5}{vc7,vc8}
\fmf{plain,tension=0.5}{vc8,vc9}
\fmf{plain,tension=0.5}{vc9,vc10}
\fmf{plain,tension=0.25,right=0.25}{v1,vc1}
\fmf{plain,tension=0.25,left=0.25}{v2,vc1}
\fmf{plain,tension=0.25,left=0.25}{v7,vc2}
\fmf{plain,tension=0.25,right=0.25}{v8,vc2}
\fmf{plain,tension=0.5}{vc1,vc2}
\fmffreeze
\fmfposition
\fmf{plain,tension=1,right=0,width=1mm}{v7,v12}
\fmfipath{p[]}
\fmfiset{p1}{vpath(__v8,__vc2)}
\fmfiset{p2}{vpath(__vc5,__vc6)}
\fmfipair{w[]}
\svertex{w1}{p1}
\svertex{w2}{p2}
\fmfi{wiggly}{w1..w2}
\fmffreeze
\end{fmfchar*}}}
}
\subfigure[$G_{4}$]{
\raisebox{\eqoff}{%
\fmfframe(3,1)(1,4){%
\begin{fmfchar*}(30,20)
\fmftop{v1}
\fmfbottom{v7}
\fmfforce{(0w,h)}{v1}
\fmfforce{(0w,0)}{v7}
\fmffixed{(0.2w,0)}{v1,v2}
\fmffixed{(0.2w,0)}{v2,v3}
\fmffixed{(0.2w,0)}{v3,v4}
\fmffixed{(0.2w,0)}{v4,v5}
\fmffixed{(0.2w,0)}{v5,v6}
\fmffixed{(0.2w,0)}{v7,v8}
\fmffixed{(0.2w,0)}{v8,v9}
\fmffixed{(0.2w,0)}{v9,v10}
\fmffixed{(0.2w,0)}{v10,v11}
\fmffixed{(0.2w,0)}{v11,v12}
\fmf{plain,tension=0.25,right=0.25}{v3,vc5}
\fmf{plain,tension=0.25,left=0.25}{v4,vc5}
\fmf{plain,left=0.25}{v9,vc6}
\fmf{plain,tension=1,left=0.25}{v5,vc7}
\fmf{plain,tension=1,left=0.25}{v6,vc9}
\fmf{plain,left=0.25}{v10,vc8}
\fmf{plain,tension=0.25,left=0.25}{v11,vc10}
\fmf{plain,tension=0.25,right=0.25}{v12,vc10}
\fmf{plain,tension=0.5}{vc5,vc6}
\fmf{plain,tension=0.5}{vc6,vc7}
\fmf{plain,tension=0.5}{vc7,vc8}
\fmf{plain,tension=0.5}{vc8,vc9}
\fmf{plain,tension=0.5}{vc9,vc10}
\fmf{plain,tension=0.25,right=0.25}{v1,vc1}
\fmf{plain,tension=0.25,left=0.25}{v2,vc1}
\fmf{plain,tension=0.25,left=0.25}{v7,vc2}
\fmf{plain,tension=0.25,right=0.25}{v8,vc2}
\fmf{plain,tension=0.5}{vc1,vc2}
\fmffreeze
\fmfposition
\fmf{plain,tension=1,right=0,width=1mm}{v7,v12}
\fmfipath{p[]}
\fmfiset{p1}{vpath(__v8,__vc2)}
\fmfiset{p2}{vpath(__v9,__vc6)}
\fmfipair{w[]}
\svertex{w1}{p1}
\svertex{w2}{p2}
\fmfi{wiggly}{w1..w2}
\fmffreeze
\end{fmfchar*}}}
}
\caption{Range-six diagrams with structure $\chi(1,3,4,5)$}
\label{r6-1345}
\end{figure}

\begin{figure}[!h]
\centering
\unitlength=0.75mm
\settoheight{\eqoff}{$\times$}%
\setlength{\eqoff}{0.5\eqoff}%
\addtolength{\eqoff}{-12.5\unitlength}%
\settoheight{\eqofftwo}{$\times$}%
\setlength{\eqofftwo}{0.5\eqofftwo}%
\addtolength{\eqofftwo}{-7.5\unitlength}%
\subfigure[$G_{1}$]{
\raisebox{\eqoff}{%
\fmfframe(3,1)(1,4){%
\begin{fmfchar*}(30,20)
\fmftop{v1}
\fmfbottom{v7}
\fmfforce{(0w,h)}{v1}
\fmfforce{(0w,0)}{v7}
\fmffixed{(0.2w,0)}{v1,v2}
\fmffixed{(0.2w,0)}{v2,v3}
\fmffixed{(0.2w,0)}{v3,v4}
\fmffixed{(0.2w,0)}{v4,v5}
\fmffixed{(0.2w,0)}{v5,v6}
\fmffixed{(0.2w,0)}{v7,v8}
\fmffixed{(0.2w,0)}{v8,v9}
\fmffixed{(0.2w,0)}{v9,v10}
\fmffixed{(0.2w,0)}{v10,v11}
\fmffixed{(0.2w,0)}{v11,v12}
\fmf{plain,tension=1,left=0.25}{v9,vc6}
\fmf{plain,tension=1,right=0.25}{v10,vc6}
\fmf{plain,tension=1,left=0.25}{v11,vc10}
\fmf{plain,tension=1,right=0.25}{v12,vc10}
\fmf{plain,tension=1,right=0.125}{v3,vc5}
\fmf{plain,tension=0.25,right=0.25}{v4,vc7}
\fmf{plain,tension=0.25,left=0.25}{v5,vc7}
\fmf{plain,tension=1,left=0.125}{v6,vc9}
\fmf{plain,tension=4}{vc5,vc6}
\fmf{plain,tension=4}{vc9,vc10}
\fmf{plain,tension=0.5}{vc5,vc8}
\fmf{plain,tension=0.5}{vc8,vc9}
\fmf{plain,tension=1}{vc7,vc8}
\fmf{plain,tension=1,right=0.25}{v1,vc1}
\fmf{plain,tension=1,left=0.25}{v2,vc1}
\fmf{plain,tension=1,left=0.25}{v7,vc2}
\fmf{plain,tension=1,right=0.25}{v8,vc2}
\fmf{plain,tension=1.5}{vc1,vc2}
\fmffreeze
\fmfposition
\fmf{plain,tension=1,right=0,width=1mm}{v7,v12}
\fmfipath{p[]}
\fmfiset{p1}{vpath(__v8,__vc2)}
\fmfiset{p2}{vpath(__v9,__vc6)}
\fmfipair{w[]}
\svertex{w1}{p1}
\svertex{w2}{p2}
\fmfi{wiggly}{w1..w2}
\fmffreeze
\end{fmfchar*}}}
}
\subfigure[$G_{2}$]{
\raisebox{\eqoff}{%
\fmfframe(3,1)(1,4){%
\begin{fmfchar*}(30,20)
\fmftop{v1}
\fmfbottom{v7}
\fmfforce{(0w,h)}{v1}
\fmfforce{(0w,0)}{v7}
\fmffixed{(0.2w,0)}{v1,v2}
\fmffixed{(0.2w,0)}{v2,v3}
\fmffixed{(0.2w,0)}{v3,v4}
\fmffixed{(0.2w,0)}{v4,v5}
\fmffixed{(0.2w,0)}{v5,v6}
\fmffixed{(0.2w,0)}{v7,v8}
\fmffixed{(0.2w,0)}{v8,v9}
\fmffixed{(0.2w,0)}{v9,v10}
\fmffixed{(0.2w,0)}{v10,v11}
\fmffixed{(0.2w,0)}{v11,v12}
\fmf{plain,tension=1,left=0.25}{v9,vc6}
\fmf{plain,tension=1,right=0.25}{v10,vc6}
\fmf{plain,tension=1,left=0.25}{v11,vc10}
\fmf{plain,tension=1,right=0.25}{v12,vc10}
\fmf{plain,tension=1,right=0.125}{v3,vc5}
\fmf{plain,tension=0.25,right=0.25}{v4,vc7}
\fmf{plain,tension=0.25,left=0.25}{v5,vc7}
\fmf{plain,tension=1,left=0.125}{v6,vc9}
\fmf{plain,tension=4}{vc5,vc6}
\fmf{plain,tension=4}{vc9,vc10}
\fmf{plain,tension=0.5}{vc5,vc8}
\fmf{plain,tension=0.5}{vc8,vc9}
\fmf{plain,tension=1}{vc7,vc8}
\fmf{plain,tension=1,right=0.25}{v1,vc1}
\fmf{plain,tension=1,left=0.25}{v2,vc1}
\fmf{plain,tension=1,left=0.25}{v7,vc2}
\fmf{plain,tension=1,right=0.25}{v8,vc2}
\fmf{plain,tension=1.5}{vc1,vc2}
\fmffreeze
\fmfposition
\fmf{plain,tension=1,right=0,width=1mm}{v7,v12}
\fmfipath{p[]}
\fmfiset{p1}{vpath(__v8,__vc2)}
\fmfiset{p2}{vpath(__vc5,__vc6)}
\fmfipair{w[]}
\svertex{w1}{p1}
\svertex{w2}{p2}
\fmfi{wiggly}{w1..w2}
\fmffreeze
\end{fmfchar*}}}
}
\subfigure[$G_{3}$]{
\raisebox{\eqoff}{%
\fmfframe(3,1)(1,4){%
\begin{fmfchar*}(30,20)
\fmftop{v1}
\fmfbottom{v7}
\fmfforce{(0w,h)}{v1}
\fmfforce{(0w,0)}{v7}
\fmffixed{(0.2w,0)}{v1,v2}
\fmffixed{(0.2w,0)}{v2,v3}
\fmffixed{(0.2w,0)}{v3,v4}
\fmffixed{(0.2w,0)}{v4,v5}
\fmffixed{(0.2w,0)}{v5,v6}
\fmffixed{(0.2w,0)}{v7,v8}
\fmffixed{(0.2w,0)}{v8,v9}
\fmffixed{(0.2w,0)}{v9,v10}
\fmffixed{(0.2w,0)}{v10,v11}
\fmffixed{(0.2w,0)}{v11,v12}
\fmf{plain,tension=1,left=0.25}{v9,vc6}
\fmf{plain,tension=1,right=0.25}{v10,vc6}
\fmf{plain,tension=1,left=0.25}{v11,vc10}
\fmf{plain,tension=1,right=0.25}{v12,vc10}
\fmf{plain,tension=1,right=0.125}{v3,vc5}
\fmf{plain,tension=0.25,right=0.25}{v4,vc7}
\fmf{plain,tension=0.25,left=0.25}{v5,vc7}
\fmf{plain,tension=1,left=0.125}{v6,vc9}
\fmf{plain,tension=4}{vc5,vc6}
\fmf{plain,tension=4}{vc9,vc10}
\fmf{plain,tension=0.5}{vc5,vc8}
\fmf{plain,tension=0.5}{vc8,vc9}
\fmf{plain,tension=1}{vc7,vc8}
\fmf{plain,tension=1,right=0.25}{v1,vc1}
\fmf{plain,tension=1,left=0.25}{v2,vc1}
\fmf{plain,tension=1,left=0.25}{v7,vc2}
\fmf{plain,tension=1,right=0.25}{v8,vc2}
\fmf{plain,tension=1.5}{vc1,vc2}
\fmffreeze
\fmfposition
\fmf{plain,tension=1,right=0,width=1mm}{v7,v12}
\fmfipath{p[]}
\fmfiset{p1}{vpath(__vc1,__vc2)}
\fmfiset{p2}{vpath(__v9,__vc6)}
\fmfipair{w[]}
\svertex{w1}{p1}
\svertex{w2}{p2}
\fmfi{wiggly}{w1..w2}
\fmffreeze
\end{fmfchar*}}}
}
\subfigure[$G_{4}$]{
\raisebox{\eqoff}{%
\fmfframe(3,1)(1,4){%
\begin{fmfchar*}(30,20)
\fmftop{v1}
\fmfbottom{v7}
\fmfforce{(0w,h)}{v1}
\fmfforce{(0w,0)}{v7}
\fmffixed{(0.2w,0)}{v1,v2}
\fmffixed{(0.2w,0)}{v2,v3}
\fmffixed{(0.2w,0)}{v3,v4}
\fmffixed{(0.2w,0)}{v4,v5}
\fmffixed{(0.2w,0)}{v5,v6}
\fmffixed{(0.2w,0)}{v7,v8}
\fmffixed{(0.2w,0)}{v8,v9}
\fmffixed{(0.2w,0)}{v9,v10}
\fmffixed{(0.2w,0)}{v10,v11}
\fmffixed{(0.2w,0)}{v11,v12}
\fmf{plain,tension=1,left=0.25}{v9,vc6}
\fmf{plain,tension=1,right=0.25}{v10,vc6}
\fmf{plain,tension=1,left=0.25}{v11,vc10}
\fmf{plain,tension=1,right=0.25}{v12,vc10}
\fmf{plain,tension=1,right=0.125}{v3,vc5}
\fmf{plain,tension=0.25,right=0.25}{v4,vc7}
\fmf{plain,tension=0.25,left=0.25}{v5,vc7}
\fmf{plain,tension=1,left=0.125}{v6,vc9}
\fmf{plain,tension=4}{vc5,vc6}
\fmf{plain,tension=4}{vc9,vc10}
\fmf{plain,tension=0.5}{vc5,vc8}
\fmf{plain,tension=0.5}{vc8,vc9}
\fmf{plain,tension=1}{vc7,vc8}
\fmf{plain,tension=1,right=0.25}{v1,vc1}
\fmf{plain,tension=1,left=0.25}{v2,vc1}
\fmf{plain,tension=1,left=0.25}{v7,vc2}
\fmf{plain,tension=1,right=0.25}{v8,vc2}
\fmf{plain,tension=1.5}{vc1,vc2}
\fmffreeze
\fmfposition
\fmf{plain,tension=1,right=0,width=1mm}{v7,v12}
\fmfipath{p[]}
\fmfiset{p1}{vpath(__vc1,__vc2)}
\fmfiset{p2}{vpath(__vc5,__vc6)}
\fmfipair{w[]}
\svertex{w1}{p1}
\svertex{w2}{p2}
\fmfi{wiggly}{w1..w2}
\fmffreeze
\end{fmfchar*}}}
}
\caption{Range-six diagrams with structure $\chi(1,4,3,5)$}
\label{r6-1435}
\end{figure}

\begin{figure}[!h]
\centering
\unitlength=0.75mm
\settoheight{\eqoff}{$\times$}%
\setlength{\eqoff}{0.5\eqoff}%
\addtolength{\eqoff}{-12.5\unitlength}%
\settoheight{\eqofftwo}{$\times$}%
\setlength{\eqofftwo}{0.5\eqofftwo}%
\addtolength{\eqofftwo}{-7.5\unitlength}%
\subfigure[$G_{1}$]{
\raisebox{\eqoff}{%
\fmfframe(3,1)(1,4){%
\begin{fmfchar*}(30,20)
\fmftop{v1}
\fmfbottom{v7}
\fmfforce{(0w,h)}{v1}
\fmfforce{(0w,0)}{v7}
\fmffixed{(0.2w,0)}{v1,v2}
\fmffixed{(0.2w,0)}{v2,v3}
\fmffixed{(0.2w,0)}{v3,v4}
\fmffixed{(0.2w,0)}{v4,v5}
\fmffixed{(0.2w,0)}{v5,v6}
\fmffixed{(0.2w,0)}{v7,v8}
\fmffixed{(0.2w,0)}{v8,v9}
\fmffixed{(0.2w,0)}{v9,v10}
\fmffixed{(0.2w,0)}{v10,v11}
\fmffixed{(0.2w,0)}{v11,v12}
\fmf{plain,tension=1,right=0.25}{v3,vc5}
\fmf{plain,tension=1,left=0.25}{v4,vc5}
\fmf{plain,tension=1,right=0.25}{v5,vc9}
\fmf{plain,tension=1,left=0.25}{v6,vc9}
\fmf{plain,tension=1,left=0.125}{v9,vc6}
\fmf{plain,tension=0.25,left=0.25}{v10,vc8}
\fmf{plain,tension=0.25,right=0.25}{v11,vc8}
\fmf{plain,tension=1,right=0.125}{v12,vc10}
\fmf{plain,tension=4}{vc5,vc6}
\fmf{plain,tension=4}{vc9,vc10}
\fmf{plain,tension=0.5}{vc6,vc7}
\fmf{plain,tension=0.5}{vc7,vc10}
\fmf{plain,tension=1}{vc7,vc8}
\fmf{plain,tension=1,right=0.25}{v1,vc1}
\fmf{plain,tension=1,left=0.25}{v2,vc1}
\fmf{plain,tension=1,left=0.25}{v7,vc2}
\fmf{plain,tension=1,right=0.25}{v8,vc2}
\fmf{plain,tension=1.5}{vc1,vc2}
\fmffreeze
\fmfposition
\fmf{plain,tension=1,right=0,width=1mm}{v7,v12}
\fmfipath{p[]}
\fmfiset{p1}{vpath(__v8,__vc2)}
\fmfiset{p2}{vpath(__v9,__vc6)}
\fmfipair{w[]}
\svertex{w1}{p1}
\svertex{w2}{p2}
\fmfi{wiggly}{w1..w2}
\fmffreeze
\end{fmfchar*}}}
}
\subfigure[$G_{2}$]{
\raisebox{\eqoff}{%
\fmfframe(3,1)(1,4){%
\begin{fmfchar*}(30,20)
\fmftop{v1}
\fmfbottom{v7}
\fmfforce{(0w,h)}{v1}
\fmfforce{(0w,0)}{v7}
\fmffixed{(0.2w,0)}{v1,v2}
\fmffixed{(0.2w,0)}{v2,v3}
\fmffixed{(0.2w,0)}{v3,v4}
\fmffixed{(0.2w,0)}{v4,v5}
\fmffixed{(0.2w,0)}{v5,v6}
\fmffixed{(0.2w,0)}{v7,v8}
\fmffixed{(0.2w,0)}{v8,v9}
\fmffixed{(0.2w,0)}{v9,v10}
\fmffixed{(0.2w,0)}{v10,v11}
\fmffixed{(0.2w,0)}{v11,v12}
\fmf{plain,tension=1,right=0.25}{v3,vc5}
\fmf{plain,tension=1,left=0.25}{v4,vc5}
\fmf{plain,tension=1,right=0.25}{v5,vc9}
\fmf{plain,tension=1,left=0.25}{v6,vc9}
\fmf{plain,tension=1,left=0.125}{v9,vc6}
\fmf{plain,tension=0.25,left=0.25}{v10,vc8}
\fmf{plain,tension=0.25,right=0.25}{v11,vc8}
\fmf{plain,tension=1,right=0.125}{v12,vc10}
\fmf{plain,tension=4}{vc5,vc6}
\fmf{plain,tension=4}{vc9,vc10}
\fmf{plain,tension=0.5}{vc6,vc7}
\fmf{plain,tension=0.5}{vc7,vc10}
\fmf{plain,tension=1}{vc7,vc8}
\fmf{plain,tension=1,right=0.25}{v1,vc1}
\fmf{plain,tension=1,left=0.25}{v2,vc1}
\fmf{plain,tension=1,left=0.25}{v7,vc2}
\fmf{plain,tension=1,right=0.25}{v8,vc2}
\fmf{plain,tension=1.5}{vc1,vc2}
\fmffreeze
\fmfposition
\fmf{plain,tension=1,right=0,width=1mm}{v7,v12}
\fmfipath{p[]}
\fmfiset{p1}{vpath(__vc1,__vc2)}
\fmfiset{p2}{vpath(__v9,__vc6)}
\fmfipair{w[]}
\svertex{w1}{p1}
\svertex{w2}{p2}
\fmfi{wiggly}{w1..w2}
\fmffreeze
\end{fmfchar*}}}
}
\subfigure[$G_{3}$]{
\raisebox{\eqoff}{%
\fmfframe(3,1)(1,4){%
\begin{fmfchar*}(30,20)
\fmftop{v1}
\fmfbottom{v7}
\fmfforce{(0w,h)}{v1}
\fmfforce{(0w,0)}{v7}
\fmffixed{(0.2w,0)}{v1,v2}
\fmffixed{(0.2w,0)}{v2,v3}
\fmffixed{(0.2w,0)}{v3,v4}
\fmffixed{(0.2w,0)}{v4,v5}
\fmffixed{(0.2w,0)}{v5,v6}
\fmffixed{(0.2w,0)}{v7,v8}
\fmffixed{(0.2w,0)}{v8,v9}
\fmffixed{(0.2w,0)}{v9,v10}
\fmffixed{(0.2w,0)}{v10,v11}
\fmffixed{(0.2w,0)}{v11,v12}
\fmf{plain,tension=1,right=0.25}{v3,vc5}
\fmf{plain,tension=1,left=0.25}{v4,vc5}
\fmf{plain,tension=1,right=0.25}{v5,vc9}
\fmf{plain,tension=1,left=0.25}{v6,vc9}
\fmf{plain,tension=1,left=0.125}{v9,vc6}
\fmf{plain,tension=0.25,left=0.25}{v10,vc8}
\fmf{plain,tension=0.25,right=0.25}{v11,vc8}
\fmf{plain,tension=1,right=0.125}{v12,vc10}
\fmf{plain,tension=4}{vc5,vc6}
\fmf{plain,tension=4}{vc9,vc10}
\fmf{plain,tension=0.5}{vc6,vc7}
\fmf{plain,tension=0.5}{vc7,vc10}
\fmf{plain,tension=1}{vc7,vc8}
\fmf{plain,tension=1,right=0.25}{v1,vc1}
\fmf{plain,tension=1,left=0.25}{v2,vc1}
\fmf{plain,tension=1,left=0.25}{v7,vc2}
\fmf{plain,tension=1,right=0.25}{v8,vc2}
\fmf{plain,tension=1.5}{vc1,vc2}
\fmffreeze
\fmfposition
\fmf{plain,tension=1,right=0,width=1mm}{v7,v12}
\fmfipath{p[]}
\fmfiset{p1}{vpath(__v8,__vc2)}
\fmfiset{p2}{vpath(__vc5,__vc6)}
\fmfipair{w[]}
\svertex{w1}{p1}
\svertex{w2}{p2}
\fmfi{wiggly}{w1..w2}
\fmffreeze
\end{fmfchar*}}}
}
\subfigure[$G_{4}$]{
\raisebox{\eqoff}{%
\fmfframe(3,1)(1,4){%
\begin{fmfchar*}(30,20)
\fmftop{v1}
\fmfbottom{v7}
\fmfforce{(0w,h)}{v1}
\fmfforce{(0w,0)}{v7}
\fmffixed{(0.2w,0)}{v1,v2}
\fmffixed{(0.2w,0)}{v2,v3}
\fmffixed{(0.2w,0)}{v3,v4}
\fmffixed{(0.2w,0)}{v4,v5}
\fmffixed{(0.2w,0)}{v5,v6}
\fmffixed{(0.2w,0)}{v7,v8}
\fmffixed{(0.2w,0)}{v8,v9}
\fmffixed{(0.2w,0)}{v9,v10}
\fmffixed{(0.2w,0)}{v10,v11}
\fmffixed{(0.2w,0)}{v11,v12}
\fmf{plain,tension=1,right=0.25}{v3,vc5}
\fmf{plain,tension=1,left=0.25}{v4,vc5}
\fmf{plain,tension=1,right=0.25}{v5,vc9}
\fmf{plain,tension=1,left=0.25}{v6,vc9}
\fmf{plain,tension=1,left=0.125}{v9,vc6}
\fmf{plain,tension=0.25,left=0.25}{v10,vc8}
\fmf{plain,tension=0.25,right=0.25}{v11,vc8}
\fmf{plain,tension=1,right=0.125}{v12,vc10}
\fmf{plain,tension=4}{vc5,vc6}
\fmf{plain,tension=4}{vc9,vc10}
\fmf{plain,tension=0.5}{vc6,vc7}
\fmf{plain,tension=0.5}{vc7,vc10}
\fmf{plain,tension=1}{vc7,vc8}
\fmf{plain,tension=1,right=0.25}{v1,vc1}
\fmf{plain,tension=1,left=0.25}{v2,vc1}
\fmf{plain,tension=1,left=0.25}{v7,vc2}
\fmf{plain,tension=1,right=0.25}{v8,vc2}
\fmf{plain,tension=1.5}{vc1,vc2}
\fmffreeze
\fmfposition
\fmf{plain,tension=1,right=0,width=1mm}{v7,v12}
\fmfipath{p[]}
\fmfiset{p1}{vpath(__vc1,__vc2)}
\fmfiset{p2}{vpath(__vc5,__vc6)}
\fmfipair{w[]}
\svertex{w1}{p1}
\svertex{w2}{p2}
\fmfi{wiggly}{w1..w2}
\fmffreeze
\end{fmfchar*}}}
}
\caption{Range-six diagrams with structure $\chi(1,3,2,5)$}
\label{r6-1325}
\end{figure}

\begin{figure}[!h]
\centering
\unitlength=0.75mm
\settoheight{\eqoff}{$\times$}%
\setlength{\eqoff}{0.5\eqoff}%
\addtolength{\eqoff}{-12.5\unitlength}%
\settoheight{\eqofftwo}{$\times$}%
\setlength{\eqofftwo}{0.5\eqofftwo}%
\addtolength{\eqofftwo}{-7.5\unitlength}%
\subfigure[$G_{1}$]{
\raisebox{\eqoff}{%
\fmfframe(3,1)(1,4){%
\begin{fmfchar*}(30,20)
\fmftop{v1}
\fmfbottom{v7}
\fmfforce{(0w,h)}{v1}
\fmfforce{(0w,0)}{v7}
\fmffixed{(0.2w,0)}{v1,v2}
\fmffixed{(0.2w,0)}{v2,v3}
\fmffixed{(0.2w,0)}{v3,v4}
\fmffixed{(0.2w,0)}{v4,v5}
\fmffixed{(0.2w,0)}{v5,v6}
\fmffixed{(0.2w,0)}{v7,v8}
\fmffixed{(0.2w,0)}{v8,v9}
\fmffixed{(0.2w,0)}{v9,v10}
\fmffixed{(0.2w,0)}{v10,v11}
\fmffixed{(0.2w,0)}{v11,v12}
\fmf{plain,tension=0.5,right=0.25}{v1,vc1}
\fmf{plain,tension=0.5,left=0.25}{v2,vc1}
\fmf{plain,left=0.25}{v3,vc3}
\fmf{plain,tension=0.5,left=0.25}{v8,vc4}
\fmf{plain,tension=0.5,right=0.25}{v9,vc4}
\fmf{plain,left=0.25}{v7,vc2}
\fmf{plain,tension=1}{vc1,vc2}
\fmf{plain,tension=0.5}{vc2,vc3}
\fmf{plain,tension=1}{vc3,vc4}
\fmf{plain,tension=0.5,right=0.25}{v5,vc9}
\fmf{plain,tension=0.5,left=0.25}{v6,vc9}
\fmf{plain,right=0.25}{v4,vc7}
\fmf{plain,tension=0.5,left=0.25}{v10,vc8}
\fmf{plain,tension=0.5,right=0.25}{v11,vc8}
\fmf{plain,right=0.25}{v12,vc10}
\fmf{plain,tension=1}{vc7,vc8}
\fmf{plain,tension=0.5}{vc7,vc10}
\fmf{plain,tension=1}{vc9,vc10}
\fmffreeze
\fmfposition
\fmf{plain,tension=1,right=0,width=1mm}{v7,v12}
\fmfipath{p[]}
\fmfiset{p1}{vpath(__v9,__vc4)}
\fmfiset{p2}{vpath(__v10,__vc8)}
\fmfipair{w[]}
\svertex{w1}{p1}
\svertex{w2}{p2}
\fmfi{wiggly}{w1..w2}
\fmffreeze
\end{fmfchar*}}}
}
\subfigure[$G_{2}$ $(\times 2)$]{
\raisebox{\eqoff}{%
\fmfframe(3,1)(1,4){%
\begin{fmfchar*}(30,20)
\fmftop{v1}
\fmfbottom{v7}
\fmfforce{(0w,h)}{v1}
\fmfforce{(0w,0)}{v7}
\fmffixed{(0.2w,0)}{v1,v2}
\fmffixed{(0.2w,0)}{v2,v3}
\fmffixed{(0.2w,0)}{v3,v4}
\fmffixed{(0.2w,0)}{v4,v5}
\fmffixed{(0.2w,0)}{v5,v6}
\fmffixed{(0.2w,0)}{v7,v8}
\fmffixed{(0.2w,0)}{v8,v9}
\fmffixed{(0.2w,0)}{v9,v10}
\fmffixed{(0.2w,0)}{v10,v11}
\fmffixed{(0.2w,0)}{v11,v12}
\fmf{plain,tension=0.5,right=0.25}{v1,vc1}
\fmf{plain,tension=0.5,left=0.25}{v2,vc1}
\fmf{plain,left=0.25}{v3,vc3}
\fmf{plain,tension=0.5,left=0.25}{v8,vc4}
\fmf{plain,tension=0.5,right=0.25}{v9,vc4}
\fmf{plain,left=0.25}{v7,vc2}
\fmf{plain,tension=1}{vc1,vc2}
\fmf{plain,tension=0.5}{vc2,vc3}
\fmf{plain,tension=1}{vc3,vc4}
\fmf{plain,tension=0.5,right=0.25}{v5,vc9}
\fmf{plain,tension=0.5,left=0.25}{v6,vc9}
\fmf{plain,right=0.25}{v4,vc7}
\fmf{plain,tension=0.5,left=0.25}{v10,vc8}
\fmf{plain,tension=0.5,right=0.25}{v11,vc8}
\fmf{plain,right=0.25}{v12,vc10}
\fmf{plain,tension=1}{vc7,vc8}
\fmf{plain,tension=0.5}{vc7,vc10}
\fmf{plain,tension=1}{vc9,vc10}
\fmffreeze
\fmfposition
\fmf{plain,tension=1,right=0,width=1mm}{v7,v12}
\fmfipath{p[]}
\fmfiset{p1}{vpath(__vc3,__vc4)}
\fmfiset{p2}{vpath(__v10,__vc8)}
\fmfipair{w[]}
\svertex{w1}{p1}
\svertex{w2}{p2}
\fmfi{wiggly}{w1..w2}
\fmffreeze
\end{fmfchar*}}}
}
\subfigure[$G_{3}$]{
\raisebox{\eqoff}{%
\fmfframe(3,1)(1,4){%
\begin{fmfchar*}(30,20)
\fmftop{v1}
\fmfbottom{v7}
\fmfforce{(0w,h)}{v1}
\fmfforce{(0w,0)}{v7}
\fmffixed{(0.2w,0)}{v1,v2}
\fmffixed{(0.2w,0)}{v2,v3}
\fmffixed{(0.2w,0)}{v3,v4}
\fmffixed{(0.2w,0)}{v4,v5}
\fmffixed{(0.2w,0)}{v5,v6}
\fmffixed{(0.2w,0)}{v7,v8}
\fmffixed{(0.2w,0)}{v8,v9}
\fmffixed{(0.2w,0)}{v9,v10}
\fmffixed{(0.2w,0)}{v10,v11}
\fmffixed{(0.2w,0)}{v11,v12}
\fmf{plain,tension=0.5,right=0.25}{v1,vc1}
\fmf{plain,tension=0.5,left=0.25}{v2,vc1}
\fmf{plain,left=0.25}{v3,vc3}
\fmf{plain,tension=0.5,left=0.25}{v8,vc4}
\fmf{plain,tension=0.5,right=0.25}{v9,vc4}
\fmf{plain,left=0.25}{v7,vc2}
\fmf{plain,tension=1}{vc1,vc2}
\fmf{plain,tension=0.5}{vc2,vc3}
\fmf{plain,tension=1}{vc3,vc4}
\fmf{plain,tension=0.5,right=0.25}{v5,vc9}
\fmf{plain,tension=0.5,left=0.25}{v6,vc9}
\fmf{plain,right=0.25}{v4,vc7}
\fmf{plain,tension=0.5,left=0.25}{v10,vc8}
\fmf{plain,tension=0.5,right=0.25}{v11,vc8}
\fmf{plain,right=0.25}{v12,vc10}
\fmf{plain,tension=1}{vc7,vc8}
\fmf{plain,tension=0.5}{vc7,vc10}
\fmf{plain,tension=1}{vc9,vc10}
\fmffreeze
\fmfposition
\fmf{plain,tension=1,right=0,width=1mm}{v7,v12}
\fmfipath{p[]}
\fmfiset{p1}{vpath(__vc3,__vc4)}
\fmfiset{p2}{vpath(__vc7,__vc8)}
\fmfipair{w[]}
\svertex{w1}{p1}
\svertex{w2}{p2}
\fmfi{wiggly}{w1..w2}
\fmffreeze
\end{fmfchar*}}}
}
\caption{Range-six diagrams with structure $\chi(1,2,5,4)$}
\label{r6-1254}
\end{figure}

\begin{figure}[!h]
\centering
\unitlength=0.75mm
\settoheight{\eqoff}{$\times$}%
\setlength{\eqoff}{0.5\eqoff}%
\addtolength{\eqoff}{-12.5\unitlength}%
\settoheight{\eqofftwo}{$\times$}%
\setlength{\eqofftwo}{0.5\eqofftwo}%
\addtolength{\eqofftwo}{-7.5\unitlength}%
\subfigure[$G_{1}$]{
\raisebox{\eqoff}{%
\fmfframe(3,1)(1,4){%
\begin{fmfchar*}(30,20)
\fmftop{v1}
\fmfbottom{v7}
\fmfforce{(0w,h)}{v1}
\fmfforce{(0w,0)}{v7}
\fmffixed{(0.2w,0)}{v1,v2}
\fmffixed{(0.2w,0)}{v2,v3}
\fmffixed{(0.2w,0)}{v3,v4}
\fmffixed{(0.2w,0)}{v4,v5}
\fmffixed{(0.2w,0)}{v5,v6}
\fmffixed{(0.2w,0)}{v7,v8}
\fmffixed{(0.2w,0)}{v8,v9}
\fmffixed{(0.2w,0)}{v9,v10}
\fmffixed{(0.2w,0)}{v10,v11}
\fmffixed{(0.2w,0)}{v11,v12}
\fmf{plain,tension=0.5,right=0.25}{v2,vc3}
\fmf{plain,tension=0.5,left=0.25}{v3,vc3}
\fmf{plain,right=0.25}{v1,vc1}
\fmf{plain,tension=0.5,left=0.25}{v7,vc2}
\fmf{plain,tension=0.5,right=0.25}{v8,vc2}
\fmf{plain,right=0.25}{v9,vc4}
\fmf{plain,tension=1}{vc1,vc2}
\fmf{plain,tension=0.5}{vc1,vc4}
\fmf{plain,tension=1}{vc3,vc4}
\fmf{plain,tension=0.5,right=0.25}{v4,vc7}
\fmf{plain,tension=0.5,left=0.25}{v5,vc7}
\fmf{plain,left=0.25}{v6,vc9}
\fmf{plain,tension=0.5,left=0.25}{v11,vc10}
\fmf{plain,tension=0.5,right=0.25}{v12,vc10}
\fmf{plain,left=0.25}{v10,vc8}
\fmf{plain,tension=1}{vc7,vc8}
\fmf{plain,tension=0.5}{vc8,vc9}
\fmf{plain,tension=1}{vc9,vc10}
\fmffreeze
\fmfposition
\fmf{plain,tension=1,right=0,width=1mm}{v7,v12}
\fmfipath{p[]}
\fmfiset{p1}{vpath(__v9,__vc4)}
\fmfiset{p2}{vpath(__v10,__vc8)}
\fmfipair{w[]}
\svertex{w1}{p1}
\svertex{w2}{p2}
\fmfi{wiggly}{w1..w2}
\fmffreeze
\end{fmfchar*}}}
}
\subfigure[$G_{2}$ $(\times 2)$]{
\raisebox{\eqoff}{%
\fmfframe(3,1)(1,4){%
\begin{fmfchar*}(30,20)
\fmftop{v1}
\fmfbottom{v7}
\fmfforce{(0w,h)}{v1}
\fmfforce{(0w,0)}{v7}
\fmffixed{(0.2w,0)}{v1,v2}
\fmffixed{(0.2w,0)}{v2,v3}
\fmffixed{(0.2w,0)}{v3,v4}
\fmffixed{(0.2w,0)}{v4,v5}
\fmffixed{(0.2w,0)}{v5,v6}
\fmffixed{(0.2w,0)}{v7,v8}
\fmffixed{(0.2w,0)}{v8,v9}
\fmffixed{(0.2w,0)}{v9,v10}
\fmffixed{(0.2w,0)}{v10,v11}
\fmffixed{(0.2w,0)}{v11,v12}
\fmf{plain,tension=0.5,right=0.25}{v2,vc3}
\fmf{plain,tension=0.5,left=0.25}{v3,vc3}
\fmf{plain,right=0.25}{v1,vc1}
\fmf{plain,tension=0.5,left=0.25}{v7,vc2}
\fmf{plain,tension=0.5,right=0.25}{v8,vc2}
\fmf{plain,right=0.25}{v9,vc4}
\fmf{plain,tension=1}{vc1,vc2}
\fmf{plain,tension=0.5}{vc1,vc4}
\fmf{plain,tension=1}{vc3,vc4}
\fmf{plain,tension=0.5,right=0.25}{v4,vc7}
\fmf{plain,tension=0.5,left=0.25}{v5,vc7}
\fmf{plain,left=0.25}{v6,vc9}
\fmf{plain,tension=0.5,left=0.25}{v11,vc10}
\fmf{plain,tension=0.5,right=0.25}{v12,vc10}
\fmf{plain,left=0.25}{v10,vc8}
\fmf{plain,tension=1}{vc7,vc8}
\fmf{plain,tension=0.5}{vc8,vc9}
\fmf{plain,tension=1}{vc9,vc10}
\fmffreeze
\fmfposition
\fmf{plain,tension=1,right=0,width=1mm}{v7,v12}
\fmfipath{p[]}
\fmfiset{p1}{vpath(__vc3,__vc4)}
\fmfiset{p2}{vpath(__v10,__vc8)}
\fmfipair{w[]}
\svertex{w1}{p1}
\svertex{w2}{p2}
\fmfi{wiggly}{w1..w2}
\fmffreeze
\end{fmfchar*}}}
}
\subfigure[$G_{3}$]{
\raisebox{\eqoff}{%
\fmfframe(3,1)(1,4){%
\begin{fmfchar*}(30,20)
\fmftop{v1}
\fmfbottom{v7}
\fmfforce{(0w,h)}{v1}
\fmfforce{(0w,0)}{v7}
\fmffixed{(0.2w,0)}{v1,v2}
\fmffixed{(0.2w,0)}{v2,v3}
\fmffixed{(0.2w,0)}{v3,v4}
\fmffixed{(0.2w,0)}{v4,v5}
\fmffixed{(0.2w,0)}{v5,v6}
\fmffixed{(0.2w,0)}{v7,v8}
\fmffixed{(0.2w,0)}{v8,v9}
\fmffixed{(0.2w,0)}{v9,v10}
\fmffixed{(0.2w,0)}{v10,v11}
\fmffixed{(0.2w,0)}{v11,v12}
\fmf{plain,tension=0.5,right=0.25}{v2,vc3}
\fmf{plain,tension=0.5,left=0.25}{v3,vc3}
\fmf{plain,right=0.25}{v1,vc1}
\fmf{plain,tension=0.5,left=0.25}{v7,vc2}
\fmf{plain,tension=0.5,right=0.25}{v8,vc2}
\fmf{plain,right=0.25}{v9,vc4}
\fmf{plain,tension=1}{vc1,vc2}
\fmf{plain,tension=0.5}{vc1,vc4}
\fmf{plain,tension=1}{vc3,vc4}
\fmf{plain,tension=0.5,right=0.25}{v4,vc7}
\fmf{plain,tension=0.5,left=0.25}{v5,vc7}
\fmf{plain,left=0.25}{v6,vc9}
\fmf{plain,tension=0.5,left=0.25}{v11,vc10}
\fmf{plain,tension=0.5,right=0.25}{v12,vc10}
\fmf{plain,left=0.25}{v10,vc8}
\fmf{plain,tension=1}{vc7,vc8}
\fmf{plain,tension=0.5}{vc8,vc9}
\fmf{plain,tension=1}{vc9,vc10}
\fmffreeze
\fmfposition
\fmf{plain,tension=1,right=0,width=1mm}{v7,v12}
\fmfipath{p[]}
\fmfiset{p1}{vpath(__vc3,__vc4)}
\fmfiset{p2}{vpath(__vc7,__vc8)}
\fmfipair{w[]}
\svertex{w1}{p1}
\svertex{w2}{p2}
\fmfi{wiggly}{w1..w2}
\fmffreeze
\end{fmfchar*}}}
}
\caption{Range-six diagrams with structure $\chi(2,1,4,5)$}
\label{r6-2145}
\end{figure}

\begin{table}
\centering
\begin{tabular}{lm{5cm}m{5cm}}
\toprule
\multirow{2}{*}{$\chi(1,2,4,5)$} 
& $G_1\rightarrow J_{10}$ & $G_2\rightarrow -J_{7}$  \\
& $G_3\rightarrow -(J_{10}+J_{1}+2J_{13})$ & $G_4\rightarrow J_{7}+J_{1}$  \\
\midrule
\multirow{2}{*}{$\chi(1,5,4,3)$}
& $G_1\rightarrow 0$ & $G_2\rightarrow J_{1}$  \\
& $G_3\rightarrow J_{4}$ & $G_4\rightarrow -(J_{1}+J_{4}+2J_{11})$  \\
\midrule
\multirow{2}{*}{$\chi(1,3,4,5)$}
& $G_1\rightarrow J_{3}$ & $G_2\rightarrow -J_{9}$  \\
& $G_3\rightarrow J_{1}+J_{9}$ & $G_4\rightarrow -(J_{3}+J_{1}+2J_{12})$  \\
\midrule
\multirow{2}{*}{$\chi(1,4,3,5)$}
& $G_1\rightarrow -(J_{9}+J_{5}+2J_{16})$ & $G_2\rightarrow J_{5}$  \\
& $G_3\rightarrow J_{9}$ & $G_4\rightarrow 0$  \\
\midrule
\multirow{2}{*}{$\chi(1,2,5,4)$}
& $G_1\rightarrow -2(J_{8}+J_{14})$ & $G_2\rightarrow J_{8}$  \\
& $G_3\rightarrow 0$ &  \\
\midrule
\multirow{2}{*}{$\chi(2,1,4,5)$}
& $G_1\rightarrow -2(J_{9}+J_{15})$ & $G_2\rightarrow J_{9}$  \\
& $G_3\rightarrow 0$ &  \\
\midrule
\multirow{2}{*}{$\chi(1,3,2,5)$}
& $G_1\rightarrow -(J_{7}+J_{8}+2J_{17})$ & $G_2\rightarrow J_{7}$  \\
& $G_3\rightarrow J_{8}$ & $G_4\rightarrow 0$  \\
\bottomrule
\end{tabular}
\caption{Results of D-algebra for diagrams with vector interactions}
\label{r6-vector}
\end{table}

\clearpage

\renewcommand{\thefigure}{B.\arabic{figure}}
\setcounter{figure}{0}
\renewcommand{\thetable}{B.\arabic{table}}
\setcounter{table}{0}

\section{Wrapping}
\label{app:wrapping}
In this appendix we list all the relevant wrapping diagrams. For the
non-symmetric chiral structures, the corresponding reflections are
indicated in the figure captions. Note that for the length-five
states, the structures $\chi(1,2,4)$ and $\chi(1,4,3)$ are the same
as $\chi(2,1,4)$ and its reflection $\chi(1,3,4)$. Similarly,
$\chi(1,3)$ is the same structure as $\chi(1,4)$.

\begin{figure}[h]
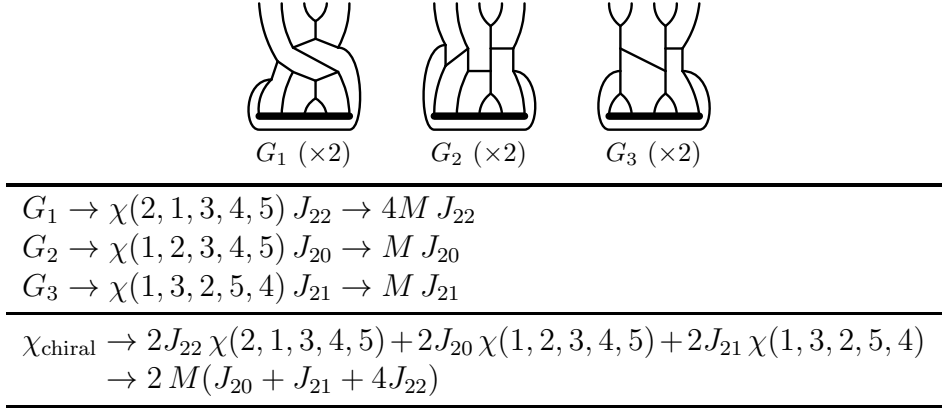

\centering
\unitlength=0.75mm
\settoheight{\eqoff}{$\times$}%
\setlength{\eqoff}{0.5\eqoff}%
\addtolength{\eqoff}{-12.5\unitlength}%
\settoheight{\eqofftwo}{$\times$}%
\setlength{\eqofftwo}{0.5\eqofftwo}%
\addtolength{\eqofftwo}{-7.5\unitlength}%
\subfigure[$G_{1}$ $(\times 2)$]{
\raisebox{\eqoff}{%
\fmfframe(3,1)(1,4){%
\begin{fmfchar*}(16,20)
\fmftop{v1}
\fmfbottom{v6}
\fmfforce{(0,h)}{v1}
\fmfforce{(0,0)}{v6}
\fmffixed{(0.25w,0)}{v1,v2}
\fmffixed{(0.25w,0)}{v2,v3}
\fmffixed{(0.25w,0)}{v3,v4}
\fmffixed{(0.25w,0)}{v4,v5}
\fmffixed{(0.25w,0)}{v6,v7}
\fmffixed{(0.25w,0)}{v7,v8}
\fmffixed{(0.25w,0)}{v8,v9}
\fmffixed{(0.25w,0)}{v9,v10}
\fmffixed{(0,whatever)}{vc1,vc5}
\fmffixed{(0,whatever)}{vc2,vc3}
\fmffixed{(0,whatever)}{vc3,vc6}
\fmffixed{(0,whatever)}{vc6,vc7}
\fmffixed{(0,whatever)}{vc4,vc8}
\fmffixed{(0.5w,0)}{vc1,vc4}
\fmffixed{(0.5w,0)}{vc5,vc8}
\fmf{plain,tension=1,right=0.125}{v2,vc1}
\fmf{plain,tension=0.25,right=0.25}{v3,vc2}
\fmf{plain,tension=0.25,left=0.25}{v4,vc2}
\fmf{plain,tension=1,left=0.125}{v5,vc4}
\fmf{plain,tension=1,left=0.125}{v7,vc5}
\fmf{plain,tension=0.25,left=0.25}{v8,vc6}
\fmf{plain,tension=0.25,right=0.25}{v9,vc6}
\fmf{plain,tension=1,right=0.125}{v10,vc8}
\fmf{plain,tension=0.5}{vc1,vc3}
\fmf{plain,tension=0.5}{vc2,vc3}
\fmf{plain,tension=0.5}{vc3,vc4}
\fmf{plain,tension=0.5}{vc5,vc7}
\fmf{plain,tension=0.5}{vc6,vc7}
\fmf{plain,tension=0.5}{vc7,vc8}
\fmf{plain,tension=2}{vc1,vc8}
\fmf{phantom,tension=2}{vc5,vc4}
\fmffreeze
\fmfposition
\fmf{plain,tension=2}{vc5,vc9}
\fmf{plain,tension=2}{vc9,vc10}
\fmf{plain,tension=1,right=0.125}{v1,vc9}
\fmf{plain,tension=1,left=0.25}{v6,vc10}
\fmffreeze
\plainwrap{vc10}{v6}{v10}{vc4}
\fmf{plain,tension=1,left=0,width=1mm}{v6,v10}
\fmffreeze
\end{fmfchar*}}}
}
\subfigspace
\subfigure[$G_{2}$ $(\times 2)$]{
\raisebox{\eqoff}{%
\fmfframe(3,1)(1,4){%
\begin{fmfchar*}(16,20)
\fmftop{v1}
\fmfbottom{v6}
\fmfforce{(0,h)}{v1}
\fmfforce{(0,0)}{v6}
\fmffixed{(0.25w,0)}{v1,v2}
\fmffixed{(0.25w,0)}{v2,v3}
\fmffixed{(0.25w,0)}{v3,v4}
\fmffixed{(0.25w,0)}{v4,v5}
\fmffixed{(0.25w,0)}{v6,v7}
\fmffixed{(0.25w,0)}{v7,v8}
\fmffixed{(0.25w,0)}{v8,v9}
\fmffixed{(0.25w,0)}{v9,v10}
\fmffixed{(whatever,0)}{vc1,vc3}
\fmffixed{(whatever,0)}{vc5,vc7}
\fmffixed{(whatever,0)}{vc3,vc4}
\fmffixed{(whatever,0)}{vc7,vc8}
\fmf{plain,tension=1,right=0.125}{v2,vc1}
\fmf{plain,tension=0.5,right=0.25}{v3,vc2}
\fmf{plain,tension=0.5,left=0.25}{v4,vc2}
\fmf{plain,tension=1,left=0.125}{v5,vc4}
\fmf{plain,tension=1,left=0.125}{v7,vc5}
\fmf{plain,tension=0.5,left=0.25}{v8,vc6}
\fmf{plain,tension=0.5,right=0.25}{v9,vc6}
\fmf{plain,tension=1,right=0.125}{v10,vc8}
\fmf{plain}{vc1,vc5}
\fmf{plain}{vc4,vc8}
\fmf{plain}{vc2,vc3}
\fmf{plain}{vc6,vc7}
\fmf{plain,tension=3}{vc3,vc7}
\fmf{plain,tension=0.5}{vc3,vc4}
\fmf{plain,tension=0.5}{vc5,vc7}
\fmf{phantom,tension=0.5}{vc7,vc8}
\fmf{phantom,tension=0.5}{vc1,vc3}
\fmf{plain,tension=0.5,right=0,width=1mm}{v6,v10}
\fmffreeze
\fmf{plain,tension=0.5,right=0.125}{v1,vc9}
\fmf{phantom,tension=0.5,left=0.125}{v2,vc9}
\fmf{plain,tension=1,left=0.125}{v6,vc10}
\fmf{plain,tension=0.5}{vc1,vc10}
\fmf{plain,tension=2}{vc9,vc10}
\fmffreeze
\fmfposition
\plainwrap{vc9}{v6}{v10}{vc8}
\fmffreeze
\end{fmfchar*}}}
}
\subfigspace
\subfigure[$G_{3}$ $(\times 2)$]{
\raisebox{\eqoff}{%
\fmfframe(3,1)(1,4){%
\begin{fmfchar*}(16,20)
\fmftop{v1}
\fmfbottom{v6}
\fmfforce{(0,h)}{v1}
\fmfforce{(0,0)}{v6}
\fmffixed{(0.25w,0)}{v1,v2}
\fmffixed{(0.25w,0)}{v2,v3}
\fmffixed{(0.25w,0)}{v3,v4}
\fmffixed{(0.25w,0)}{v4,v5}
\fmffixed{(0.25w,0)}{v6,v7}
\fmffixed{(0.25w,0)}{v7,v8}
\fmffixed{(0.25w,0)}{v8,v9}
\fmffixed{(0.25w,0)}{v9,v10}
\fmffixed{(0,0.9w)}{v6,vh1}
\fmf{plain,tension=0.5,right=0.25}{v1,vc1}
\fmf{plain,tension=0.5,left=0.25}{v2,vc1}
\fmf{plain,tension=0.5,right=0.25}{v3,vc2}
\fmf{plain,tension=0.5,left=0.25}{v4,vc2}
\fmf{plain}{vc1,vc3}
\fmf{plain}{vc3,vc7}
\fmf{plain}{vc7,vc5}
\fmf{plain}{vc2,vc8}
\fmf{plain}{vc8,vc4}
\fmf{plain}{vc4,vc6}
\fmf{plain,tension=0}{vc3,vc4}
\fmf{plain,tension=0.5,left=0.25}{v6,vc5}
\fmf{plain,tension=0.5,right=0.25}{v7,vc5}
\fmf{plain,tension=0.5,left=0.25}{v8,vc6}
\fmf{plain,tension=0.5,right=0.25}{v9,vc6}
\fmf{plain,tension=0.5,right=0,width=1mm}{v6,v10}
\fmffreeze
\fmfposition
\fmf{plain,tension=0.5,left=0.125}{v5,vc9}
\fmf{plain,tension=0.5,right=0.125}{v10,vc10}
\fmf{plain,tension=0.5}{vc8,vc9}
\fmf{phantom,tension=0.5}{vc4,vc10}
\fmf{plain}{vc9,vc10}
\fmffreeze
\plainwrap{vc7}{v6}{v10}{vc10}
\end{fmfchar*}}}
}

\begin{tabular}{m{12cm}}
\toprule
$G_1\rightarrow \chi(2,1,3,4,5)\,J_{22} \rightarrow 4M\,J_{22}$  \\
$G_2\rightarrow \chi(1,2,3,4,5)\,J_{20} \rightarrow M\,J_{20}$ \\
$G_3\rightarrow \chi(1,3,2,5,4)\,J_{21} \rightarrow M\,J_{21}$ \\
\midrule
$\chi_{\mathrm{chiral}}\rightarrow 2J_{22}\,\chi(2,1,3,4,5)+2J_{20}\,\chi(1,2,3,4,5)+2J_{21}\,\chi(1,3,2,5,4)$  
$\phantom{\chi_{\mathrm{chiral}}}\rightarrow 2\,M(J_{20}+J_{21}+4J_{22})$  \\
\bottomrule
\end{tabular}

\caption{Wrapping diagrams with completely chiral structure}
\label{wrap-chiral}
\end{figure}

\begin{figure}[h]
\centering
\unitlength=0.75mm
\settoheight{\eqoff}{$\times$}%
\setlength{\eqoff}{0.5\eqoff}%
\addtolength{\eqoff}{-12.5\unitlength}%
\settoheight{\eqofftwo}{$\times$}%
\setlength{\eqofftwo}{0.5\eqofftwo}%
\addtolength{\eqofftwo}{-7.5\unitlength}%
\subfigure[$G_{1}$]{
\raisebox{\eqoff}{%
\fmfframe(3,1)(1,4){%
\begin{fmfchar*}(16,20)
\Wuqtd
\fmfipair{wa[]}
\fmfipair{wb[]}
\fmfipair{wc[]}
\fmfipair{wd[]}
\fmfiequ{wa0}{point 1*length(p0)/2 of p0}
\fmfiv{d.shape=circle,d.size=2}{wa0}
\fmfiequ{wa3}{point 1*length(p3)/2 of p3}
\fmfiv{d.shape=circle,d.size=2}{wa3}
\fmfforce{(-0w,-0h)}{va0}
\fmfforce{(w,-0h)}{vb0}
\wigglywrap{wa0}{va0}{vb0}{wa3}
\end{fmfchar*}}}
}
\subfigspace
\subfigure[$G_{2}$]{
\raisebox{\eqoff}{%
\fmfframe(3,1)(1,4){%
\begin{fmfchar*}(16,20)
\Wuqtd
\fmfipair{wa[]}
\fmfipair{wb[]}
\fmfipair{wc[]}
\fmfipair{wd[]}
\fmfiequ{wa0}{point 1*length(p0)/2 of p0}
\fmfiv{d.shape=circle,d.size=2}{wa0}
\fmfiequ{wa4}{point 1*length(p4)/2 of p4}
\fmfiv{d.shape=circle,d.size=2}{wa4}
\fmfforce{(-0w,-0h)}{va0}
\fmfforce{(w,-0h)}{vb0}
\wigglywrap{wa0}{va0}{vb0}{wa4}
\end{fmfchar*}}}
}
\subfigspace
\subfigure[$G_{3}$]{
\raisebox{\eqoff}{%
\fmfframe(3,1)(1,4){%
\begin{fmfchar*}(16,20)
\Wuqtd
\fmfipair{wa[]}
\fmfipair{wb[]}
\fmfipair{wc[]}
\fmfipair{wd[]}
\fmfiequ{wa1}{point 1*length(p1)/2 of p1}
\fmfiv{d.shape=circle,d.size=2}{wa1}
\fmfiequ{wa3}{point 1*length(p3)/2 of p3}
\fmfiv{d.shape=circle,d.size=2}{wa3}
\fmfforce{(-0w,-0h)}{va0}
\fmfforce{(w,-0h)}{vb0}
\wigglywrap{wa1}{va0}{vb0}{wa3}
\end{fmfchar*}}}
}
\subfigspace
\subfigure[$G_{4}$]{
\raisebox{\eqoff}{%
\fmfframe(3,1)(1,4){%
\begin{fmfchar*}(16,20)
\Wuqtd
\fmfipair{wa[]}
\fmfipair{wb[]}
\fmfipair{wc[]}
\fmfipair{wd[]}
\fmfiequ{wa1}{point 1*length(p1)/2 of p1}
\fmfiv{d.shape=circle,d.size=2}{wa1}
\fmfiequ{wa4}{point 1*length(p4)/2 of p4}
\fmfiv{d.shape=circle,d.size=2}{wa4}
\fmfforce{(-0w,-0h)}{va0}
\fmfforce{(w,-0h)}{vb0}
\wigglywrap{wa1}{va0}{vb0}{wa4}
\end{fmfchar*}}}
}

\begin{tabular}{m{12cm}}
\toprule
$G_1\rightarrow \chi(1,4,3,2)(-J_{20}-J_{22}-2J_{27}) \rightarrow -2M(-J_{20}-J_{22}-2J_{27})$  \\
$G_2\rightarrow -\chi(1,4,3,2)(-J_{20}) \rightarrow 2M(-J_{20})$ \\
$G_3\rightarrow -\chi(1,4,3,2)(-J_{22}) \rightarrow 2M(-J_{22})$ \\
$G_4\rightarrow 0$  \\
\midrule
$\chi(1,4,3,2)\rightarrow -4J_{27}\,\chi(1,4,3,2) \rightarrow 8\,M\,J_{27}$   \\
\bottomrule
\end{tabular}

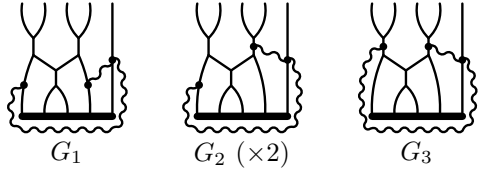
\caption{Wrapping diagrams with structure $\chi(1,4,3,2)$ or $\chi(1,2,4,3)$}
\label{wrap-1432}
\end{figure}

\begin{figure}[h]
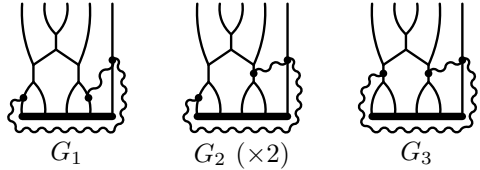

\centering
\unitlength=0.75mm
\settoheight{\eqoff}{$\times$}%
\setlength{\eqoff}{0.5\eqoff}%
\addtolength{\eqoff}{-12.5\unitlength}%
\settoheight{\eqofftwo}{$\times$}%
\setlength{\eqofftwo}{0.5\eqofftwo}%
\addtolength{\eqofftwo}{-7.5\unitlength}%
\subfigure[$G_{1}$]{
\raisebox{\eqoff}{%
\fmfframe(3,1)(1,4){%
\begin{fmfchar*}(16,20)
\Wutd
\fmfipair{wa[]}
\fmfipair{wb[]}
\fmfipair{wc[]}
\fmfipair{wd[]}
\fmfiequ{wa0}{point 1*length(p0)/2 of p0}
\fmfiv{d.shape=circle,d.size=2}{wa0}
\fmfiequ{wa3}{point 1*length(p3)/2 of p3}
\fmfiv{d.shape=circle,d.size=2}{wa3}
\fmfiequ{wa6}{point 1*length(p6)/2 of p6}
\fmfiv{d.shape=circle,d.size=2}{wa6}
\fmfforce{(-0w,-0h)}{va0}
\fmfforce{(1w,-0h)}{vb0}
\wigglywrap{wa0}{va0}{vb0}{wa6}
\fmfi{wiggly}{wa3..wa6}
\end{fmfchar*}}}
}
\subfigspace
\subfigure[$G_{2}$ $(\times 2)$]{
\raisebox{\eqoff}{%
\fmfframe(3,1)(1,4){%
\begin{fmfchar*}(16,20)
\Wutd
\fmfipair{wa[]}
\fmfipair{wb[]}
\fmfipair{wc[]}
\fmfipair{wd[]}
\fmfiequ{wa0}{point 1*length(p0)/2 of p0}
\fmfiv{d.shape=circle,d.size=2}{wa0}
\fmfiequ{wa4}{point 1*length(p4)/2 of p4}
\fmfiv{d.shape=circle,d.size=2}{wa4}
\fmfiequ{wa6}{point 1*length(p6)/2 of p6}
\fmfiv{d.shape=circle,d.size=2}{wa6}
\fmfforce{(-0w,-0h)}{va0}
\fmfforce{(1w,-0h)}{vb0}
\wigglywrap{wa0}{va0}{vb0}{wa6}
\fmfi{wiggly}{wa4..wa6}
\end{fmfchar*}}}
}
\subfigspace
\subfigure[$G_{3}$]{
\raisebox{\eqoff}{%
\fmfframe(3,1)(1,4){%
\begin{fmfchar*}(16,20)
\Wutd
\fmfipair{wa[]}
\fmfipair{wb[]}
\fmfipair{wc[]}
\fmfipair{wd[]}
\fmfiequ{wa1}{point 1*length(p1)/2 of p1}
\fmfiv{d.shape=circle,d.size=2}{wa1}
\fmfiequ{wa4}{point 1*length(p4)/2 of p4}
\fmfiv{d.shape=circle,d.size=2}{wa4}
\fmfiequ{wa6}{point 1*length(p6)/2 of p6}
\fmfiv{d.shape=circle,d.size=2}{wa6}
\fmfforce{(-0w,-0h)}{va0}
\fmfforce{(1w,-0h)}{vb0}
\wigglywrap{wa1}{va0}{vb0}{wa6}
\fmfi{wiggly}{wa4..wa6}
\end{fmfchar*}}}
}

\begin{tabular}{m{12cm}}
\toprule
$G_1\rightarrow -\chi(1,3,2)(2J_{20}+2J_{29}) \rightarrow 2M(2J_{20}+2J_{29})$  \\
$G_2\rightarrow \chi(1,3,2)\,J_{20} \rightarrow -2M\,J_{20}$ \\
$G_3\rightarrow 0$  \\
\midrule
$\chi(1,3,2)\rightarrow -2J_{29}\chi(1,3,2) \rightarrow 4\,M\,J_{29}$   \\
\bottomrule
\end{tabular}

\caption{Wrapping diagrams with structure $\chi(1,3,2)$}
\label{wrap-132}
\end{figure}

\begin{figure}[h]
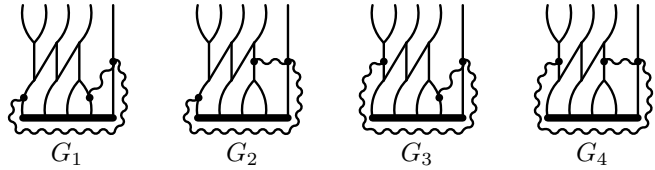

\centering
\unitlength=0.75mm
\settoheight{\eqoff}{$\times$}%
\setlength{\eqoff}{0.5\eqoff}%
\addtolength{\eqoff}{-12.5\unitlength}%
\settoheight{\eqofftwo}{$\times$}%
\setlength{\eqofftwo}{0.5\eqofftwo}%
\addtolength{\eqofftwo}{-7.5\unitlength}%
\subfigure[$G_{1}$]{
\raisebox{\eqoff}{%
\fmfframe(3,1)(1,4){%
\begin{fmfchar*}(16,20)
\Wdut
\fmfipair{wa[]}
\fmfipair{wb[]}
\fmfipair{wc[]}
\fmfipair{wd[]}
\fmfiequ{wa0}{point 1*length(p0)/2 of p0}
\fmfiv{d.shape=circle,d.size=2}{wa0}
\fmfiequ{wa3}{point 1*length(p3)/2 of p3}
\fmfiv{d.shape=circle,d.size=2}{wa3}
\fmfiequ{wa6}{point 1*length(p6)/2 of p6}
\fmfiv{d.shape=circle,d.size=2}{wa6}
\fmfforce{(-0w,-0h)}{va0}
\fmfforce{(1w,-0h)}{vb0}
\wigglywrap{wa0}{va0}{vb0}{wa6}
\fmfi{wiggly}{wa3..wa6}
\end{fmfchar*}}}
}
\subfigspace
\subfigure[$G_{2}$ $(\times 2)$]{
\raisebox{\eqoff}{%
\fmfframe(3,1)(1,4){%
\begin{fmfchar*}(16,20)
\Wdut
\fmfipair{wa[]}
\fmfipair{wb[]}
\fmfipair{wc[]}
\fmfipair{wd[]}
\fmfiequ{wa0}{point 1*length(p0)/2 of p0}
\fmfiv{d.shape=circle,d.size=2}{wa0}
\fmfiequ{wa4}{point 1*length(p4)/2 of p4}
\fmfiv{d.shape=circle,d.size=2}{wa4}
\fmfiequ{wa6}{point 1*length(p6)/2 of p6}
\fmfiv{d.shape=circle,d.size=2}{wa6}
\fmfforce{(-0w,-0h)}{va0}
\fmfforce{(1w,-0h)}{vb0}
\wigglywrap{wa0}{va0}{vb0}{wa6}
\fmfi{wiggly}{wa4..wa6}
\end{fmfchar*}}}
}
\subfigspace
\subfigure[$G_{3}$]{
\raisebox{\eqoff}{%
\fmfframe(3,1)(1,4){%
\begin{fmfchar*}(16,20)
\Wdut
\fmfipair{wa[]}
\fmfipair{wb[]}
\fmfipair{wc[]}
\fmfipair{wd[]}
\fmfiequ{wa1}{point 1*length(p1)/2 of p1}
\fmfiv{d.shape=circle,d.size=2}{wa1}
\fmfiequ{wa4}{point 1*length(p4)/2 of p4}
\fmfiv{d.shape=circle,d.size=2}{wa4}
\fmfiequ{wa6}{point 1*length(p6)/2 of p6}
\fmfiv{d.shape=circle,d.size=2}{wa6}
\fmfforce{(-0w,-0h)}{va0}
\fmfforce{(1w,-0h)}{vb0}
\wigglywrap{wa1}{va0}{vb0}{wa6}
\fmfi{wiggly}{wa4..wa6}
\end{fmfchar*}}}
}

\begin{tabular}{m{12cm}}
\toprule
$G_1\rightarrow -\chi(2,1,3)(2J_{20}+2J_{28}) \rightarrow 2M(2J_{20}+2J_{28})$  \\
$G_2\rightarrow \chi(2,1,3)\,J_{20} \rightarrow -2M\,J_{20}$ \\
$G_3\rightarrow 0$  \\
\midrule
$\chi(2,1,3)\rightarrow -2J_{28}\chi(2,1,3) \rightarrow 4\,M\,J_{28}$   \\
\bottomrule
\end{tabular}

\caption{Wrapping diagrams with structure $\chi(2,1,3)$}
\label{wrap-213}
\end{figure}

\begin{figure}[h]
\centering
\unitlength=0.75mm
\settoheight{\eqoff}{$\times$}%
\setlength{\eqoff}{0.5\eqoff}%
\addtolength{\eqoff}{-12.5\unitlength}%
\settoheight{\eqofftwo}{$\times$}%
\setlength{\eqofftwo}{0.5\eqofftwo}%
\addtolength{\eqofftwo}{-7.5\unitlength}%
\subfigure[$G_{1}$]{
\raisebox{\eqoff}{%
\fmfframe(3,1)(1,4){%
\begin{fmfchar*}(16,20)
\Wudt
\fmfipair{wa[]}
\fmfipair{wb[]}
\fmfipair{wc[]}
\fmfipair{wd[]}
\fmfiequ{wa0}{point 1*length(p0)/2 of p0}
\fmfiv{d.shape=circle,d.size=2}{wa0}
\fmfiequ{wa3}{point 1*length(p3)/2 of p3}
\fmfiv{d.shape=circle,d.size=2}{wa3}
\fmfiequ{wa6}{point 1*length(p6)/2 of p6}
\fmfiv{d.shape=circle,d.size=2}{wa6}
\fmfforce{(-0w,-0h)}{va0}
\fmfforce{(1w,-0h)}{vb0}
\wigglywrap{wa0}{va0}{vb0}{wa6}
\fmfi{wiggly}{wa3..wa6}
\end{fmfchar*}}}
}
\subfigspace
\subfigure[$G_{2}$]{
\raisebox{\eqoff}{%
\fmfframe(3,1)(1,4){%
\begin{fmfchar*}(16,20)
\Wudt
\fmfipair{wa[]}
\fmfipair{wb[]}
\fmfipair{wc[]}
\fmfipair{wd[]}
\fmfiequ{wa0}{point 1*length(p0)/2 of p0}
\fmfiv{d.shape=circle,d.size=2}{wa0}
\fmfiequ{wa4}{point 1*length(p4)/2 of p4}
\fmfiv{d.shape=circle,d.size=2}{wa4}
\fmfiequ{wa6}{point 1*length(p6)/2 of p6}
\fmfiv{d.shape=circle,d.size=2}{wa6}
\fmfforce{(-0w,-0h)}{va0}
\fmfforce{(1w,-0h)}{vb0}
\wigglywrap{wa0}{va0}{vb0}{wa6}
\fmfi{wiggly}{wa4..wa6}
\end{fmfchar*}}}
}
\subfigspace
\subfigure[$G_{3}$]{
\raisebox{\eqoff}{%
\fmfframe(3,1)(1,4){%
\begin{fmfchar*}(16,20)
\Wudt
\fmfipair{wa[]}
\fmfipair{wb[]}
\fmfipair{wc[]}
\fmfipair{wd[]}
\fmfiequ{wa1}{point 1*length(p1)/2 of p1}
\fmfiv{d.shape=circle,d.size=2}{wa1}
\fmfiequ{wa3}{point 1*length(p3)/2 of p3}
\fmfiv{d.shape=circle,d.size=2}{wa3}
\fmfiequ{wa6}{point 1*length(p6)/2 of p6}
\fmfiv{d.shape=circle,d.size=2}{wa6}
\fmfforce{(-0w,-0h)}{va0}
\fmfforce{(1w,-0h)}{vb0}
\wigglywrap{wa1}{va0}{vb0}{wa6}
\fmfi{wiggly}{wa3..wa6}
\end{fmfchar*}}}
}
\subfigspace
\subfigure[$G_{4}$]{
\raisebox{\eqoff}{%
\fmfframe(3,1)(1,4){%
\begin{fmfchar*}(16,20)
\Wudt
\fmfipair{wa[]}
\fmfipair{wb[]}
\fmfipair{wc[]}
\fmfipair{wd[]}
\fmfiequ{wa1}{point 1*length(p1)/2 of p1}
\fmfiv{d.shape=circle,d.size=2}{wa1}
\fmfiequ{wa4}{point 1*length(p4)/2 of p4}
\fmfiv{d.shape=circle,d.size=2}{wa4}
\fmfiequ{wa6}{point 1*length(p6)/2 of p6}
\fmfiv{d.shape=circle,d.size=2}{wa6}
\fmfforce{(-0w,-0h)}{va0}
\fmfforce{(1w,-0h)}{vb0}
\wigglywrap{wa1}{va0}{vb0}{wa6}
\fmfi{wiggly}{wa4..wa6}
\end{fmfchar*}}}
}

\begin{tabular}{m{12cm}}
\toprule
$G_1\rightarrow -\chi(1,2,3)(J_{22}+J_{23}+2J_{26}) \rightarrow -M(J_{22}+J_{23}+2J_{26})$  \\
$G_2\rightarrow \chi(1,2,3)\,J_{22} \rightarrow M\,J_{22}$ \\
$G_3\rightarrow \chi(1,2,3)(J_{22}+J_{23}+2J_{26}) \rightarrow M(J_{22}+J_{23}+2J_{26})$ \\
$G_4\rightarrow -\chi(1,2,3)\,J_{22} \rightarrow -M\,J_{22}$  \\
\midrule
$\chi(1,2,3)\rightarrow 0$   \\
\bottomrule
\end{tabular}

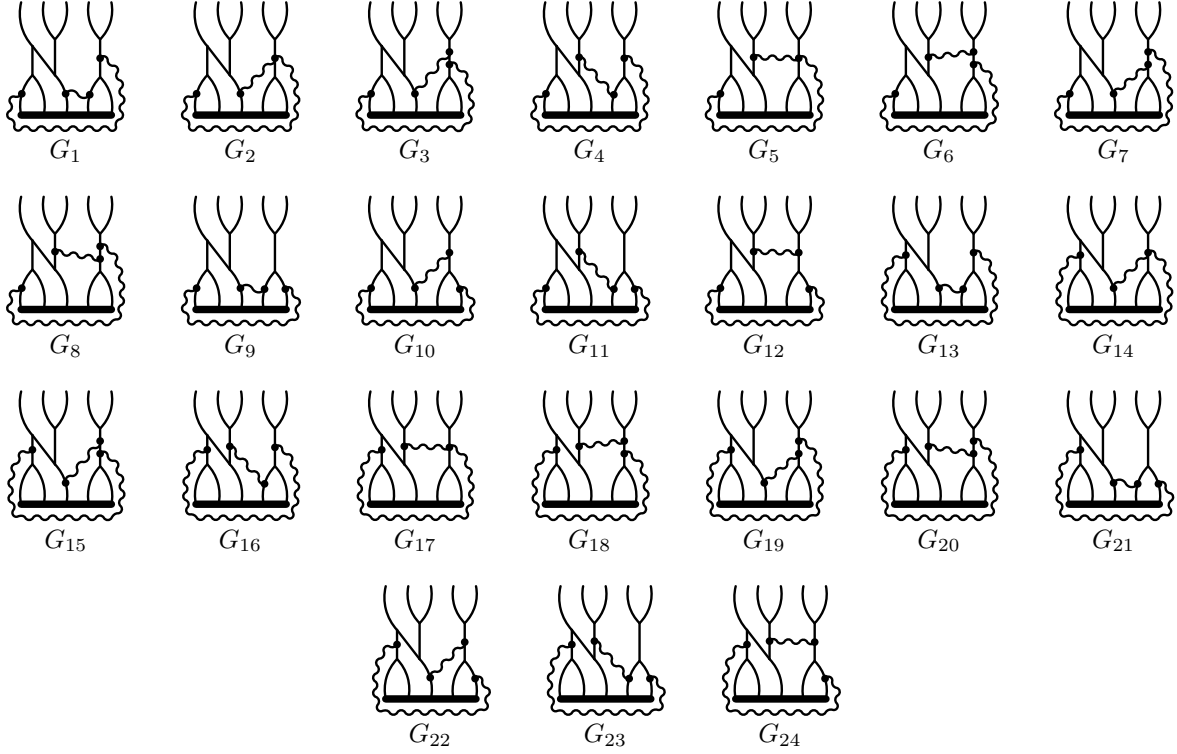
\caption{Wrapping diagrams with structure $\chi(1,2,3)$ or $\chi(3,2,1)$}
\label{wrap-123}
\end{figure}

\begin{figure}[h]
\vspace{-1cm}
\centering
\unitlength=0.75mm
\settoheight{\eqoff}{$\times$}%
\setlength{\eqoff}{0.5\eqoff}%
\addtolength{\eqoff}{-12.5\unitlength}%
\settoheight{\eqofftwo}{$\times$}%
\setlength{\eqofftwo}{0.5\eqofftwo}%
\addtolength{\eqofftwo}{-7.5\unitlength}%
\subfigure[$G_{1}$]{
\raisebox{\eqoff}{%
\fmfframe(3,1)(1,4){%
\begin{fmfchar*}(16,20)
\Wduq
\fmfipair{wa[]}
\fmfipair{wb[]}
\fmfipair{wc[]}
\fmfipair{wd[]}
\fmfiequ{wa0}{point 1*length(p0)/2 of p0}
\fmfiv{d.shape=circle,d.size=2}{wa0}
\fmfiequ{wa3}{point 1*length(p3)/2 of p3}
\fmfiv{d.shape=circle,d.size=2}{wa3}
\fmfiequ{wa6}{point 1*length(p6)/2 of p6}
\fmfiv{d.shape=circle,d.size=2}{wa6}
\fmfiequ{wa7}{point 1*length(p7)/2 of p7}
\fmfiv{d.shape=circle,d.size=2}{wa7}
\fmfforce{(-0w,-0h)}{va0}
\fmfforce{(1w,-0h)}{vb0}
\wigglywrap{wa0}{va0}{vb0}{wa7}
\fmfi{wiggly}{wa3..wa6}
\end{fmfchar*}}}
}
\subfigspace
\subfigure[$G_{2}$]{
\raisebox{\eqoff}{%
\fmfframe(3,1)(1,4){%
\begin{fmfchar*}(16,20)
\Wduq
\fmfipair{wa[]}
\fmfipair{wb[]}
\fmfipair{wc[]}
\fmfipair{wd[]}
\fmfiequ{wa0}{point 1*length(p0)/2 of p0}
\fmfiv{d.shape=circle,d.size=2}{wa0}
\fmfiequ{wa3}{point 1*length(p3)/2 of p3}
\fmfiv{d.shape=circle,d.size=2}{wa3}
\fmfiequ{wa7}{point 1*length(p7)/2 of p7}
\fmfiv{d.shape=circle,d.size=2}{wa7}
\fmfforce{(-0w,-0h)}{va0}
\fmfforce{(1w,-0h)}{vb0}
\wigglywrap{wa0}{va0}{vb0}{wa7}
\fmfi{wiggly}{wa3..wa7}
\end{fmfchar*}}}
}
\subfigspace
\subfigure[$G_{3}$]{
\raisebox{\eqoff}{%
\fmfframe(3,1)(1,4){%
\begin{fmfchar*}(16,20)
\Wduq
\fmfipair{wa[]}
\fmfipair{wb[]}
\fmfipair{wc[]}
\fmfipair{wd[]}
\fmfiequ{wa0}{point 1*length(p0)/2 of p0}
\fmfiv{d.shape=circle,d.size=2}{wa0}
\fmfiequ{wa3}{point 1*length(p3)/2 of p3}
\fmfiv{d.shape=circle,d.size=2}{wa3}
\fmfiequ{wa7}{point 1*length(p7)/3 of p7}
\fmfiv{d.shape=circle,d.size=2}{wa7}
\fmfiequ{wb7}{point 2*length(p7)/3 of p7}
\fmfiv{d.shape=circle,d.size=2}{wb7}
\fmfforce{(-0w,-0h)}{va0}
\fmfforce{(1w,-0h)}{vb0}
\wigglywrap{wa0}{va0}{vb0}{wa7}
\fmfi{wiggly}{wa3..wb7}
\end{fmfchar*}}}
}
\subfigspace
\subfigure[$G_{4}$]{
\raisebox{\eqoff}{%
\fmfframe(3,1)(1,4){%
\begin{fmfchar*}(16,20)
\Wduq
\fmfipair{wa[]}
\fmfipair{wb[]}
\fmfipair{wc[]}
\fmfipair{wd[]}
\fmfiequ{wa0}{point 1*length(p0)/2 of p0}
\fmfiv{d.shape=circle,d.size=2}{wa0}
\fmfiequ{wa4}{point 1*length(p4)/2 of p4}
\fmfiv{d.shape=circle,d.size=2}{wa4}
\fmfiequ{wa6}{point 1*length(p6)/2 of p6}
\fmfiv{d.shape=circle,d.size=2}{wa6}
\fmfiequ{wa7}{point 1*length(p7)/2 of p7}
\fmfiv{d.shape=circle,d.size=2}{wa7}
\fmfforce{(-0w,-0h)}{va0}
\fmfforce{(1w,-0h)}{vb0}
\wigglywrap{wa0}{va0}{vb0}{wa7}
\fmfi{wiggly}{wa4..wa6}
\end{fmfchar*}}}
}
\subfigspace
\subfigure[$G_{5}$]{
\raisebox{\eqoff}{%
\fmfframe(3,1)(1,4){%
\begin{fmfchar*}(16,20)
\Wduq
\fmfipair{wa[]}
\fmfipair{wb[]}
\fmfipair{wc[]}
\fmfipair{wd[]}
\fmfiequ{wa0}{point 1*length(p0)/2 of p0}
\fmfiv{d.shape=circle,d.size=2}{wa0}
\fmfiequ{wa4}{point 1*length(p4)/2 of p4}
\fmfiv{d.shape=circle,d.size=2}{wa4}
\fmfiequ{wa7}{point 1*length(p7)/2 of p7}
\fmfiv{d.shape=circle,d.size=2}{wa7}
\fmfforce{(-0w,-0h)}{va0}
\fmfforce{(1w,-0h)}{vb0}
\wigglywrap{wa0}{va0}{vb0}{wa7}
\fmfi{wiggly}{wa4..wa7}
\end{fmfchar*}}}
}
\subfigspace
\subfigure[$G_{6}$]{
\raisebox{\eqoff}{%
\fmfframe(3,1)(1,4){%
\begin{fmfchar*}(16,20)
\Wduq
\fmfipair{wa[]}
\fmfipair{wb[]}
\fmfipair{wc[]}
\fmfipair{wd[]}
\fmfiequ{wa0}{point 1*length(p0)/2 of p0}
\fmfiv{d.shape=circle,d.size=2}{wa0}
\fmfiequ{wa4}{point 1*length(p4)/2 of p4}
\fmfiv{d.shape=circle,d.size=2}{wa4}
\fmfiequ{wa7}{point 1*length(p7)/3 of p7}
\fmfiv{d.shape=circle,d.size=2}{wa7}
\fmfiequ{wb7}{point 2*length(p7)/3 of p7}
\fmfiv{d.shape=circle,d.size=2}{wb7}
\fmfforce{(-0w,-0h)}{va0}
\fmfforce{(1w,-0h)}{vb0}
\wigglywrap{wa0}{va0}{vb0}{wa7}
\fmfi{wiggly}{wa4..wb7}
\end{fmfchar*}}}
}
\subfigspace
\subfigure[$G_{7}$]{
\raisebox{\eqoff}{%
\fmfframe(3,1)(1,4){%
\begin{fmfchar*}(16,20)
\Wduq
\fmfipair{wa[]}
\fmfipair{wb[]}
\fmfipair{wc[]}
\fmfipair{wd[]}
\fmfiequ{wa0}{point 1*length(p0)/2 of p0}
\fmfiv{d.shape=circle,d.size=2}{wa0}
\fmfiequ{wa3}{point 1*length(p3)/2 of p3}
\fmfiv{d.shape=circle,d.size=2}{wa3}
\fmfiequ{wa7}{point 1*length(p7)/3 of p7}
\fmfiv{d.shape=circle,d.size=2}{wa7}
\fmfiequ{wb7}{point 2*length(p7)/3 of p7}
\fmfiv{d.shape=circle,d.size=2}{wb7}
\fmfforce{(-0w,-0h)}{va0}
\fmfforce{(1w,-0h)}{vb0}
\wigglywrap{wa0}{va0}{vb0}{wb7}
\fmfi{wiggly}{wa3..wa7}
\end{fmfchar*}}}
}
\\
\subfigure[$G_{8}$]{
\raisebox{\eqoff}{%
\fmfframe(3,1)(1,4){%
\begin{fmfchar*}(16,20)
\Wduq
\fmfipair{wa[]}
\fmfipair{wb[]}
\fmfipair{wc[]}
\fmfipair{wd[]}
\fmfiequ{wa0}{point 1*length(p0)/2 of p0}
\fmfiv{d.shape=circle,d.size=2}{wa0}
\fmfiequ{wa4}{point 1*length(p4)/2 of p4}
\fmfiv{d.shape=circle,d.size=2}{wa4}
\fmfiequ{wa7}{point 1*length(p7)/3 of p7}
\fmfiv{d.shape=circle,d.size=2}{wa7}
\fmfiequ{wb7}{point 2*length(p7)/3 of p7}
\fmfiv{d.shape=circle,d.size=2}{wb7}
\fmfforce{(-0w,-0h)}{va0}
\fmfforce{(1w,-0h)}{vb0}
\wigglywrap{wa0}{va0}{vb0}{wb7}
\fmfi{wiggly}{wa4..wa7}
\end{fmfchar*}}}
}
\subfigspace
\subfigure[$G_{9}$]{
\raisebox{\eqoff}{%
\fmfframe(3,1)(1,4){%
\begin{fmfchar*}(16,20)
\Wduq
\fmfipair{wa[]}
\fmfipair{wb[]}
\fmfipair{wc[]}
\fmfipair{wd[]}
\fmfiequ{wa0}{point 1*length(p0)/2 of p0}
\fmfiv{d.shape=circle,d.size=2}{wa0}
\fmfiequ{wa3}{point 1*length(p3)/2 of p3}
\fmfiv{d.shape=circle,d.size=2}{wa3}
\fmfiequ{wa6}{point 1*length(p6)/2 of p6}
\fmfiv{d.shape=circle,d.size=2}{wa6}
\fmfiequ{wa9}{point 1*length(p9)/2 of p9}
\fmfiv{d.shape=circle,d.size=2}{wa9}
\fmfforce{(-0w,-0h)}{va0}
\fmfforce{(1w,-0h)}{vb0}
\wigglywrap{wa0}{va0}{vb0}{wa9}
\fmfi{wiggly}{wa3..wa6}
\end{fmfchar*}}}
}
\subfigspace
\subfigure[$G_{10}$]{
\raisebox{\eqoff}{%
\fmfframe(3,1)(1,4){%
\begin{fmfchar*}(16,20)
\Wduq
\fmfipair{wa[]}
\fmfipair{wb[]}
\fmfipair{wc[]}
\fmfipair{wd[]}
\fmfiequ{wa0}{point 1*length(p0)/2 of p0}
\fmfiv{d.shape=circle,d.size=2}{wa0}
\fmfiequ{wa3}{point 1*length(p3)/2 of p3}
\fmfiv{d.shape=circle,d.size=2}{wa3}
\fmfiequ{wa7}{point 1*length(p7)/2 of p7}
\fmfiv{d.shape=circle,d.size=2}{wa7}
\fmfiequ{wa9}{point 1*length(p9)/2 of p9}
\fmfiv{d.shape=circle,d.size=2}{wa9}
\fmfforce{(-0w,-0h)}{va0}
\fmfforce{(1w,-0h)}{vb0}
\wigglywrap{wa0}{va0}{vb0}{wa9}
\fmfi{wiggly}{wa3..wa7}
\end{fmfchar*}}}
}
\subfigspace
\subfigure[$G_{11}$]{
\raisebox{\eqoff}{%
\fmfframe(3,1)(1,4){%
\begin{fmfchar*}(16,20)
\Wduq
\fmfipair{wa[]}
\fmfipair{wb[]}
\fmfipair{wc[]}
\fmfipair{wd[]}
\fmfiequ{wa0}{point 1*length(p0)/2 of p0}
\fmfiv{d.shape=circle,d.size=2}{wa0}
\fmfiequ{wa4}{point 1*length(p4)/2 of p4}
\fmfiv{d.shape=circle,d.size=2}{wa4}
\fmfiequ{wa6}{point 1*length(p6)/2 of p6}
\fmfiv{d.shape=circle,d.size=2}{wa6}
\fmfiequ{wa9}{point 1*length(p9)/2 of p9}
\fmfiv{d.shape=circle,d.size=2}{wa9}
\fmfforce{(-0w,-0h)}{va0}
\fmfforce{(1w,-0h)}{vb0}
\wigglywrap{wa0}{va0}{vb0}{wa9}
\fmfi{wiggly}{wa4..wa6}
\end{fmfchar*}}}
}
\subfigspace
\subfigure[$G_{12}$]{
\raisebox{\eqoff}{%
\fmfframe(3,1)(1,4){%
\begin{fmfchar*}(16,20)
\Wduq
\fmfipair{wa[]}
\fmfipair{wb[]}
\fmfipair{wc[]}
\fmfipair{wd[]}
\fmfiequ{wa0}{point 1*length(p0)/2 of p0}
\fmfiv{d.shape=circle,d.size=2}{wa0}
\fmfiequ{wa4}{point 1*length(p4)/2 of p4}
\fmfiv{d.shape=circle,d.size=2}{wa4}
\fmfiequ{wa7}{point 1*length(p7)/2 of p7}
\fmfiv{d.shape=circle,d.size=2}{wa7}
\fmfiequ{wa9}{point 1*length(p9)/2 of p9}
\fmfiv{d.shape=circle,d.size=2}{wa9}
\fmfforce{(-0w,-0h)}{va0}
\fmfforce{(1w,-0h)}{vb0}
\wigglywrap{wa0}{va0}{vb0}{wa9}
\fmfi{wiggly}{wa4..wa7}
\end{fmfchar*}}}
}
\subfigspace
\subfigure[$G_{13}$]{
\raisebox{\eqoff}{%
\fmfframe(3,1)(1,4){%
\begin{fmfchar*}(16,20)
\Wduq
\fmfipair{wa[]}
\fmfipair{wb[]}
\fmfipair{wc[]}
\fmfipair{wd[]}
\fmfiequ{wa1}{point 1*length(p1)/2 of p1}
\fmfiv{d.shape=circle,d.size=2}{wa1}
\fmfiequ{wa3}{point 1*length(p3)/2 of p3}
\fmfiv{d.shape=circle,d.size=2}{wa3}
\fmfiequ{wa6}{point 1*length(p6)/2 of p6}
\fmfiv{d.shape=circle,d.size=2}{wa6}
\fmfiequ{wa7}{point 1*length(p7)/2 of p7}
\fmfiv{d.shape=circle,d.size=2}{wa7}
\fmfforce{(-0w,-0h)}{va0}
\fmfforce{(1w,-0h)}{vb0}
\wigglywrap{wa1}{va0}{vb0}{wa7}
\fmfi{wiggly}{wa3..wa6}
\end{fmfchar*}}}
}
\subfigspace
\subfigure[$G_{14}$]{
\raisebox{\eqoff}{%
\fmfframe(3,1)(1,4){%
\begin{fmfchar*}(16,20)
\Wduq
\fmfipair{wa[]}
\fmfipair{wb[]}
\fmfipair{wc[]}
\fmfipair{wd[]}
\fmfiequ{wa1}{point 1*length(p1)/2 of p1}
\fmfiv{d.shape=circle,d.size=2}{wa1}
\fmfiequ{wa3}{point 1*length(p3)/2 of p3}
\fmfiv{d.shape=circle,d.size=2}{wa3}
\fmfiequ{wa7}{point 1*length(p7)/2 of p7}
\fmfiv{d.shape=circle,d.size=2}{wa7}
\fmfforce{(-0w,-0h)}{va0}
\fmfforce{(1w,-0h)}{vb0}
\wigglywrap{wa1}{va0}{vb0}{wa7}
\fmfi{wiggly}{wa3..wa7}
\end{fmfchar*}}}
}
\\
\subfigure[$G_{15}$]{
\raisebox{\eqoff}{%
\fmfframe(3,1)(1,4){%
\begin{fmfchar*}(16,20)
\Wduq
\fmfipair{wa[]}
\fmfipair{wb[]}
\fmfipair{wc[]}
\fmfipair{wd[]}
\fmfiequ{wa1}{point 1*length(p1)/2 of p1}
\fmfiv{d.shape=circle,d.size=2}{wa1}
\fmfiequ{wa3}{point 1*length(p3)/2 of p3}
\fmfiv{d.shape=circle,d.size=2}{wa3}
\fmfiequ{wa7}{point 1*length(p7)/3 of p7}
\fmfiv{d.shape=circle,d.size=2}{wa7}
\fmfiequ{wb7}{point 2*length(p7)/3 of p7}
\fmfiv{d.shape=circle,d.size=2}{wb7}
\fmfforce{(-0w,-0h)}{va0}
\fmfforce{(1w,-0h)}{vb0}
\wigglywrap{wa1}{va0}{vb0}{wa7}
\fmfi{wiggly}{wa3..wb7}
\end{fmfchar*}}}
}
\subfigspace
\subfigure[$G_{16}$]{
\raisebox{\eqoff}{%
\fmfframe(3,1)(1,4){%
\begin{fmfchar*}(16,20)
\Wduq
\fmfipair{wa[]}
\fmfipair{wb[]}
\fmfipair{wc[]}
\fmfipair{wd[]}
\fmfiequ{wa1}{point 1*length(p1)/2 of p1}
\fmfiv{d.shape=circle,d.size=2}{wa1}
\fmfiequ{wa4}{point 1*length(p4)/2 of p4}
\fmfiv{d.shape=circle,d.size=2}{wa4}
\fmfiequ{wa6}{point 1*length(p6)/2 of p6}
\fmfiv{d.shape=circle,d.size=2}{wa6}
\fmfiequ{wa7}{point 1*length(p7)/2 of p7}
\fmfiv{d.shape=circle,d.size=2}{wa7}
\fmfforce{(-0w,-0h)}{va0}
\fmfforce{(1w,-0h)}{vb0}
\wigglywrap{wa1}{va0}{vb0}{wa7}
\fmfi{wiggly}{wa4..wa6}
\end{fmfchar*}}}
}
\subfigspace
\subfigure[$G_{17}$]{
\raisebox{\eqoff}{%
\fmfframe(3,1)(1,4){%
\begin{fmfchar*}(16,20)
\Wduq
\fmfipair{wa[]}
\fmfipair{wb[]}
\fmfipair{wc[]}
\fmfipair{wd[]}
\fmfiequ{wa1}{point 1*length(p1)/2 of p1}
\fmfiv{d.shape=circle,d.size=2}{wa1}
\fmfiequ{wa4}{point 1*length(p4)/2 of p4}
\fmfiv{d.shape=circle,d.size=2}{wa4}
\fmfiequ{wa7}{point 1*length(p7)/2 of p7}
\fmfiv{d.shape=circle,d.size=2}{wa7}
\fmfforce{(-0w,-0h)}{va0}
\fmfforce{(1w,-0h)}{vb0}
\wigglywrap{wa1}{va0}{vb0}{wa7}
\fmfi{wiggly}{wa4..wa7}
\end{fmfchar*}}}
}
\subfigspace
\subfigure[$G_{18}$]{
\raisebox{\eqoff}{%
\fmfframe(3,1)(1,4){%
\begin{fmfchar*}(16,20)
\Wduq
\fmfipair{wa[]}
\fmfipair{wb[]}
\fmfipair{wc[]}
\fmfipair{wd[]}
\fmfiequ{wa1}{point 1*length(p1)/2 of p1}
\fmfiv{d.shape=circle,d.size=2}{wa1}
\fmfiequ{wa4}{point 1*length(p4)/2 of p4}
\fmfiv{d.shape=circle,d.size=2}{wa4}
\fmfiequ{wa7}{point 1*length(p7)/3 of p7}
\fmfiv{d.shape=circle,d.size=2}{wa7}
\fmfiequ{wb7}{point 2*length(p7)/3 of p7}
\fmfiv{d.shape=circle,d.size=2}{wb7}
\fmfforce{(-0w,-0h)}{va0}
\fmfforce{(1w,-0h)}{vb0}
\wigglywrap{wa1}{va0}{vb0}{wa7}
\fmfi{wiggly}{wa4..wb7}
\end{fmfchar*}}}
}
\subfigspace
\subfigure[$G_{19}$]{
\raisebox{\eqoff}{%
\fmfframe(3,1)(1,4){%
\begin{fmfchar*}(16,20)
\Wduq
\fmfipair{wa[]}
\fmfipair{wb[]}
\fmfipair{wc[]}
\fmfipair{wd[]}
\fmfiequ{wa1}{point 1*length(p1)/2 of p1}
\fmfiv{d.shape=circle,d.size=2}{wa1}
\fmfiequ{wa3}{point 1*length(p3)/2 of p3}
\fmfiv{d.shape=circle,d.size=2}{wa3}
\fmfiequ{wa7}{point 1*length(p7)/3 of p7}
\fmfiv{d.shape=circle,d.size=2}{wa7}
\fmfiequ{wb7}{point 2*length(p7)/3 of p7}
\fmfiv{d.shape=circle,d.size=2}{wb7}
\fmfforce{(-0w,-0h)}{va0}
\fmfforce{(1w,-0h)}{vb0}
\wigglywrap{wa1}{va0}{vb0}{wb7}
\fmfi{wiggly}{wa3..wa7}
\end{fmfchar*}}}
}
\subfigspace
\subfigure[$G_{20}$]{
\raisebox{\eqoff}{%
\fmfframe(3,1)(1,4){%
\begin{fmfchar*}(16,20)
\Wduq
\fmfipair{wa[]}
\fmfipair{wb[]}
\fmfipair{wc[]}
\fmfipair{wd[]}
\fmfiequ{wa1}{point 1*length(p1)/2 of p1}
\fmfiv{d.shape=circle,d.size=2}{wa1}
\fmfiequ{wa4}{point 1*length(p4)/2 of p4}
\fmfiv{d.shape=circle,d.size=2}{wa4}
\fmfiequ{wa7}{point 1*length(p7)/3 of p7}
\fmfiv{d.shape=circle,d.size=2}{wa7}
\fmfiequ{wb7}{point 2*length(p7)/3 of p7}
\fmfiv{d.shape=circle,d.size=2}{wb7}
\fmfforce{(-0w,-0h)}{va0}
\fmfforce{(1w,-0h)}{vb0}
\wigglywrap{wa1}{va0}{vb0}{wb7}
\fmfi{wiggly}{wa4..wa7}
\end{fmfchar*}}}
}
\subfigspace
\subfigure[$G_{21}$]{
\raisebox{\eqoff}{%
\fmfframe(3,1)(1,4){%
\begin{fmfchar*}(16,20)
\Wduq
\fmfipair{wa[]}
\fmfipair{wb[]}
\fmfipair{wc[]}
\fmfipair{wd[]}
\fmfiequ{wa1}{point 1*length(p1)/2 of p1}
\fmfiv{d.shape=circle,d.size=2}{wa1}
\fmfiequ{wa3}{point 1*length(p3)/2 of p3}
\fmfiv{d.shape=circle,d.size=2}{wa3}
\fmfiequ{wa6}{point 1*length(p6)/2 of p6}
\fmfiv{d.shape=circle,d.size=2}{wa6}
\fmfiequ{wa9}{point 1*length(p9)/2 of p9}
\fmfiv{d.shape=circle,d.size=2}{wa9}
\fmfforce{(-0w,-0h)}{va0}
\fmfforce{(1w,-0h)}{vb0}
\wigglywrap{wa1}{va0}{vb0}{wa9}
\fmfi{wiggly}{wa3..wa6}
\end{fmfchar*}}}
}
\subfigspace
\subfigure[$G_{22}$]{
\raisebox{\eqoff}{%
\fmfframe(3,1)(1,4){%
\begin{fmfchar*}(16,20)
\Wduq
\fmfipair{wa[]}
\fmfipair{wb[]}
\fmfipair{wc[]}
\fmfipair{wd[]}
\fmfiequ{wa1}{point 1*length(p1)/2 of p1}
\fmfiv{d.shape=circle,d.size=2}{wa1}
\fmfiequ{wa3}{point 1*length(p3)/2 of p3}
\fmfiv{d.shape=circle,d.size=2}{wa3}
\fmfiequ{wa7}{point 1*length(p7)/2 of p7}
\fmfiv{d.shape=circle,d.size=2}{wa7}
\fmfiequ{wa9}{point 1*length(p9)/2 of p9}
\fmfiv{d.shape=circle,d.size=2}{wa9}
\fmfforce{(-0w,-0h)}{va0}
\fmfforce{(1w,-0h)}{vb0}
\wigglywrap{wa1}{va0}{vb0}{wa9}
\fmfi{wiggly}{wa3..wa7}
\end{fmfchar*}}}
}
\subfigspace
\subfigure[$G_{23}$]{
\raisebox{\eqoff}{%
\fmfframe(3,1)(1,4){%
\begin{fmfchar*}(16,20)
\Wduq
\fmfipair{wa[]}
\fmfipair{wb[]}
\fmfipair{wc[]}
\fmfipair{wd[]}
\fmfiequ{wa1}{point 1*length(p1)/2 of p1}
\fmfiv{d.shape=circle,d.size=2}{wa1}
\fmfiequ{wa4}{point 1*length(p4)/2 of p4}
\fmfiv{d.shape=circle,d.size=2}{wa4}
\fmfiequ{wa6}{point 1*length(p6)/2 of p6}
\fmfiv{d.shape=circle,d.size=2}{wa6}
\fmfiequ{wa9}{point 1*length(p9)/2 of p9}
\fmfiv{d.shape=circle,d.size=2}{wa9}
\fmfforce{(-0w,-0h)}{va0}
\fmfforce{(1w,-0h)}{vb0}
\wigglywrap{wa1}{va0}{vb0}{wa9}
\fmfi{wiggly}{wa4..wa6}
\end{fmfchar*}}}
}
\subfigspace
\subfigure[$G_{24}$]{
\raisebox{\eqoff}{%
\fmfframe(3,1)(1,4){%
\begin{fmfchar*}(16,20)
\Wduq
\fmfipair{wa[]}
\fmfipair{wb[]}
\fmfipair{wc[]}
\fmfipair{wd[]}
\fmfiequ{wa1}{point 1*length(p1)/2 of p1}
\fmfiv{d.shape=circle,d.size=2}{wa1}
\fmfiequ{wa4}{point 1*length(p4)/2 of p4}
\fmfiv{d.shape=circle,d.size=2}{wa4}
\fmfiequ{wa7}{point 1*length(p7)/2 of p7}
\fmfiv{d.shape=circle,d.size=2}{wa7}
\fmfiequ{wa9}{point 1*length(p9)/2 of p9}
\fmfiv{d.shape=circle,d.size=2}{wa9}
\fmfforce{(-0w,-0h)}{va0}
\fmfforce{(1w,-0h)}{vb0}
\wigglywrap{wa1}{va0}{vb0}{wa9}
\fmfi{wiggly}{wa4..wa7}
\end{fmfchar*}}}
}

\begin{tabular}{m{12cm}}
\toprule
$G_1\rightarrow \chi(2,1,4)(-J_{22}-J_{23}-2J_{26}) \rightarrow M(-J_{22}-J_{23}-2J_{26})$  \\
$G_2\rightarrow -\chi(2,1,4)\,J_{22} \rightarrow -M\,J_{22}$ \\
$G_3\rightarrow -\chi(2,1,4)(-J_{22}) \rightarrow -M(-J_{22})$ \\
$G_4\rightarrow -\chi(2,1,4)(-J_{21}-J_{23}+(-2J_{32}+2J_{33}+2J_{34}-2i\,\epsilon_{\mu\nu\rho\sigma}J_{35}^{\mu\sigma\rho\nu}))$ \\
$\phantom{G_4}\rightarrow -M(-J_{21}-J_{23}+(-2J_{32}+2J_{33}+2J_{34}-2i\,\epsilon_{\mu\nu\rho\sigma}J_{35}^{\mu\sigma\rho\nu}))$ \\
$G_5\rightarrow \chi(2,1,4)\,J_{20} \rightarrow M\,J_{20}$ \\
$G_6\rightarrow \chi(2,1,4)(-J_{20}) \rightarrow M(-J_{20})$ \\
$G_7\rightarrow -\chi(2,1,4)(-J_{22}) \rightarrow -M(-J_{22})$ \\
$G_8\rightarrow \chi(2,1,4)(-J_{21}) \rightarrow M(-J_{21})$ \\
$G_9\rightarrow -\chi(2,1,4)(-J_{20}-2J_{22}-J_{23}-2J_{25}-4J_{26}-2J_{27}-4J_{30})$ \\
$\phantom{G_9}\rightarrow -M(-J_{20}-2J_{22}-J_{23}-2J_{25}-4J_{26}-2J_{27}-4J_{30})$ \\
$G_{10}\rightarrow \chi(2,1,4)(-J_{20}-J_{22}-2J_{25}) \rightarrow M(-J_{20}-J_{22}-2J_{25})$ \\
$G_{11}\rightarrow \chi(2,1,4)(-J_{22}-J_{23}-2J_{26}) \rightarrow M(-J_{22}-J_{23}-2J_{26})$ \\
$G_{12}=G_{13}=G_{14}=G_{15}\rightarrow 0$ \\
$G_{16}\rightarrow \chi(2,1,4)(-J_{21}) \rightarrow M(-J_{21})$ \\
$G_{17}=G_{18}=G_{19}\rightarrow 0$ \\
$G_{20}\rightarrow -\chi(2,1,4)(-J_{21}) \rightarrow -M(-J_{21})$ \\
$G_{21}\rightarrow \chi(2,1,4)(-J_{20}-J_{22}-2J_{27}) \rightarrow M(-J_{20}-J_{22}-2J_{27})$ \\
$G_{22}\rightarrow -\chi(2,1,4)(-J_{20}) \rightarrow -M(-J_{20})$ \\
$G_{23}\rightarrow -\chi(2,1,4)(-J_{22}) \rightarrow -M(-J_{22})$ \\
$G_{24}\rightarrow 0$ \\
\midrule
$\chi(2,1,4)\rightarrow 4(2J_{30}+J_{32}-J_{33}-J_{34}+i\,\epsilon_{\mu\nu\rho\sigma}J_{35}^{\mu\sigma\rho\nu})\chi(2,1,4) $  \\
$\phantom{\chi(2,1,4)}\rightarrow 4\,M\,(2J_{30}+J_{32}-J_{33}-J_{34}+i\,\epsilon_{\mu\nu\rho\sigma}J_{35}^{\mu\sigma\rho\nu})$  \\
\bottomrule
\end{tabular}

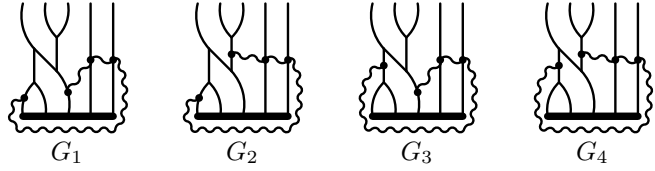
\caption{Wrapping diagrams with structure $\chi(2,1,4)$ or $\chi(1,3,4)$}
\label{wrap-214}
\end{figure}

\begin{figure}[h]
\centering
\unitlength=0.75mm
\settoheight{\eqoff}{$\times$}%
\setlength{\eqoff}{0.5\eqoff}%
\addtolength{\eqoff}{-12.5\unitlength}%
\settoheight{\eqofftwo}{$\times$}%
\setlength{\eqofftwo}{0.5\eqofftwo}%
\addtolength{\eqofftwo}{-7.5\unitlength}%
\subfigure[$G_{1}$ $(\times 2)$]{
\raisebox{\eqoff}{%
\fmfframe(3,1)(1,4){%
\begin{fmfchar*}(16,20)
\Wu
\fmfipair{wa[]}
\fmfipair{wb[]}
\fmfipair{wc[]}
\fmfipair{wd[]}
\fmfiequ{wa0}{point 1*length(p0)/2 of p0}
\fmfiv{d.shape=circle,d.size=2}{wa0}
\fmfiequ{wa1}{point 1*length(p1)/2 of p1}
\fmfiv{d.shape=circle,d.size=2}{wa1}
\fmfiequ{wa2}{point 1*length(p2)/2 of p2}
\fmfiv{d.shape=circle,d.size=2}{wa2}
\fmfiequ{wa6}{point 1*length(p6)/2 of p6}
\fmfiv{d.shape=circle,d.size=2}{wa6}
\fmfiequ{wa7}{point 1*length(p7)/2 of p7}
\fmfiv{d.shape=circle,d.size=2}{wa7}
\fmfi{wiggly}{wa0..wa1}
\fmfforce{(-0w,-0h)}{va0}
\fmfforce{(1w,-0h)}{vb0}
\wigglywrap{wa0}{va0}{vb0}{wa7}
\fmfi{wiggly}{wa2..wa6}
\fmfi{wiggly}{wa6..wa7}
\end{fmfchar*}}}
}
\subfigspace
\subfigure[$G_{2}$ $(\times 2)$]{
\raisebox{\eqoff}{%
\fmfframe(3,1)(1,4){%
\begin{fmfchar*}(16,20)
\Wu
\fmfipair{wa[]}
\fmfipair{wb[]}
\fmfipair{wc[]}
\fmfipair{wd[]}
\fmfiequ{wa0}{point 1*length(p0)/2 of p0}
\fmfiv{d.shape=circle,d.size=2}{wa0}
\fmfiequ{wa2}{point 1*length(p2)/3 of p2}
\fmfiv{d.shape=circle,d.size=2}{wa2}
\fmfiequ{wb2}{point 2*length(p2)/3 of p2}
\fmfiv{d.shape=circle,d.size=2}{wb2}
\fmfiequ{wa6}{point 1*length(p6)/2 of p6}
\fmfiv{d.shape=circle,d.size=2}{wa6}
\fmfiequ{wa7}{point 1*length(p7)/2 of p7}
\fmfiv{d.shape=circle,d.size=2}{wa7}
\fmfi{wiggly}{wa0..wa2}
\fmfforce{(-0w,-0h)}{va0}
\fmfforce{(1w,-0h)}{vb0}
\wigglywrap{wa0}{va0}{vb0}{wa7}
\fmfi{wiggly}{wb2..wa6}
\fmfi{wiggly}{wa6..wa7}
\end{fmfchar*}}}
}

\begin{tabular}{m{12cm}}
\toprule
$G_1\rightarrow \chi(1)\,J_1 \rightarrow-2M\,J_1$  \\
$G_2\rightarrow -\chi(1)\,J_{20} \rightarrow 2M\,J_{20}$  \\
\midrule
$\chi(1)\rightarrow 2\chi(1)(J_1-J_{20}) \rightarrow -4M(J_1-J_{20})$   \\ 
\bottomrule
\end{tabular}

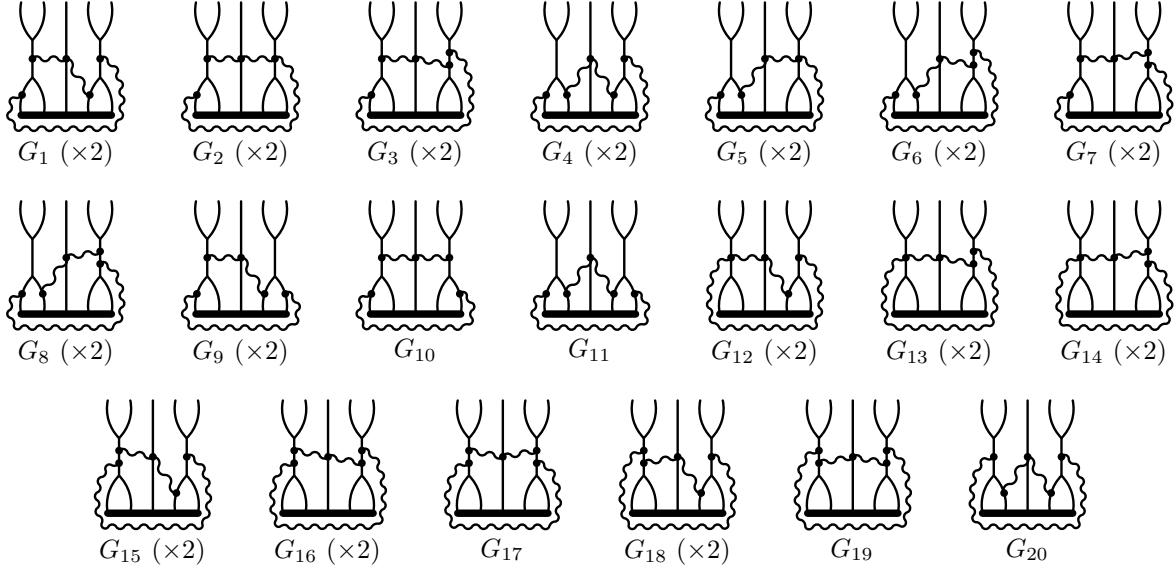
\caption{Wrapping diagrams with structure $\chi(1)$}
\label{wrap-1}
\end{figure}

The wrapping diagrams we need to compute and the corresponding
contributions are listed in Figs.~\ref{wrap-chiral}-\ref{wrap-1}. In
each case, the final result for the whole structure already contains
all the symmetry factors and the contribution for the possible
reflected structure.
The symmetry factor of a diagram is explicitly shown if different from
1. In each diagram, the color factor $(g^2 N)^5$ combines with the
$1/(4\pi)^{10}$ from the momentum integral to produce the coupling
$\lambda^5$ as defined in \eqref{lambdadef}, and thus it is not shown
explicitly.

\begin{figure}[h]
\centering
\unitlength=0.75mm
\settoheight{\eqoff}{$\times$}%
\setlength{\eqoff}{0.5\eqoff}%
\addtolength{\eqoff}{-12.5\unitlength}%
\settoheight{\eqofftwo}{$\times$}%
\setlength{\eqofftwo}{0.5\eqofftwo}%
\addtolength{\eqofftwo}{-7.5\unitlength}%
\subfigure[$G_{1}$]{
\raisebox{\eqoff}{%
\fmfframe(3,1)(1,4){%
\begin{fmfchar*}(16,20)
\Wdqut
\fmfipair{wa[]}
\fmfipair{wb[]}
\fmfipair{wc[]}
\fmfipair{wd[]}
\fmfiequ{wa0}{point 1*length(p0)/2 of p0}
\fmfiv{d.shape=circle,d.size=2}{wa0}
\fmfiequ{wa3}{point 1*length(p3)/2 of p3}
\fmfiv{d.shape=circle,d.size=2}{wa3}
\fmfforce{(-0w,-0h)}{va0}
\fmfforce{(w,-0h)}{vb0}
\wigglywrap{wa0}{va0}{vb0}{wa3}
\end{fmfchar*}}}
}
\subfigspace
\subfigure[$G_{2}$]{
\raisebox{\eqoff}{%
\fmfframe(3,1)(1,4){%
\begin{fmfchar*}(16,20)
\Wdqut
\fmfipair{wa[]}
\fmfipair{wb[]}
\fmfipair{wc[]}
\fmfipair{wd[]}
\fmfiequ{wa0}{point 1*length(p0)/2 of p0}
\fmfiv{d.shape=circle,d.size=2}{wa0}
\fmfiequ{wa4}{point 1*length(p4)/2 of p4}
\fmfiv{d.shape=circle,d.size=2}{wa4}
\fmfforce{(-0w,-0h)}{va0}
\fmfforce{(w,-0h)}{vb0}
\wigglywrap{wa0}{va0}{vb0}{wa4}
\end{fmfchar*}}}
}
\subfigspace
\subfigure[$G_{3}$]{
\raisebox{\eqoff}{%
\fmfframe(3,1)(1,4){%
\begin{fmfchar*}(16,20)
\Wdqut
\fmfipair{wa[]}
\fmfipair{wb[]}
\fmfipair{wc[]}
\fmfipair{wd[]}
\fmfiequ{wa1}{point 1*length(p1)/2 of p1}
\fmfiv{d.shape=circle,d.size=2}{wa1}
\fmfiequ{wa3}{point 1*length(p3)/2 of p3}
\fmfiv{d.shape=circle,d.size=2}{wa3}
\fmfforce{(-0w,-0h)}{va0}
\fmfforce{(w,-0h)}{vb0}
\wigglywrap{wa1}{va0}{vb0}{wa3}
\end{fmfchar*}}}
}
\subfigspace
\subfigure[$G_{4}$]{
\raisebox{\eqoff}{%
\fmfframe(3,1)(1,4){%
\begin{fmfchar*}(16,20)
\Wdqut
\fmfipair{wa[]}
\fmfipair{wb[]}
\fmfipair{wc[]}
\fmfipair{wd[]}
\fmfiequ{wa1}{point 1*length(p1)/2 of p1}
\fmfiv{d.shape=circle,d.size=2}{wa1}
\fmfiequ{wa4}{point 1*length(p4)/2 of p4}
\fmfiv{d.shape=circle,d.size=2}{wa4}
\fmfforce{(-0w,-0h)}{va0}
\fmfforce{(w,-0h)}{vb0}
\wigglywrap{wa1}{va0}{vb0}{wa4}
\end{fmfchar*}}}
}
\begin{tabular}{m{12cm}}
\toprule
$G_1\rightarrow \chi(2,4,1,3)(-J_{20}-J_{21}-2J_{24}) \rightarrow M(-J_{20}-J_{21}-2J_{24})$  \\
$G_2\rightarrow \chi(2,4,1,3)\,J_{20} \rightarrow  M\,J_{20}$ \\
$G_3\rightarrow \chi(2,4,1,3)\,J_{21} \rightarrow M\,J_{21}$ \\
$G_4\rightarrow0$  \\
\midrule
$\chi(2,4,1,3)\rightarrow -4\,J_{24}\,\chi(2,4,1,3) \rightarrow -4\,M\,J_{24}$   \\
\bottomrule
\end{tabular}
\caption{Wrapping diagrams with structure $\chi(2,4,1,3)$ or $\chi(1,3,2,4)$}
\label{wrap-2413}
\end{figure}



\begin{figure}[h]
\centering
\unitlength=0.75mm
\settoheight{\eqoff}{$\times$}%
\setlength{\eqoff}{0.5\eqoff}%
\addtolength{\eqoff}{-12.5\unitlength}%
\settoheight{\eqofftwo}{$\times$}%
\setlength{\eqofftwo}{0.5\eqofftwo}%
\addtolength{\eqofftwo}{-7.5\unitlength}%
\subfigure[$G_{1}$]{
\raisebox{\eqoff}{%
\fmfframe(3,1)(1,4){%
\begin{fmfchar*}(16,20)
\Wtduq
\fmfipair{wa[]}
\fmfipair{wb[]}
\fmfipair{wc[]}
\fmfipair{wd[]}
\fmfiequ{wa0}{point 1*length(p0)/2 of p0}
\fmfiv{d.shape=circle,d.size=2}{wa0}
\fmfiequ{wa3}{point 1*length(p3)/2 of p3}
\fmfiv{d.shape=circle,d.size=2}{wa3}
\fmfforce{(-0w,-0h)}{va0}
\fmfforce{(w,-0h)}{vb0}
\wigglywrap{wa0}{va0}{vb0}{wa3}
\end{fmfchar*}}}
}
\subfigspace
\subfigure[$G_{2}$]{
\raisebox{\eqoff}{%
\fmfframe(3,1)(1,4){%
\begin{fmfchar*}(16,20)
\Wtduq
\fmfipair{wa[]}
\fmfipair{wb[]}
\fmfipair{wc[]}
\fmfipair{wd[]}
\fmfiequ{wa0}{point 1*length(p0)/2 of p0}
\fmfiv{d.shape=circle,d.size=2}{wa0}
\fmfiequ{wa4}{point 1*length(p4)/2 of p4}
\fmfiv{d.shape=circle,d.size=2}{wa4}
\fmfforce{(-0w,-0h)}{va0}
\fmfforce{(w,-0h)}{vb0}
\wigglywrap{wa0}{va0}{vb0}{wa4}
\end{fmfchar*}}}
}
\subfigspace
\subfigure[$G_{3}$]{
\raisebox{\eqoff}{%
\fmfframe(3,1)(1,4){%
\begin{fmfchar*}(16,20)
\Wtduq
\fmfipair{wa[]}
\fmfipair{wb[]}
\fmfipair{wc[]}
\fmfipair{wd[]}
\fmfiequ{wa1}{point 1*length(p1)/2 of p1}
\fmfiv{d.shape=circle,d.size=2}{wa1}
\fmfiequ{wa3}{point 1*length(p3)/2 of p3}
\fmfiv{d.shape=circle,d.size=2}{wa3}
\fmfforce{(-0w,-0h)}{va0}
\fmfforce{(w,-0h)}{vb0}
\wigglywrap{wa1}{va0}{vb0}{wa3}
\end{fmfchar*}}}
}
\subfigspace
\subfigure[$G_{4}$]{
\raisebox{\eqoff}{%
\fmfframe(3,1)(1,4){%
\begin{fmfchar*}(16,20)
\Wtduq
\fmfipair{wa[]}
\fmfipair{wb[]}
\fmfipair{wc[]}
\fmfipair{wd[]}
\fmfiequ{wa1}{point 1*length(p1)/2 of p1}
\fmfiv{d.shape=circle,d.size=2}{wa1}
\fmfiequ{wa4}{point 1*length(p4)/2 of p4}
\fmfiv{d.shape=circle,d.size=2}{wa4}
\fmfforce{(-0w,-0h)}{va0}
\fmfforce{(w,-0h)}{vb0}
\wigglywrap{wa1}{va0}{vb0}{wa4}
\end{fmfchar*}}}
}
\begin{tabular}{m{12cm}}
\toprule
$G_1\rightarrow \chi(3,2,1,4)(-J_{20}-J_{22}-2J_{25}) \rightarrow -2M(-J_{20}-J_{22}-2J_{25})$  \\
$G_2\rightarrow -\chi(3,2,1,4)(-J_{20}) \rightarrow 2M(-J_{20})$ \\
$G_3\rightarrow -\chi(3,2,1,4)(-J_{22}) \rightarrow 2M(-J_{22})$ \\
$G_4\rightarrow0$  \\
\midrule
$\chi(3,2,1,4)\rightarrow -4J_{25}\chi(3,2,1,4) \rightarrow 8M\,J_{25}$   \\
\bottomrule
\end{tabular}

\caption{Wrapping diagrams with structure $\chi(3,2,1,4)$ or $\chi(2,1,3,4)$}
\label{wrap-3214}
\end{figure}



\begin{figure}[h]
\centering
\unitlength=0.75mm
\settoheight{\eqoff}{$\times$}%
\setlength{\eqoff}{0.5\eqoff}%
\addtolength{\eqoff}{-12.5\unitlength}%
\settoheight{\eqofftwo}{$\times$}%
\setlength{\eqofftwo}{0.5\eqofftwo}%
\addtolength{\eqofftwo}{-7.5\unitlength}%
\subfigure[$G_{1}$]{
\raisebox{\eqoff}{%
\fmfframe(3,1)(1,4){%
\begin{fmfchar*}(16,20)
\Wudtq
\fmfipair{wa[]}
\fmfipair{wb[]}
\fmfipair{wc[]}
\fmfipair{wd[]}
\fmfiequ{wa0}{point 1*length(p0)/2 of p0}
\fmfiv{d.shape=circle,d.size=2}{wa0}
\fmfiequ{wa3}{point 1*length(p3)/2 of p3}
\fmfiv{d.shape=circle,d.size=2}{wa3}
\fmfforce{(-0w,-0h)}{va0}
\fmfforce{(w,-0h)}{vb0}
\wigglywrap{wa0}{va0}{vb0}{wa3}
\end{fmfchar*}}}
}
\subfigspace
\subfigure[$G_{2}$]{
\raisebox{\eqoff}{%
\fmfframe(3,1)(1,4){%
\begin{fmfchar*}(16,20)
\Wudtq
\fmfipair{wa[]}
\fmfipair{wb[]}
\fmfipair{wc[]}
\fmfipair{wd[]}
\fmfiequ{wa0}{point 1*length(p0)/2 of p0}
\fmfiv{d.shape=circle,d.size=2}{wa0}
\fmfiequ{wa4}{point 1*length(p4)/2 of p4}
\fmfiv{d.shape=circle,d.size=2}{wa4}
\fmfforce{(-0w,-0h)}{va0}
\fmfforce{(w,-0h)}{vb0}
\wigglywrap{wa0}{va0}{vb0}{wa4}
\end{fmfchar*}}}
}
\subfigspace
\subfigure[$G_{3}$]{
\raisebox{\eqoff}{%
\fmfframe(3,1)(1,4){%
\begin{fmfchar*}(16,20)
\Wudtq
\fmfipair{wa[]}
\fmfipair{wb[]}
\fmfipair{wc[]}
\fmfipair{wd[]}
\fmfiequ{wa1}{point 1*length(p1)/2 of p1}
\fmfiv{d.shape=circle,d.size=2}{wa1}
\fmfiequ{wa3}{point 1*length(p3)/2 of p3}
\fmfiv{d.shape=circle,d.size=2}{wa3}
\fmfforce{(-0w,-0h)}{va0}
\fmfforce{(w,-0h)}{vb0}
\wigglywrap{wa1}{va0}{vb0}{wa3}
\end{fmfchar*}}}
}
\subfigspace
\subfigure[$G_{4}$]{
\raisebox{\eqoff}{%
\fmfframe(3,1)(1,4){%
\begin{fmfchar*}(16,20)
\Wudtq
\fmfipair{wa[]}
\fmfipair{wb[]}
\fmfipair{wc[]}
\fmfipair{wd[]}
\fmfiequ{wa1}{point 1*length(p1)/2 of p1}
\fmfiv{d.shape=circle,d.size=2}{wa1}
\fmfiequ{wa4}{point 1*length(p4)/2 of p4}
\fmfiv{d.shape=circle,d.size=2}{wa4}
\fmfforce{(-0w,-0h)}{va0}
\fmfforce{(w,-0h)}{vb0}
\wigglywrap{wa1}{va0}{vb0}{wa4}
\end{fmfchar*}}}
}
\begin{tabular}{m{12cm}}
\toprule
$G_1\rightarrow \chi(1,2,3,4)(-J_{22}-J_{23}-2J_{26}) \rightarrow -2M(-J_{22}-J_{23}-2J_{26})$  \\
$G_2\rightarrow -\chi(1,2,3,4)(-J_{22}) \rightarrow 2M(-J_{22})$ \\
$G_3\rightarrow -\chi(1,2,3,4)(-J_{1}) \rightarrow 2M(-J_{1})$ \\
$G_4\rightarrow \chi(1,2,3,4)(-J_{20}) \rightarrow -2M(-J_{20})$  \\
\midrule
$\chi(1,2,3,4)\rightarrow -2\chi(1,2,3,4)(-J_1+J_{20}+J_{23}+2J_{26})$  \\
$\phantom{\chi(1,2,3,4)}\rightarrow 4M(-J_1+J_{20}+J_{23}+2J_{26})$  \\
\bottomrule
\end{tabular}

\caption{Wrapping diagrams with structure $\chi(1,2,3,4)$ or $\chi(4,3,2,1)$}
\label{wrap-1234}
\end{figure}



\newpage


\newpage


\begin{figure}[h]
\centering
\unitlength=0.75mm
\settoheight{\eqoff}{$\times$}%
\setlength{\eqoff}{0.5\eqoff}%
\addtolength{\eqoff}{-12.5\unitlength}%
\settoheight{\eqofftwo}{$\times$}%
\setlength{\eqofftwo}{0.5\eqofftwo}%
\addtolength{\eqofftwo}{-7.5\unitlength}%
\subfigure[$G_{1}$]{
\raisebox{\eqoff}{%
\fmfframe(3,1)(1,4){%
\begin{fmfchar*}(16,20)
\Wdu
\fmfipair{wa[]}
\fmfipair{wb[]}
\fmfipair{wc[]}
\fmfipair{wd[]}
\fmfiequ{wa0}{point 1*length(p0)/2 of p0}
\fmfiv{d.shape=circle,d.size=2}{wa0}
\fmfiequ{wa3}{point 1*length(p3)/2 of p3}
\fmfiv{d.shape=circle,d.size=2}{wa3}
\fmfiequ{wa6}{point 1*length(p6)/2 of p6}
\fmfiv{d.shape=circle,d.size=2}{wa6}
\fmfiequ{wa7}{point 1*length(p7)/2 of p7}
\fmfiv{d.shape=circle,d.size=2}{wa7}
\fmfforce{(-0w,-0h)}{va0}
\fmfforce{(1w,-0h)}{vb0}
\wigglywrap{wa0}{va0}{vb0}{wa7}
\fmfi{wiggly}{wa3..wa6}
\fmfi{wiggly}{wa6..wa7}
\end{fmfchar*}}}
}
\subfigspace
\subfigure[$G_{2}$]{
\raisebox{\eqoff}{%
\fmfframe(3,1)(1,4){%
\begin{fmfchar*}(16,20)
\Wdu
\fmfipair{wa[]}
\fmfipair{wb[]}
\fmfipair{wc[]}
\fmfipair{wd[]}
\fmfiequ{wa0}{point 1*length(p0)/2 of p0}
\fmfiv{d.shape=circle,d.size=2}{wa0}
\fmfiequ{wa4}{point 1*length(p4)/2 of p4}
\fmfiv{d.shape=circle,d.size=2}{wa4}
\fmfiequ{wa6}{point 1*length(p6)/2 of p6}
\fmfiv{d.shape=circle,d.size=2}{wa6}
\fmfiequ{wa7}{point 1*length(p7)/2 of p7}
\fmfiv{d.shape=circle,d.size=2}{wa7}
\fmfforce{(-0w,-0h)}{va0}
\fmfforce{(1w,-0h)}{vb0}
\wigglywrap{wa0}{va0}{vb0}{wa7}
\fmfi{wiggly}{wa4..wa6}
\fmfi{wiggly}{wa6..wa7}
\end{fmfchar*}}}
}
\subfigspace
\subfigure[$G_{3}$]{
\raisebox{\eqoff}{%
\fmfframe(3,1)(1,4){%
\begin{fmfchar*}(16,20)
\Wdu
\fmfipair{wa[]}
\fmfipair{wb[]}
\fmfipair{wc[]}
\fmfipair{wd[]}
\fmfiequ{wa1}{point 1*length(p1)/2 of p1}
\fmfiv{d.shape=circle,d.size=2}{wa1}
\fmfiequ{wa3}{point 1*length(p3)/2 of p3}
\fmfiv{d.shape=circle,d.size=2}{wa3}
\fmfiequ{wa6}{point 1*length(p6)/2 of p6}
\fmfiv{d.shape=circle,d.size=2}{wa6}
\fmfiequ{wa7}{point 1*length(p7)/2 of p7}
\fmfiv{d.shape=circle,d.size=2}{wa7}
\fmfforce{(-0w,-0h)}{va0}
\fmfforce{(1w,-0h)}{vb0}
\wigglywrap{wa1}{va0}{vb0}{wa7}
\fmfi{wiggly}{wa3..wa6}
\fmfi{wiggly}{wa6..wa7}
\end{fmfchar*}}}
}
\subfigspace
\subfigure[$G_{4}$]{
\raisebox{\eqoff}{%
\fmfframe(3,1)(1,4){%
\begin{fmfchar*}(16,20)
\Wdu
\fmfipair{wa[]}
\fmfipair{wb[]}
\fmfipair{wc[]}
\fmfipair{wd[]}
\fmfiequ{wa1}{point 1*length(p1)/2 of p1}
\fmfiv{d.shape=circle,d.size=2}{wa1}
\fmfiequ{wa4}{point 1*length(p4)/2 of p4}
\fmfiv{d.shape=circle,d.size=2}{wa4}
\fmfiequ{wa6}{point 1*length(p6)/2 of p6}
\fmfiv{d.shape=circle,d.size=2}{wa6}
\fmfiequ{wa7}{point 1*length(p7)/2 of p7}
\fmfiv{d.shape=circle,d.size=2}{wa7}
\fmfforce{(-0w,-0h)}{va0}
\fmfforce{(1w,-0h)}{vb0}
\wigglywrap{wa1}{va0}{vb0}{wa7}
\fmfi{wiggly}{wa4..wa6}
\fmfi{wiggly}{wa6..wa7}
\end{fmfchar*}}}
}

\begin{tabular}{m{12cm}}
\toprule
$G_1\rightarrow \chi(2,1)(-J_1) \rightarrow M(-J_1)$  \\
$G_2\rightarrow -\chi(2,1)(-J_{22}-J_{23}-2J_{26}) \rightarrow-M(-J_{22}-J_{23}-2J_{26})$ \\
$G_3\rightarrow -\chi(2,1)(-J_{20}) \rightarrow-M(-J_{20})$ \\
$G_4\rightarrow \chi(2,1)(-J_{22}) \rightarrow M(-J_{22})$  \\
\midrule
$\chi(2,1)\rightarrow -2\chi(2,1)(J_1-J_{20}-J_{23}-2J_{26})$  \\
$\phantom{\chi(2,1)}\rightarrow -2M(J_1-J_{20}-J_{23}-2J_{26})$  \\
\bottomrule
\end{tabular}

\caption{Wrapping diagrams with structure $\chi(2,1)$ or $\chi(1,2)$}
\label{wrap-21}
\end{figure}



\newpage


\begin{figure}[h]
\centering
\unitlength=0.75mm
\settoheight{\eqoff}{$\times$}%
\setlength{\eqoff}{0.5\eqoff}%
\addtolength{\eqoff}{-12.5\unitlength}%
\settoheight{\eqofftwo}{$\times$}%
\setlength{\eqofftwo}{0.5\eqofftwo}%
\addtolength{\eqofftwo}{-7.5\unitlength}%
\subfigure[$G_{1}$ $(\times 2)$]{
\raisebox{\eqoff}{%
\fmfframe(3,1)(1,4){%
\begin{fmfchar*}(16,20)
\Wuq
\fmfipair{wa[]}
\fmfipair{wb[]}
\fmfipair{wc[]}
\fmfipair{wd[]}
\fmfiequ{wa0}{point 1*length(p0)/2 of p0}
\fmfiv{d.shape=circle,d.size=2}{wa0}
\fmfiequ{wa1}{point 1*length(p1)/2 of p1}
\fmfiv{d.shape=circle,d.size=2}{wa1}
\fmfiequ{wa5}{point 1*length(p5)/2 of p5}
\fmfiv{d.shape=circle,d.size=2}{wa5}
\fmfiequ{wa6}{point 1*length(p6)/2 of p6}
\fmfiv{d.shape=circle,d.size=2}{wa6}
\fmfiequ{wa7}{point 1*length(p7)/2 of p7}
\fmfiv{d.shape=circle,d.size=2}{wa7}
\fmfforce{(-0w,-0h)}{va0}
\fmfforce{(1w,-0h)}{vb0}
\wigglywrap{wa0}{va0}{vb0}{wa7}
\fmfi{wiggly}{wa1..wa5}
\fmfi{wiggly}{wa5..wa6}
\end{fmfchar*}}}
}
\subfigspace
\subfigure[$G_{2}$ $(\times 2)$]{
\raisebox{\eqoff}{%
\fmfframe(3,1)(1,4){%
\begin{fmfchar*}(16,20)
\Wuq
\fmfipair{wa[]}
\fmfipair{wb[]}
\fmfipair{wc[]}
\fmfipair{wd[]}
\fmfiequ{wa0}{point 1*length(p0)/2 of p0}
\fmfiv{d.shape=circle,d.size=2}{wa0}
\fmfiequ{wa1}{point 1*length(p1)/2 of p1}
\fmfiv{d.shape=circle,d.size=2}{wa1}
\fmfiequ{wa5}{point 1*length(p5)/2 of p5}
\fmfiv{d.shape=circle,d.size=2}{wa5}
\fmfiequ{wa7}{point 1*length(p7)/2 of p7}
\fmfiv{d.shape=circle,d.size=2}{wa7}
\fmfforce{(-0w,-0h)}{va0}
\fmfforce{(1w,-0h)}{vb0}
\wigglywrap{wa0}{va0}{vb0}{wa7}
\fmfi{wiggly}{wa1..wa5}
\fmfi{wiggly}{wa5..wa7}
\end{fmfchar*}}}
}
\subfigspace
\subfigure[$G_{3}$ $(\times 2)$]{
\raisebox{\eqoff}{%
\fmfframe(3,1)(1,4){%
\begin{fmfchar*}(16,20)
\Wuq
\fmfipair{wa[]}
\fmfipair{wb[]}
\fmfipair{wc[]}
\fmfipair{wd[]}
\fmfiequ{wa0}{point 1*length(p0)/2 of p0}
\fmfiv{d.shape=circle,d.size=2}{wa0}
\fmfiequ{wa1}{point 1*length(p1)/2 of p1}
\fmfiv{d.shape=circle,d.size=2}{wa1}
\fmfiequ{wa5}{point 1*length(p5)/2 of p5}
\fmfiv{d.shape=circle,d.size=2}{wa5}
\fmfiequ{wa7}{point 1*length(p7)/3 of p7}
\fmfiv{d.shape=circle,d.size=2}{wa7}
\fmfiequ{wb7}{point 2*length(p7)/3 of p7}
\fmfiv{d.shape=circle,d.size=2}{wb7}
\fmfforce{(-0w,-0h)}{va0}
\fmfforce{(1w,-0h)}{vb0}
\wigglywrap{wa0}{va0}{vb0}{wa7}
\fmfi{wiggly}{wa1..wa5}
\fmfi{wiggly}{wa5..wb7}
\end{fmfchar*}}}
}
\subfigspace
\subfigure[$G_{4}$ $(\times 2)$]{
\raisebox{\eqoff}{%
\fmfframe(3,1)(1,4){%
\begin{fmfchar*}(16,20)
\Wuq
\fmfipair{wa[]}
\fmfipair{wb[]}
\fmfipair{wc[]}
\fmfipair{wd[]}
\fmfiequ{wa0}{point 1*length(p0)/2 of p0}
\fmfiv{d.shape=circle,d.size=2}{wa0}
\fmfiequ{wa3}{point 1*length(p3)/2 of p3}
\fmfiv{d.shape=circle,d.size=2}{wa3}
\fmfiequ{wa5}{point 1*length(p5)/2 of p5}
\fmfiv{d.shape=circle,d.size=2}{wa5}
\fmfiequ{wa6}{point 1*length(p6)/2 of p6}
\fmfiv{d.shape=circle,d.size=2}{wa6}
\fmfiequ{wa7}{point 1*length(p7)/2 of p7}
\fmfiv{d.shape=circle,d.size=2}{wa7}
\fmfforce{(-0w,-0h)}{va0}
\fmfforce{(1w,-0h)}{vb0}
\wigglywrap{wa0}{va0}{vb0}{wa7}
\fmfi{wiggly}{wa3..wa5}
\fmfi{wiggly}{wa5..wa6}
\end{fmfchar*}}}
}
\subfigspace
\subfigure[$G_{5}$ $(\times 2)$]{
\raisebox{\eqoff}{%
\fmfframe(3,1)(1,4){%
\begin{fmfchar*}(16,20)
\Wuq
\fmfipair{wa[]}
\fmfipair{wb[]}
\fmfipair{wc[]}
\fmfipair{wd[]}
\fmfiequ{wa0}{point 1*length(p0)/2 of p0}
\fmfiv{d.shape=circle,d.size=2}{wa0}
\fmfiequ{wa3}{point 1*length(p3)/2 of p3}
\fmfiv{d.shape=circle,d.size=2}{wa3}
\fmfiequ{wa5}{point 1*length(p5)/2 of p5}
\fmfiv{d.shape=circle,d.size=2}{wa5}
\fmfiequ{wa7}{point 1*length(p7)/2 of p7}
\fmfiv{d.shape=circle,d.size=2}{wa7}
\fmfforce{(-0w,-0h)}{va0}
\fmfforce{(1w,-0h)}{vb0}
\wigglywrap{wa0}{va0}{vb0}{wa7}
\fmfi{wiggly}{wa3..wa5}
\fmfi{wiggly}{wa5..wa7}
\end{fmfchar*}}}
}
\subfigspace
\subfigure[$G_{6}$ $(\times 2)$]{
\raisebox{\eqoff}{%
\fmfframe(3,1)(1,4){%
\begin{fmfchar*}(16,20)
\Wuq
\fmfipair{wa[]}
\fmfipair{wb[]}
\fmfipair{wc[]}
\fmfipair{wd[]}
\fmfiequ{wa0}{point 1*length(p0)/2 of p0}
\fmfiv{d.shape=circle,d.size=2}{wa0}
\fmfiequ{wa3}{point 1*length(p3)/2 of p3}
\fmfiv{d.shape=circle,d.size=2}{wa3}
\fmfiequ{wa5}{point 1*length(p5)/2 of p5}
\fmfiv{d.shape=circle,d.size=2}{wa5}
\fmfiequ{wa7}{point 1*length(p7)/3 of p7}
\fmfiv{d.shape=circle,d.size=2}{wa7}
\fmfiequ{wb7}{point 2*length(p7)/3 of p7}
\fmfiv{d.shape=circle,d.size=2}{wb7}
\fmfforce{(-0w,-0h)}{va0}
\fmfforce{(1w,-0h)}{vb0}
\wigglywrap{wa0}{va0}{vb0}{wa7}
\fmfi{wiggly}{wa3..wa5}
\fmfi{wiggly}{wa5..wb7}
\end{fmfchar*}}}
}
\subfigspace
\subfigure[$G_{7}$ $(\times 2)$]{
\raisebox{\eqoff}{%
\fmfframe(3,1)(1,4){%
\begin{fmfchar*}(16,20)
\Wuq
\fmfipair{wa[]}
\fmfipair{wb[]}
\fmfipair{wc[]}
\fmfipair{wd[]}
\fmfiequ{wa0}{point 1*length(p0)/2 of p0}
\fmfiv{d.shape=circle,d.size=2}{wa0}
\fmfiequ{wa1}{point 1*length(p1)/2 of p1}
\fmfiv{d.shape=circle,d.size=2}{wa1}
\fmfiequ{wa5}{point 1*length(p5)/2 of p5}
\fmfiv{d.shape=circle,d.size=2}{wa5}
\fmfiequ{wa7}{point 1*length(p7)/3 of p7}
\fmfiv{d.shape=circle,d.size=2}{wa7}
\fmfiequ{wb7}{point 2*length(p7)/3 of p7}
\fmfiv{d.shape=circle,d.size=2}{wb7}
\fmfforce{(-0w,-0h)}{va0}
\fmfforce{(1w,-0h)}{vb0}
\wigglywrap{wa0}{va0}{vb0}{wb7}
\fmfi{wiggly}{wa1..wa5}
\fmfi{wiggly}{wa5..wa7}
\end{fmfchar*}}}
}
\\
\subfigure[$G_{8}$ $(\times 2)$]{
\raisebox{\eqoff}{%
\fmfframe(3,1)(1,4){%
\begin{fmfchar*}(16,20)
\Wuq
\fmfipair{wa[]}
\fmfipair{wb[]}
\fmfipair{wc[]}
\fmfipair{wd[]}
\fmfiequ{wa0}{point 1*length(p0)/2 of p0}
\fmfiv{d.shape=circle,d.size=2}{wa0}
\fmfiequ{wa3}{point 1*length(p3)/2 of p3}
\fmfiv{d.shape=circle,d.size=2}{wa3}
\fmfiequ{wa5}{point 1*length(p5)/2 of p5}
\fmfiv{d.shape=circle,d.size=2}{wa5}
\fmfiequ{wa7}{point 1*length(p7)/3 of p7}
\fmfiv{d.shape=circle,d.size=2}{wa7}
\fmfiequ{wb7}{point 2*length(p7)/3 of p7}
\fmfiv{d.shape=circle,d.size=2}{wb7}
\fmfforce{(-0w,-0h)}{va0}
\fmfforce{(1w,-0h)}{vb0}
\wigglywrap{wa0}{va0}{vb0}{wb7}
\fmfi{wiggly}{wa3..wa5}
\fmfi{wiggly}{wa5..wa7}
\end{fmfchar*}}}
}
\subfigspace
\subfigure[$G_{9}$ $(\times 2)$]{
\raisebox{\eqoff}{%
\fmfframe(3,1)(1,4){%
\begin{fmfchar*}(16,20)
\Wuq
\fmfipair{wa[]}
\fmfipair{wb[]}
\fmfipair{wc[]}
\fmfipair{wd[]}
\fmfiequ{wa0}{point 1*length(p0)/2 of p0}
\fmfiv{d.shape=circle,d.size=2}{wa0}
\fmfiequ{wa1}{point 1*length(p1)/2 of p1}
\fmfiv{d.shape=circle,d.size=2}{wa1}
\fmfiequ{wa5}{point 1*length(p5)/2 of p5}
\fmfiv{d.shape=circle,d.size=2}{wa5}
\fmfiequ{wa6}{point 1*length(p6)/2 of p6}
\fmfiv{d.shape=circle,d.size=2}{wa6}
\fmfiequ{wa9}{point 1*length(p9)/2 of p9}
\fmfiv{d.shape=circle,d.size=2}{wa9}
\fmfforce{(-0w,-0h)}{va0}
\fmfforce{(1w,-0h)}{vb0}
\wigglywrap{wa0}{va0}{vb0}{wa9}
\fmfi{wiggly}{wa1..wa5}
\fmfi{wiggly}{wa5..wa6}
\end{fmfchar*}}}
}
\subfigspace
\subfigure[$G_{10}$]{
\raisebox{\eqoff}{%
\fmfframe(3,1)(1,4){%
\begin{fmfchar*}(16,20)
\Wuq
\fmfipair{wa[]}
\fmfipair{wb[]}
\fmfipair{wc[]}
\fmfipair{wd[]}
\fmfiequ{wa0}{point 1*length(p0)/2 of p0}
\fmfiv{d.shape=circle,d.size=2}{wa0}
\fmfiequ{wa1}{point 1*length(p1)/2 of p1}
\fmfiv{d.shape=circle,d.size=2}{wa1}
\fmfiequ{wa5}{point 1*length(p5)/2 of p5}
\fmfiv{d.shape=circle,d.size=2}{wa5}
\fmfiequ{wa7}{point 1*length(p7)/2 of p7}
\fmfiv{d.shape=circle,d.size=2}{wa7}
\fmfiequ{wa9}{point 1*length(p9)/2 of p9}
\fmfiv{d.shape=circle,d.size=2}{wa9}
\fmfforce{(-0w,-0h)}{va0}
\fmfforce{(1w,-0h)}{vb0}
\wigglywrap{wa0}{va0}{vb0}{wa9}
\fmfi{wiggly}{wa1..wa5}
\fmfi{wiggly}{wa5..wa7}
\end{fmfchar*}}}
}
\subfigspace
\subfigure[$G_{11}$]{
\raisebox{\eqoff}{%
\fmfframe(3,1)(1,4){%
\begin{fmfchar*}(16,20)
\Wuq
\fmfipair{wa[]}
\fmfipair{wb[]}
\fmfipair{wc[]}
\fmfipair{wd[]}
\fmfiequ{wa0}{point 1*length(p0)/2 of p0}
\fmfiv{d.shape=circle,d.size=2}{wa0}
\fmfiequ{wa3}{point 1*length(p3)/2 of p3}
\fmfiv{d.shape=circle,d.size=2}{wa3}
\fmfiequ{wa5}{point 1*length(p5)/2 of p5}
\fmfiv{d.shape=circle,d.size=2}{wa5}
\fmfiequ{wa6}{point 1*length(p6)/2 of p6}
\fmfiv{d.shape=circle,d.size=2}{wa6}
\fmfiequ{wa9}{point 1*length(p9)/2 of p9}
\fmfiv{d.shape=circle,d.size=2}{wa9}
\fmfforce{(-0w,-0h)}{va0}
\fmfforce{(1w,-0h)}{vb0}
\wigglywrap{wa0}{va0}{vb0}{wa9}
\fmfi{wiggly}{wa3..wa5}
\fmfi{wiggly}{wa5..wa6}
\end{fmfchar*}}}
}
\subfigspace
\subfigure[$G_{12}$ $(\times 2)$]{
\raisebox{\eqoff}{%
\fmfframe(3,1)(1,4){%
\begin{fmfchar*}(16,20)
\Wuq
\fmfipair{wa[]}
\fmfipair{wb[]}
\fmfipair{wc[]}
\fmfipair{wd[]}
\fmfiequ{wa1}{point 1*length(p1)/2 of p1}
\fmfiv{d.shape=circle,d.size=2}{wa1}
\fmfiequ{wa5}{point 1*length(p5)/2 of p5}
\fmfiv{d.shape=circle,d.size=2}{wa5}
\fmfiequ{wa6}{point 1*length(p6)/2 of p6}
\fmfiv{d.shape=circle,d.size=2}{wa6}
\fmfiequ{wa7}{point 1*length(p7)/2 of p7}
\fmfiv{d.shape=circle,d.size=2}{wa7}
\fmfi{wiggly}{wa1..wa5}
\fmfforce{(-0w,-0h)}{va0}
\fmfforce{(1w,-0h)}{vb0}
\wigglywrap{wa1}{va0}{vb0}{wa7}
\fmfi{wiggly}{wa5..wa6}
\end{fmfchar*}}}
}
\subfigspace
\subfigure[$G_{13}$ $(\times 2)$]{
\raisebox{\eqoff}{%
\fmfframe(3,1)(1,4){%
\begin{fmfchar*}(16,20)
\Wuq
\fmfipair{wa[]}
\fmfipair{wb[]}
\fmfipair{wc[]}
\fmfipair{wd[]}
\fmfiequ{wa1}{point 1*length(p1)/2 of p1}
\fmfiv{d.shape=circle,d.size=2}{wa1}
\fmfiequ{wa5}{point 1*length(p5)/2 of p5}
\fmfiv{d.shape=circle,d.size=2}{wa5}
\fmfiequ{wa7}{point 1*length(p7)/3 of p7}
\fmfiv{d.shape=circle,d.size=2}{wa7}
\fmfiequ{wb7}{point 2*length(p7)/3 of p7}
\fmfiv{d.shape=circle,d.size=2}{wb7}
\fmfi{wiggly}{wa1..wa5}
\fmfforce{(-0w,-0h)}{va0}
\fmfforce{(1w,-0h)}{vb0}
\wigglywrap{wa1}{va0}{vb0}{wa7}
\fmfi{wiggly}{wa5..wb7}
\end{fmfchar*}}}
}
\subfigspace
\subfigure[$G_{14}$ $(\times 2)$]{
\raisebox{\eqoff}{%
\fmfframe(3,1)(1,4){%
\begin{fmfchar*}(16,20)
\Wuq
\fmfipair{wa[]}
\fmfipair{wb[]}
\fmfipair{wc[]}
\fmfipair{wd[]}
\fmfiequ{wa1}{point 1*length(p1)/2 of p1}
\fmfiv{d.shape=circle,d.size=2}{wa1}
\fmfiequ{wa5}{point 1*length(p5)/2 of p5}
\fmfiv{d.shape=circle,d.size=2}{wa5}
\fmfiequ{wa7}{point 1*length(p7)/3 of p7}
\fmfiv{d.shape=circle,d.size=2}{wa7}
\fmfiequ{wb7}{point 2*length(p7)/3 of p7}
\fmfiv{d.shape=circle,d.size=2}{wb7}
\fmfi{wiggly}{wa1..wa5}
\fmfforce{(-0w,-0h)}{va0}
\fmfforce{(1w,-0h)}{vb0}
\wigglywrap{wa1}{va0}{vb0}{wb7}
\fmfi{wiggly}{wa5..wa7}
\end{fmfchar*}}}
}
\\
\subfigure[$G_{15}$ $(\times 2)$]{
\raisebox{\eqoff}{%
\fmfframe(3,1)(1,4){%
\begin{fmfchar*}(16,20)
\Wuq
\fmfipair{wa[]}
\fmfipair{wb[]}
\fmfipair{wc[]}
\fmfipair{wd[]}
\fmfiequ{wa1}{point 1*length(p1)/3 of p1}
\fmfiv{d.shape=circle,d.size=2}{wa1}
\fmfiequ{wb1}{point 2*length(p1)/3 of p1}
\fmfiv{d.shape=circle,d.size=2}{wb1}
\fmfiequ{wa5}{point 1*length(p5)/2 of p5}
\fmfiv{d.shape=circle,d.size=2}{wa5}
\fmfiequ{wa6}{point 1*length(p6)/2 of p6}
\fmfiv{d.shape=circle,d.size=2}{wa6}
\fmfiequ{wa7}{point 1*length(p7)/2 of p7}
\fmfiv{d.shape=circle,d.size=2}{wa7}
\fmfi{wiggly}{wa1..wa5}
\fmfforce{(-0w,-0h)}{va0}
\fmfforce{(1w,-0h)}{vb0}
\wigglywrap{wb1}{va0}{vb0}{wa7}
\fmfi{wiggly}{wa5..wa6}
\end{fmfchar*}}}
}
\subfigspace
\subfigure[$G_{16}$ $(\times 2)$]{
\raisebox{\eqoff}{%
\fmfframe(3,1)(1,4){%
\begin{fmfchar*}(16,20)
\Wuq
\fmfipair{wa[]}
\fmfipair{wb[]}
\fmfipair{wc[]}
\fmfipair{wd[]}
\fmfiequ{wa1}{point 1*length(p1)/3 of p1}
\fmfiv{d.shape=circle,d.size=2}{wa1}
\fmfiequ{wb1}{point 2*length(p1)/3 of p1}
\fmfiv{d.shape=circle,d.size=2}{wb1}
\fmfiequ{wa5}{point 1*length(p5)/2 of p5}
\fmfiv{d.shape=circle,d.size=2}{wa5}
\fmfiequ{wa7}{point 1*length(p7)/3 of p7}
\fmfiv{d.shape=circle,d.size=2}{wa7}
\fmfiequ{wb7}{point 2*length(p7)/3 of p7}
\fmfiv{d.shape=circle,d.size=2}{wb7}
\fmfi{wiggly}{wa1..wa5}
\fmfforce{(-0w,-0h)}{va0}
\fmfforce{(1w,-0h)}{vb0}
\wigglywrap{wb1}{va0}{vb0}{wa7}
\fmfi{wiggly}{wa5..wb7}
\end{fmfchar*}}}
}
\subfigspace
\subfigure[$G_{17}$]{
\raisebox{\eqoff}{%
\fmfframe(3,1)(1,4){%
\begin{fmfchar*}(16,20)
\Wuq
\fmfipair{wa[]}
\fmfipair{wb[]}
\fmfipair{wc[]}
\fmfipair{wd[]}
\fmfiequ{wa1}{point 1*length(p1)/3 of p1}
\fmfiv{d.shape=circle,d.size=2}{wa1}
\fmfiequ{wb1}{point 2*length(p1)/3 of p1}
\fmfiv{d.shape=circle,d.size=2}{wb1}
\fmfiequ{wa5}{point 1*length(p5)/2 of p5}
\fmfiv{d.shape=circle,d.size=2}{wa5}
\fmfiequ{wa7}{point 1*length(p7)/3 of p7}
\fmfiv{d.shape=circle,d.size=2}{wa7}
\fmfiequ{wb7}{point 2*length(p7)/3 of p7}
\fmfiv{d.shape=circle,d.size=2}{wb7}
\fmfi{wiggly}{wa1..wa5}
\fmfforce{(-0w,-0h)}{va0}
\fmfforce{(1w,-0h)}{vb0}
\wigglywrap{wb1}{va0}{vb0}{wb7}
\fmfi{wiggly}{wa5..wa7}
\end{fmfchar*}}}
}
\subfigspace
\subfigure[$G_{18}$ $(\times 2)$]{
\raisebox{\eqoff}{%
\fmfframe(3,1)(1,4){%
\begin{fmfchar*}(16,20)
\Wuq
\fmfipair{wa[]}
\fmfipair{wb[]}
\fmfipair{wc[]}
\fmfipair{wd[]}
\fmfiequ{wa1}{point 1*length(p1)/3 of p1}
\fmfiv{d.shape=circle,d.size=2}{wa1}
\fmfiequ{wb1}{point 2*length(p1)/3 of p1}
\fmfiv{d.shape=circle,d.size=2}{wb1}
\fmfiequ{wa5}{point 1*length(p5)/2 of p5}
\fmfiv{d.shape=circle,d.size=2}{wa5}
\fmfiequ{wa6}{point 1*length(p6)/2 of p6}
\fmfiv{d.shape=circle,d.size=2}{wa6}
\fmfiequ{wa7}{point 1*length(p7)/2 of p7}
\fmfiv{d.shape=circle,d.size=2}{wa7}
\fmfi{wiggly}{wb1..wa5}
\fmfforce{(-0w,-0h)}{va0}
\fmfforce{(1w,-0h)}{vb0}
\wigglywrap{wa1}{va0}{vb0}{wa7}
\fmfi{wiggly}{wa5..wa6}
\end{fmfchar*}}}
}
\subfigspace
\subfigure[$G_{19}$]{
\raisebox{\eqoff}{%
\fmfframe(3,1)(1,4){%
\begin{fmfchar*}(16,20)
\Wuq
\fmfipair{wa[]}
\fmfipair{wb[]}
\fmfipair{wc[]}
\fmfipair{wd[]}
\fmfiequ{wa1}{point 1*length(p1)/3 of p1}
\fmfiv{d.shape=circle,d.size=2}{wa1}
\fmfiequ{wb1}{point 2*length(p1)/3 of p1}
\fmfiv{d.shape=circle,d.size=2}{wb1}
\fmfiequ{wa5}{point 1*length(p5)/2 of p5}
\fmfiv{d.shape=circle,d.size=2}{wa5}
\fmfiequ{wa7}{point 1*length(p7)/3 of p7}
\fmfiv{d.shape=circle,d.size=2}{wa7}
\fmfiequ{wb7}{point 2*length(p7)/3 of p7}
\fmfiv{d.shape=circle,d.size=2}{wb7}
\fmfi{wiggly}{wb1..wa5}
\fmfforce{(-0w,-0h)}{va0}
\fmfforce{(1w,-0h)}{vb0}
\wigglywrap{wa1}{va0}{vb0}{wa7}
\fmfi{wiggly}{wa5..wb7}
\end{fmfchar*}}}
}
\subfigspace
\subfigure[$G_{20}$]{
\raisebox{\eqoff}{%
\fmfframe(3,1)(1,4){%
\begin{fmfchar*}(16,20)
\Wuq
\fmfipair{wa[]}
\fmfipair{wb[]}
\fmfipair{wc[]}
\fmfipair{wd[]}
\fmfiequ{wa1}{point 1*length(p1)/2 of p1}
\fmfiv{d.shape=circle,d.size=2}{wa1}
\fmfiequ{wa3}{point 1*length(p3)/2 of p3}
\fmfiv{d.shape=circle,d.size=2}{wa3}
\fmfiequ{wa5}{point 1*length(p5)/2 of p5}
\fmfiv{d.shape=circle,d.size=2}{wa5}
\fmfiequ{wa6}{point 1*length(p6)/2 of p6}
\fmfiv{d.shape=circle,d.size=2}{wa6}
\fmfiequ{wa7}{point 1*length(p7)/2 of p7}
\fmfiv{d.shape=circle,d.size=2}{wa7}
\fmfforce{(-0w,-0h)}{va0}
\fmfforce{(1w,-0h)}{vb0}
\wigglywrap{wa1}{va0}{vb0}{wa7}
\fmfi{wiggly}{wa3..wa5}
\fmfi{wiggly}{wa5..wa6}
\end{fmfchar*}}}
}

\begin{tabular}{m{12cm}}
\toprule
$G_1\rightarrow \chi(1,4)(J_{21}+J_{23}-(-2J_{32}+2J_{33}+2J_{34}-2i\,\epsilon_{\mu\nu\rho\sigma}J_{35}^{\mu\nu\rho\sigma}))$  \\
$\phantom{G_1}\rightarrow M(J_{21}+J_{23}-(-2J_{32}+2J_{33}+2J_{34}-2i\,\epsilon_{\mu\nu\rho\sigma}J_{35}^{\mu\nu\rho\sigma}))$  \\
$G_2\rightarrow 0$ \\
$G_3\rightarrow -\chi(1,4)\,J_{21} \rightarrow -M\,J_{21}$ \\
$G_4\rightarrow -\chi(1,4)(J_{22}+J_{23}+2J_{26}) \rightarrow -M(J_{22}+J_{23}+2J_{26})$ \\
$G_5\rightarrow \chi(1,4)(-J_{22}) \rightarrow M(-J_{22})$ \\
$G_6\rightarrow \chi(1,4)\,J_{22} \rightarrow M\,J_{22}$ \\
$G_7\rightarrow 0$ \\
$G_8\rightarrow \chi(1,4)\,J_{22} \rightarrow M\,J_{22}$ \\
$G_9\rightarrow -\chi(1,4)(J_{22}+J_{23}+2J_{26}) \rightarrow -M(J_{22}+J_{23}+2J_{26})$ \\
$G_{10}\rightarrow 0$ \\
$G_{11}\rightarrow \chi(1,4)(2J_{22}+2J_{23}+8J_{26}+4J_{31}) $ \\
$\phantom{G_11}\rightarrow M(2J_{22}+2J_{23}+8J_{26}+4J_{31})$ \\
$G_{12}=G_{13}=G_{14}\rightarrow 0$ \\
$G_{15}\rightarrow -\chi(1,4)\,J_{21} \rightarrow -M\,J_{21}$ \\
$G_{16}\rightarrow \chi(1,4)\,J_{21} \rightarrow M\,J_{21}$ \\
$G_{17}\rightarrow 0$ \\
$G_{18}=G_{19}=G_{20}\rightarrow 0$  \\
\midrule
$\chi(1,4)\rightarrow 4(J_{31}+J_{32}-J_{33}-J_{34}+i\,\epsilon_{\mu\nu\rho\sigma}J_{35}^{\mu\nu\rho\sigma})\,\chi(1,4)$  \\
$\phantom{\chi(1,4)}\rightarrow 4\,M\,(J_{31}+J_{32}-J_{33}-J_{34}+i\,\epsilon_{\mu\nu\rho\sigma}J_{35}^{\mu\nu\rho\sigma})$  \\
\bottomrule
\end{tabular}

\caption{Wrapping diagrams with structure $\chi(1,4)$} 
\label{wrap-14}
\end{figure}


\clearpage
\newpage

\renewcommand{\thefigure}{C.\arabic{figure}}
\setcounter{figure}{0}
\renewcommand{\thetable}{C.\arabic{table}}
\setcounter{table}{0}

\clearpage
\newpage
\section{Integrals}
\label{app:integrals}
In this appendix we list the pole parts of all the required 
logarithmically divergent momentum integrals. 
The computations have been performed using the
GPXT~\cite{Chetyrkin:1980pr,Kotikov:1995cw,uslong}.
The factor $1/(4\pi)^{10}$ in each integral has been omitted.
Note that the explicit value of integral $J_{35}^{\mu\nu\rho\sigma}$
is not needed since this integral appears only for structures
$\chi(2,1,4)$ and $\chi(1,4)$, and the two contributions exactly
cancel each other.

\settoheight{\eqoff}{$\times$}%
\setlength{\eqoff}{0.5\eqoff}%
\addtolength{\eqoff}{-7.5\unitlength}
\begin{align*}
J_{1}=
\raisebox{\eqoff}{%
\begin{fmfchar*}(20,15)
\fmfleft{in}
\fmfright{out}
\fmf{plain}{in,v1}
\fmf{plain,left=0.25}{v1,v2}
\fmf{plain,left=0}{v2,v5}
\fmf{plain,left=0}{v5,v4}
\fmf{plain,left=0.25}{v4,v3}
\fmf{plain,tension=0.5,right=0.25}{v1,v0,v1}
\fmf{plain,right=0.25}{v0,v3}
\fmf{plain}{v0,v2}
\fmf{plain}{v0,v4}
\fmf{plain}{v0,v5}
\fmf{plain}{v3,out}
\fmffixed{(0.9w,0)}{v1,v3}
\fmffixed{(0.4w,0)}{v2,v4}
\fmffixed{(0.2w,0)}{v2,v5}
\fmffixed{(0.2w,0)}{v5,v4}
\fmfpoly{phantom}{v4,v2,v0}
\fmffreeze
\end{fmfchar*}}
&=
\frac{1}{120\varepsilon^5}-\frac{1}{12\varepsilon^4}+\frac{11}{24\varepsilon^3}-\frac{19}{12\varepsilon^2}+\frac{14}{5\varepsilon}
\\
J_{2}=\raisebox{\eqoff}{%
\begin{fmfchar*}(20,15)
\fmfleft{in}
\fmfright{out}
\fmf{plain}{in,v1}
\fmf{plain}{v5,out}
\fmf{plain}{v1,v2}
\fmf{plain}{v2,v3}
\fmf{plain}{v3,v4}
\fmf{plain}{v4,v5}
\fmffixed{(0,0.5h)}{v0,v3}
\fmffixed{(0.9w,0)}{v1,v5}
\fmffixed{(0.8w,0)}{v2,v4}
\fmffixed{(0,0.5h)}{v0,v3}
\fmf{plain,tension=0.25,right=0.25}{v2,v0,v2}
\fmf{plain,tension=0.25,right=0.25}{v4,v0,v4}
\fmf{plain,tension=0.25,right=0.25}{v3,v0,v3}
\fmffreeze
\end{fmfchar*}}
&=
\frac{2}{15\varepsilon^5}-\frac{4}{15\varepsilon^4}-\frac{1}{6\varepsilon^3}+\frac{3}{10\varepsilon^2}+\frac{4}{15\varepsilon}
\\
J_{3}=\raisebox{\eqoff}{%
\begin{fmfchar*}(20,15)
\fmfleft{in}
\fmfright{out}
\fmf{plain}{in,v1}
\fmf{plain}{v5,out}
\fmf{plain,tension=0.25}{v1,v2}
\fmf{plain}{v2,v3}
\fmf{plain}{v3,v4}
\fmf{plain,tension=0.25}{v4,v5}
\fmf{plain}{v1,v0}
\fmf{plain}{v0,v5}
\fmffixed{(0,0.5h)}{v0,v3}
\fmffixed{(0.9w,0)}{v1,v5}
\fmffixed{(0.6w,0)}{v2,v4}
\fmf{plain,tension=0.25,right=0.25}{v2,v0,v2}
\fmf{plain,tension=0.25,right=0.25}{v4,v0,v4}
\fmffreeze
\end{fmfchar*}}
&=
\frac{2}{15\varepsilon^5}-\frac{2}{5\varepsilon^4}+\frac{1}{5\varepsilon^3}+\frac{4}{15\varepsilon^2}-\frac{1}{15\varepsilon}
\\
J_{4}=\raisebox{\eqoff}{%
\begin{fmfchar*}(20,15)
\fmfleft{in}
\fmfright{out}
\fmf{plain}{in,v1}
\fmf{plain}{v5,out}
\fmf{plain}{v1,v2}
\fmf{plain}{v2,v3}
\fmf{plain}{v3,v4}
\fmf{plain}{v4,v5}
\fmf{plain}{v1,v0}
\fmf{plain}{v0,v5}
\fmffixed{(0,0.5h)}{v0,v3}
\fmffixed{(0.9w,0)}{v1,v5}
\fmffixed{(0.6w,0)}{v2,v4}
\fmf{plain,tension=0.25}{v0,v2}
\fmf{plain,tension=0.25}{v0,v4}
\fmf{plain,tension=0.5,right=0.25}{v3,v0,v3}
\fmffreeze
\end{fmfchar*}}
&=
\frac{1}{20\varepsilon^5}-\frac{3}{10\varepsilon^4}+\frac{17}{20\varepsilon^3}-\frac{14}{15\varepsilon^2}+\frac{2}{5\varepsilon}
\\
J_{5}=\raisebox{\eqoff}{%
\begin{fmfchar*}(20,15)
\fmfleft{in}
\fmfright{out}
\fmf{plain}{in,v1}
\fmf{plain}{v4,out}
\fmf{plain}{v1,v2}
\fmf{plain}{v2,v3}
\fmf{plain}{v3,v4}
\fmf{plain}{v4,v5}
\fmf{plain}{v0,v1}
\fmf{plain,tension=0.5}{v0,v2}
\fmffixed{(0.9w,0)}{v1,v4}
\fmffixed{(whatever,0)}{v1,in}
\fmffixed{(whatever,0)}{v4,out}
\fmf{plain,tension=0.5,right=0.25}{v3,v0,v3}
\fmf{plain,tension=0.5,right=0.25}{v5,v0,v5}
\fmffixed{(whatever,0.3h)}{v0,v5}
\fmffixed{(whatever,-0.3h)}{v0,v3}
\fmffixed{(0.5w,0)}{v1,v0}
\fmffixed{(0.25w,whatever)}{v1,v2}
\fmffreeze
\end{fmfchar*}}
&=
\frac{3}{40\varepsilon^5}-\frac{3}{10\varepsilon^4}+\frac{17}{40\varepsilon^3}+\frac{1}{30\varepsilon^2}-\frac{1}{2\varepsilon}
\\
J_{6}=\raisebox{\eqoff}{%
\begin{fmfchar*}(20,15)
\fmfleft{in}
\fmfright{out}
\fmf{plain}{in,v0}
\fmf{phantom,tension=0.5,left=0.25}{v1,v2}
\fmf{plain,tension=0.5,left=0.25}{v2,v3}
\fmf{plain,tension=0.5,left=0.25}{v3,v4}
\fmf{phantom,tension=0.5,left=0.25}{v4,v1}
\fmf{plain,tension=0.5,right=0.5}{v2,v0,v2}
\fmf{plain,tension=0.5,right=0.5}{v0,v4,v0}
\fmf{plain}{v3,out}
\fmffixed{(0.9w,0)}{v1,v3}
\fmffixed{(0,0.45w)}{v4,v2}
\fmffreeze
\fmfipath{px}
\fmfipath{py}
\fmfipair{w[]}
\fmfiset{px}{vpath(__v2,__v3)}
\fmfiset{py}{vpath(__v4,__v3)}
\fmfiequ{w1}{point length(px)/2 of px}
\fmfiequ{w2}{(xpart(vloc(__v0)),ypart(vloc(__v0)))}
\fmfiequ{w3}{point length(py)/2 of py}
\fmfi{plain}{w1..w2}
\fmfi{plain}{w3..w2}
\end{fmfchar*}}
&=
\frac{1}{20\varepsilon^5}-\frac{1}{5\varepsilon^4}+\frac{17}{60\varepsilon^3}-\frac{2}{15\varepsilon^2}+\frac{4}{15\varepsilon}
\\
J_{7}=\raisebox{\eqoff}{%
\begin{fmfchar*}(20,15)
\fmfleft{in}
\fmfright{out}
\fmf{plain}{in,v1}
\fmf{phantom,left=0.25}{v1,v2}
\fmf{plain,left=0.25}{v2,v3}
\fmf{plain,left=0.25}{v3,v4}
\fmf{plain,left=0.25}{v4,v1}
\fmf{plain,tension=0.5,right=0.5}{v2,v0,v2}
\fmf{plain,tension=0.5,right=0.5}{v0,v4,v0}
\fmf{plain}{v3,out}
\fmffixed{(0.9w,0)}{v1,v3}
\fmffixed{(0,0.45w)}{v4,v2}
\fmffreeze
\fmf{plain}{v1,v0}
\fmfipath{px}
\fmfipair{w[]}
\fmfiset{px}{vpath(__v2,__v3)}
\fmfiequ{w1}{point length(px)/2 of px}
\fmfiequ{w2}{(xpart(vloc(__v0)),ypart(vloc(__v0)))}
\fmfi{plain,left=0.25}{w1..w2}
\end{fmfchar*}}
&=
\frac{3}{40\varepsilon^5}-\frac{17}{60\varepsilon^4}+\frac{43}{120\varepsilon^3}+\frac{7}{60\varepsilon^2}-\frac{8}{15\varepsilon}
\\
J_8=
\raisebox{\eqoff}{%
\begin{fmfchar*}(20,15)
\fmfleft{in}
\fmfright{out}
\fmf{plain}{in,v1}
\fmf{plain,left=0}{v1,v2}
\fmf{plain,tension=0.35,left=0}{v2,v3}
\fmf{plain,left=0}{v3,v4}
\fmf{plain,left=0}{v0,v1}
\fmf{plain,tension=0.35,left=0}{v0,v2}
\fmf{plain,left=0}{v0,v4}
\fmf{plain,tension=0.35,right=0.25}{v3,v0,v3}
\fmf{plain}{v4,out}
\fmffixed{(0.9w,0)}{v1,v4}
\fmffixed{(whatever,0.5h)}{v0,v3}
\fmffixed{(whatever,0.4h)}{v0,v2}
\fmffreeze
\fmfipath{px}
\fmfipair{w[]}
\fmfiset{px}{vpath(__v2,__v3)}
\fmfiequ{w1}{point length(px)/2 of px}
\fmfiequ{w2}{(xpart(vloc(__v0)),ypart(vloc(__v0)))}
\fmfi{plain}{w1..w2}
\end{fmfchar*}}
&=
\frac{1}{30\varepsilon^5}-\frac{7}{30\varepsilon^4}+\frac{5}{6\varepsilon^3}-\frac{3}{2\varepsilon^2}+\frac{11}{30\varepsilon}
\\
J_9=
\raisebox{\eqoff}{%
\begin{fmfchar*}(20,15)
\fmfleft{in}
\fmfright{out}
\fmf{plain}{in,v1}
\fmf{plain,left=0}{v1,v2}
\fmf{plain,left=0}{v2,v4}
\fmf{plain,left=0}{v3,v4}
\fmf{plain,left=0}{v0,v2}
\fmf{plain,tension=0.35,right=0.25}{v1,v0,v1}
\fmf{plain,tension=0.25,right=0.25}{v3,v0,v3}
\fmf{plain}{v4,out}
\fmffixed{(0.9w,0)}{v1,v4}
\fmffixed{(whatever,0.5h)}{v0,v2}
\fmffreeze
\fmfipath{px}
\fmfipair{w[]}
\fmfiset{px}{vpath(__v2,__v4)}
\fmfiequ{w1}{point length(px)/3 of px}
\fmfiequ{w2}{(xpart(vloc(__v0)),ypart(vloc(__v0)))}
\fmfi{plain}{w1..w2}
\end{fmfchar*}}
&=
\frac{1}{30\varepsilon^5}-\frac{11}{60\varepsilon^4}+\frac{7}{15\varepsilon^3}-\frac{19}{60\varepsilon^2}-\frac{14}{15\varepsilon}
\\
J_{10}=\raisebox{\eqoff}{%
\begin{fmfchar*}(20,15)
\fmfleft{in}
\fmfright{out}
\fmf{plain}{in,v1}
\fmf{plain,left=0.25}{v1,v2}
\fmf{plain,left=0.25}{v2,v3}
\fmf{plain,left=0.25}{v3,v4}
\fmf{phantom,left=0.25}{v4,v1}
\fmf{plain,tension=0.5,right=0.5}{v2,v0,v2}
\fmf{plain,tension=0.5,right=0.5}{v0,v4,v0}
\fmf{plain}{v3,out}
\fmffixed{(0.9w,0)}{v1,v3}
\fmffixed{(0,0.45w)}{v4,v2}
\fmffreeze
\fmf{plain}{v1,v0}
\fmfipath{px}
\fmfipair{w[]}
\fmfiset{px}{vpath(__v2,__v3)}
\fmfiequ{w1}{point length(px)/2 of px}
\fmfiequ{w2}{(xpart(vloc(__v0)),ypart(vloc(__v0)))}
\fmfi{plain,left=0.25}{w1..w2}
\end{fmfchar*}}
&=
\frac{11}{120\varepsilon^5}-\frac{1}{3\varepsilon^4}+\frac{3}{8\varepsilon^3}+\frac{1}{6\varepsilon^2}-\frac{1}{30\varepsilon}
\\
J_{11}=\raisebox{\eqoff}{%
\begin{fmfchar*}(20,15)
\fmfleft{in}
\fmfright{out}
\fmf{plain}{in,v1}
\fmf{plain}{v6,out}
\fmffixed{(0.9w,0)}{v1,v6}
\fmffixed{(whatever,0)}{v1,in}
\fmffixed{(whatever,0)}{out,v6}
\fmf{derplain}{v2,v1}
\fmf{plain}{v2,v3}
\fmf{derplain}{v3,v4}
\fmf{plain}{v4,v5}
\fmf{plain}{v5,v6}
\fmf{plain}{v0,v1}
\fmf{plain,tension=0.5}{v0,v2}
\fmf{plain}{v0,v3}
\fmf{plain}{v0,v4}
\fmf{plain,tension=0.5}{v0,v5}
\fmf{plain}{v0,v6}
\fmffixed{(whatever,0.5h)}{v0,v3}
\fmffixed{(0.25w,0)}{v3,v4}
\fmffixed{(0.65w,0)}{v2,v5}
\fmffixed{(whatever,0.3h)}{v0,v2}
\fmffreeze
\end{fmfchar*}}
&=
-\frac{1}{10\varepsilon^3}+\frac{13}{30\varepsilon^2}-\frac{23}{30\varepsilon}
\\
J_{12}=\raisebox{\eqoff}{%
\begin{fmfchar*}(20,15)
\fmfleft{in}
\fmfright{out}
\fmf{plain}{in,v1}
\fmf{plain}{v4,out}
\fmffixed{(0.9w,0)}{v1,v4}
\fmffixed{(whatever,0)}{v1,in}
\fmffixed{(whatever,0)}{out,v4}
\fmf{derplain}{v2,v1}
\fmf{plain}{v2,v3}
\fmf{derplain}{v3,v4}
\fmf{plain}{v4,v5}
\fmf{plain,tension=0.5}{v5,v6}
\fmf{plain}{v0,v1}
\fmf{plain,tension=0.5}{v0,v2}
\fmf{plain}{v0,v3}
\fmf{plain,tension=0.5}{v0,v5}
\fmf{plain,tension=0.25,right=0.25}{v6,v0,v6}
\fmffixed{(whatever,0.5h)}{v0,v3}
\fmffixed{(whatever,0.3h)}{v6,v0}
\fmffreeze
\end{fmfchar*}}
&=
-\frac{1}{60\varepsilon^3}+\frac{1}{12\varepsilon^2}-\frac{2}{5\varepsilon}
\\
J_{13}=
\settoheight{\eqoff}{$\times$}%
\setlength{\eqoff}{0.5\eqoff}%
\addtolength{\eqoff}{-7.5\unitlength}%
\raisebox{\eqoff}{%
\begin{fmfchar*}(20,15)
  \fmfleft{in}
  \fmfright{out}
  \fmf{plain}{in,v1}
  \fmf{phantom,tension=2,left=0.25}{v1,v2}
  \fmf{plain,tension=2,left=0.25}{v2,v3}
  \fmf{plain,left=0.25}{v4,v1}
  \fmf{plain,left=0.25}{v0,v4}
  \fmf{plain,right=0}{v0,v1}
  \fmf{plain,right=0.25}{v0,v5}
  \fmf{derplain,left=0.25}{v6,v4}
  \fmf{plain,right=0.25}{v6,v5}
  \fmf{phantom,right=0}{v3,v0}
  \fmf{derplain,right=0.25}{v5,v3}
  \fmf{plain}{v3,out}
\fmffixed{(0.9w,0)}{v1,v3}
\fmfpoly{phantom}{v2,v4,v5}
\fmffixed{(0.5w,0)}{v4,v5}
\fmf{plain}{v0,v6}
\fmf{plain,tension=0.25,right=0.25}{v2,v0}
\fmf{plain,tension=0.25,right=0.25}{v0,v2}
\fmffreeze
\fmffreeze
\fmfshift{(0,0.15w)}{in,out,v1,v2,v3,v4,v5,v0}
\end{fmfchar*}}
&=
-\frac{1}{20\varepsilon^3}+\frac{3}{20\varepsilon^2}-\frac{13}{30\varepsilon}
\\
J_{14}=\raisebox{\eqoff}{%
\begin{fmfchar*}(20,15)
\fmfleft{in}
\fmfright{out}
\fmf{plain}{in,v1}
\fmf{plain}{v6,out}
\fmffixed{(0.9w,0)}{v1,v6}
\fmffixed{(whatever,0)}{v1,in}
\fmffixed{(whatever,0)}{out,v6}
\fmf{plain}{v2,v1}
\fmf{derplain}{v3,v2}
\fmf{plain}{v3,v4}
\fmf{derplain}{v4,v5}
\fmf{plain}{v5,v6}
\fmf{plain}{v0,v1}
\fmf{plain,tension=0.5}{v0,v2}
\fmf{plain}{v0,v3}
\fmf{plain}{v0,v4}
\fmf{plain,tension=0.5}{v0,v5}
\fmf{plain}{v0,v6}
\fmffixed{(whatever,0.5h)}{v0,v3}
\fmffixed{(0.25w,0)}{v3,v4}
\fmffixed{(0.65w,0)}{v2,v5}
\fmffixed{(whatever,0.3h)}{v0,v2}
\fmffreeze
\end{fmfchar*}}
&=
-\frac{1}{5\varepsilon^3}+\frac{2}{3\varepsilon^2}-\frac{4}{15\varepsilon}
\\
J_{15}=\raisebox{\eqoff}{%
\begin{fmfchar*}(20,15)
\fmfleft{in}
\fmfright{out}
\fmf{plain}{in,v1}
\fmf{plain}{v4,out}
\fmffixed{(0.9w,0)}{v1,v4}
\fmffixed{(whatever,0)}{v1,in}
\fmffixed{(whatever,0)}{out,v4}
\fmf{derplain}{v2,v1}
\fmf{plain}{v2,v3}
\fmf{derplain}{v3,v4}
\fmf{plain}{v4,v5}
\fmf{plain}{v6,v1}
\fmf{plain}{v0,v2}
\fmf{plain}{v0,v3}
\fmf{plain,tension=0.25,right=0.25}{v5,v0,v5}
\fmf{plain,tension=0.25,right=0.25}{v6,v0,v6}
\fmffixed{(whatever,0.5h)}{v0,v2}
\fmffixed{(whatever,0)}{v2,v3}
\fmffreeze
\end{fmfchar*}}
&=
-\frac{1}{30\varepsilon^3}+\frac{1}{30\varepsilon^2}+\frac{1}{3\varepsilon}
\\
J_{16}=\raisebox{\eqoff}{%
\begin{fmfchar*}(20,15)
\fmfleft{in}
\fmfright{out}
\fmf{plain}{in,v1}
\fmf{plain}{v5,out}
\fmffixed{(0.9w,0)}{v1,v5}
\fmffixed{(whatever,0)}{v1,in}
\fmffixed{(whatever,0)}{out,v5}
\fmf{derplain}{v2,v1}
\fmf{plain}{v2,v3}
\fmf{derplain,tension=0.5}{v3,v4}
\fmf{plain}{v4,v5}
\fmf{plain}{v5,v6}
\fmf{plain}{v0,v1}
\fmf{plain}{v0,v2}
\fmf{plain}{v0,v3}
\fmf{plain}{v0,v4}
\fmf{plain,tension=0.25,right=0.25}{v6,v0,v6}
\fmffixed{(whatever,0.5h)}{v0,v3}
\fmffixed{(whatever,0.4h)}{v0,v2}
\fmffixed{(whatever,0.4h)}{v0,v4}
\fmffreeze
\end{fmfchar*}}
&=
-\frac{3}{20\varepsilon^3}+\frac{11}{60\varepsilon^2}+\frac{3}{10\varepsilon}
\\
J_{17}=\raisebox{\eqoff}{%
\begin{fmfchar*}(20,15)
\fmfleft{in}
\fmfright{out}
\fmf{plain}{in,v1}
\fmf{plain}{v4,out}
\fmffixed{(0.9w,0)}{v1,v4}
\fmffixed{(whatever,0)}{v1,in}
\fmffixed{(whatever,0)}{out,v4}
\fmf{derplain}{v2,v1}
\fmf{plain}{v2,v3}
\fmf{derplain}{v3,v4}
\fmf{plain}{v5,v6}
\fmf{plain}{v6,v1}
\fmf{plain}{v0,v2}
\fmf{plain}{v0,v3}
\fmf{plain}{v0,v4}
\fmf{plain}{v0,v5}
\fmf{plain,tension=0.25,right=0.25}{v6,v0,v6}
\fmffixed{(whatever,0.5h)}{v0,v3}
\fmffixed{(whatever,0.4h)}{v0,v2}
\fmffixed{(whatever,0.2h)}{v5,v0}
\fmffreeze
\end{fmfchar*}}
&=
-\frac{1}{15\varepsilon^3}+\frac{2}{15\varepsilon^2}+\frac{1}{15\varepsilon}
\\
J_{18}=\raisebox{\eqoff}{%
\begin{fmfchar*}(20,15)
\fmfleft{in}
\fmfright{out}
\fmf{plain}{in,v1}
\fmf{plain}{v3,out}
\fmffixed{(0.9w,0)}{v1,v3}
\fmffixed{(whatever,0)}{v1,in}
\fmffixed{(whatever,0)}{out,v3}
\fmf{plain}{v1,v2}
\fmf{plain}{v2,v3}
\fmf{plain}{v3,v4}
\fmf{plain}{v4,v5}
\fmf{plain}{v5,v1}
\fmf{plain}{v5,v2}
\fmf{plain}{v0,v2}
\fmf{plain}{v0,v4}
\fmf{plain,tension=0.25,right=0.25}{v5,v0,v5}
\fmffixed{(whatever,0.5h)}{v2,v4}
\fmffixed{(0.3w,whatever)}{v5,v0}
\fmffreeze
\end{fmfchar*}}
&=
\frac{1}{24\varepsilon^5}-\frac{1}{4\varepsilon^4}+\frac{5}{8\varepsilon^3}-\frac{1}{\varepsilon^2}\Big(\frac{1}{12}+\frac{4\zeta(3)}{5}\Big)-\frac{1}{\varepsilon}\Big(\frac{5}{6}-\frac{4\zeta(3)}{5}-\frac{\pi^4}{300}\Big)
\\
J_{19}=\raisebox{\eqoff}{%
\begin{fmfchar*}(20,15)
\fmfleft{in}
\fmfright{out}
\fmf{plain}{in,v1}
\fmf{plain}{v3,out}
\fmffixed{(0.9w,0)}{v1,v3}
\fmffixed{(whatever,0)}{in,v1}
\fmffixed{(whatever,0)}{out,v3}
\fmf{plain}{v1,v2}
\fmf{plain}{v2,v3}
\fmf{plain}{v1,v4}
\fmf{plain}{v4,v5}
\fmf{plain}{v4,v0}
\fmf{plain}{v2,v0}
\fmf{plain,tension=0.25,right=0.25}{v5,v0,v5}
\fmf{plain,tension=0.25,right=0.25}{v5,v3,v5}
\fmffixed{(whatever,0.5h)}{v2,v4}
\fmffixed{(0.28w,whatever)}{v0,v5}
\fmffreeze
\end{fmfchar*}}
&=
\frac{1}{24\varepsilon^5}-\frac{1}{6\varepsilon^4}+\frac{1}{8\varepsilon^3}+\frac{1}{\varepsilon^2}\Big(\frac{1}{3}-\frac{\zeta(3)}{5}\Big)-\frac{1}{\varepsilon}\Big(\frac{1}{3}-\frac{4\zeta(5)}{5}+\frac{\pi^4}{300}\Big)
\\
J_{20}=\raisebox{\eqoff}{%
\begin{fmfchar*}(20,15)
\fmfleft{in}
\fmfright{out}
\fmf{plain}{in,v1}
\fmf{plain,left=0.25}{v1,v2}
\fmf{plain,left=0.25}{v2,v3}
\fmffixed{(0.9w,0)}{v1,v3}
\fmffixed{(0,0.45w)}{v6,v2}
\fmffixed{(0.15w,0)}{v4,v6}
\fmffixed{(0.45w,0)}{v1,v0}
\fmf{plain,left=0.25}{v4,v1}
\fmf{plain,tension=0.5,right=0.5}{v2,v0,v2}
\fmf{plain}{v4,v5}
\fmf{plain,left=0.25}{v3,v5}
\fmffixed{(0.3w,0)}{v4,v5}
\fmf{phantom}{v0,v3}
\fmf{plain}{v1,v0}
\fmf{plain}{v0,v4}
\fmf{plain}{v0,v5}
\fmf{plain}{v3,out}
\fmffreeze
\end{fmfchar*}}
&=
\frac{1}{120\varepsilon^5}-\frac{1}{12\varepsilon^4}+\frac{11}{24\varepsilon^3}-\frac{19}{12\varepsilon^2}+\frac{1}{\varepsilon}\Big(\frac{14}{5}-4\zeta(5)\Big)
\\
J_{21}=\raisebox{\eqoff}{%
\begin{fmfchar*}(20,15)
\fmfleft{in}
\fmfright{out}
\fmf{plain}{in,v1}
\fmf{plain,left=0.25}{v1,v2}
\fmf{plain,left=0.25}{v2,v3}
\fmf{plain,left=0.25}{v3,v4}
\fmf{plain,left=0.25}{v4,v1}
\fmf{plain,tension=0.5,right=0.5}{v2,v0,v2}
\fmf{plain,tension=0.5,right=0.5}{v0,v4,v0}
\fmf{plain}{v3,out}
\fmffixed{(0.9w,0)}{v1,v3}
\fmffixed{(0,0.45w)}{v4,v2}
\fmffreeze
\fmfipath{px}
\fmfipair{w[]}
\fmfiset{px}{vpath(__v2,__v3)}
\fmfiequ{w1}{point length(px)/2 of px}
\fmfiequ{w2}{(xpart(vloc(__v0)),ypart(vloc(__v0)))}
\fmfi{plain,left=0.25}{w1..w2}
\end{fmfchar*}}
&=
\frac{1}{15\varepsilon^5}-\frac{1}{4\varepsilon^4}+\frac{1}{5\varepsilon^3}+\frac{29}{60\varepsilon^2}-\frac{1}{\varepsilon}\Big(\frac{19}{30}-\frac{4\zeta(3)}{5}\Big)
\\
J_{22}=\raisebox{\eqoff}{%
\begin{fmfchar*}(20,15)
\fmfleft{in}
\fmfright{out}
\fmf{plain}{in,v1}
\fmf{plain,left=0.25}{v1,v2}
\fmf{plain,left=0.25}{v2,v3}
\fmf{plain,left=0.25}{v3,v4}
\fmf{plain,left=0.25}{v4,v1}
\fmf{plain,tension=0.5,right=0.25}{v1,v0,v1}
\fmf{phantom}{v0,v3}
\fmf{plain}{v2,v0}
\fmf{plain}{v0,v4}
\fmf{plain}{v3,out}
\fmffixed{(0.9w,0)}{v1,v3}
\fmffixed{(0,0.45w)}{v4,v2}
\fmffreeze
\fmfipath{px}
\fmfipair{w[]}
\fmfiset{px}{vpath(__v3,__v4)}
\fmfiequ{w1}{point length(px)/2 of px}
\fmfiequ{w2}{(xpart(vloc(__v0)),ypart(vloc(__v0)))}
\fmfi{plain}{w1..w2}
\end{fmfchar*}}
&=
\frac{1}{40\varepsilon^5}-\frac{1}{6\varepsilon^4}+\frac{61}{120\varepsilon^3}-\frac{17}{30\varepsilon^2}-\frac{1}{\varepsilon}\Big(\frac{1}{6}-\frac{4\zeta(3)}{5}\Big)
\\
J_{23}=\raisebox{\eqoff}{%
\begin{fmfchar*}(20,15)
\fmfleft{in}
\fmfright{out}
\fmfpoly{plain}{v5,v4,v3,v2,v1}
\fmf{plain}{in,v1}
\fmf{plain}{v3,out}
\fmffixed{(0.8w,0)}{v1,v3}
\fmf{plain}{v0,v1}
\fmf{plain}{v0,v2}
\fmf{plain}{v0,v3}
\fmf{plain}{v0,v4}
\fmf{plain}{v0,v5}
\fmffreeze
\end{fmfchar*}}
&=
\frac{14}{\varepsilon}\zeta(7)
\\
J_{24}=\raisebox{\eqoff}{%
\begin{fmfchar*}(20,15)
\fmfleft{in}
\fmfright{out}
\fmf{plain}{in,v1}
\fmf{plain}{v5,out}
\fmffixed{(0.9w,0)}{v1,v5}
\fmffixed{(whatever,0)}{v1,in}
\fmffixed{(whatever,0)}{out,v5}
\fmf{derplain}{v2,v1}
\fmf{plain,tension=0.5}{v2,v3}
\fmf{derplain,tension=0.5}{v3,v4}
\fmf{plain}{v4,v5}
\fmf{plain}{v5,v6}
\fmf{plain}{v6,v1}
\fmf{plain}{v0,v2}
\fmf{plain}{v0,v3}
\fmf{plain}{v0,v4}
\fmf{plain,tension=0.25,right=0.25}{v6,v0,v6}
\fmffixed{(whatever,0.5h)}{v6,v4}
\fmffixed{(whatever,0)}{v4,v2}
\fmffixed{(whatever,0.3h)}{v0,v3}
\fmffixed{(whatever,0.3h)}{v6,v0}
\fmffreeze
\end{fmfchar*}}
&=
-\frac{1}{20\varepsilon^3}+\frac{3}{20\varepsilon^2}+\frac{1}{\varepsilon}\Big(\frac{1}{6}+\frac{6\zeta(3)}{5}-2\zeta(5)\Big)
\\
J_{25}=\raisebox{\eqoff}{%
\begin{fmfchar*}(20,15)
\fmfleft{in}
\fmfright{out}
\fmf{plain}{in,v1}
\fmf{plain}{v4,out}
\fmffixed{(0.8w,0)}{v1,v4}
\fmffixed{(whatever,0)}{v1,in}
\fmffixed{(whatever,0)}{out,v4}
\fmfpoly{phantom}{v1,v2,v3,v4,v5,v6}
\fmf{plain}{v2,v1}
\fmf{derplain}{v2,v3}
\fmf{plain}{v3,v4}
\fmf{plain}{v5,v4}
\fmf{plain}{v5,v6}
\fmf{derplain}{v1,v6}
\fmf{plain}{v0,v1}
\fmf{plain}{v0,v2}
\fmf{plain}{v0,v3}
\fmf{plain}{v0,v5}
\fmf{plain}{v0,v6}
\fmffixed{(0.4w,0)}{v1,v0}
\fmffreeze
\end{fmfchar*}}
&=
-\frac{1}{10\varepsilon^3}+\frac{13}{30\varepsilon^2}-\frac{1}{\varepsilon}\Big(\frac{1}{6}-\frac{6\zeta(3)}{5}+2\zeta(5)\Big)
\\
J_{26}=\raisebox{\eqoff}{%
\begin{fmfchar*}(20,15)
\fmfleft{in}
\fmfright{out}
\fmf{plain}{in,v1}
\fmf{plain}{v4,out}
\fmffixed{(0.8w,0)}{v1,v4}
\fmffixed{(whatever,0)}{v1,in}
\fmffixed{(whatever,0)}{out,v4}
\fmfpoly{phantom}{v1,v2,v3,v4,v5,v6}
\fmf{plain}{v2,v1}
\fmf{plain}{v2,v3}
\fmf{plain}{v3,v4}
\fmf{derplain}{v5,v4}
\fmf{plain}{v5,v6}
\fmf{derplain}{v6,v1}
\fmf{plain}{v0,v1}
\fmf{plain}{v0,v2}
\fmf{plain}{v0,v3}
\fmf{plain}{v0,v5}
\fmf{plain}{v0,v6}
\fmffixed{(0.4w,0)}{v1,v0}
\fmffreeze
\end{fmfchar*}}
&=
-\frac{2\zeta(5)}{\varepsilon}
\\
J_{27}=\raisebox{\eqoff}{%
\begin{fmfchar*}(20,15)
\fmfleft{in}
\fmfright{out}
\fmf{plain}{in,v1}
\fmf{plain}{v4,out}
\fmffixed{(0.8w,0)}{v1,v4}
\fmffixed{(whatever,0)}{v1,in}
\fmffixed{(whatever,0)}{out,v4}
\fmf{plain}{v1,v2}
\fmf{plain,tension=0.5}{v2,v3}
\fmf{plain}{v3,v4}
\fmf{plain}{v4,v5}
\fmf{plain}{v5,v6}
\fmf{plain}{v6,v1}
\fmf{plain,tension=0.4}{v0,v3}
\fmf{plain}{v0,v5}
\fmf{plain}{v0,v6}
\fmf{plain,tension=0.25,right=0.25}{v2,v0,v2}
\fmffixed{(whatever,0)}{v5,v6}
\fmffixed{(whatever,0.6h)}{v5,v2}
\fmffixed{(0,whatever)}{v0,v2}
\fmffreeze
\end{fmfchar*}}
&=
-\frac{1}{60\varepsilon^3}+\frac{1}{12\varepsilon^2}+\frac{1}{\varepsilon}\Big(\frac{1}{5}+\frac{6\zeta(3)}{5}-2\zeta(5)\Big)
\\
J_{28}=\raisebox{\eqoff}{%
\begin{fmfchar*}(20,15)
\fmfleft{in}
\fmfright{out}
\fmf{plain}{in,v1}
\fmf{plain}{v4,out}
\fmffixed{(0.8w,0)}{v1,v4}
\fmffixed{(whatever,0)}{v1,in}
\fmffixed{(whatever,0)}{out,v4}
\fmfpoly{phantom}{v1,v2,v3,v4,v5,v6}
\fmf{plain}{v2,v1}
\fmf{derplain}{v2,v3}
\fmf{plain}{v3,v4}
\fmf{plain}{v5,v4}
\fmf{derplain}{v6,v5}
\fmf{plain}{v6,v1}
\fmf{plain}{v0,v1}
\fmf{plain}{v0,v2}
\fmf{plain}{v0,v3}
\fmf{plain}{v0,v5}
\fmf{plain}{v0,v6}
\fmffixed{(0.4w,0)}{v1,v0}
\fmffreeze
\end{fmfchar*}}
&=
\frac{1}{5\varepsilon^2}-\frac{1}{\varepsilon}\Big(\frac{19}{20}+\frac{4\zeta(3)}{5}\Big)
\\
J_{29}=\raisebox{\eqoff}{%
\begin{fmfchar*}(20,15)
\fmfleft{in}
\fmfright{out}
\fmf{plain}{in,v1}
\fmf{plain}{v3,out}
\fmffixed{(0.8w,0)}{v1,v3}
\fmffixed{(whatever,0)}{v1,in}
\fmffixed{(whatever,0)}{out,v3}
\fmf{plain}{v1,v2}
\fmf{plain}{v2,v3}
\fmf{derplain}{v4,v3}
\fmf{plain}{v4,v5}
\fmf{plain}{v5,v6}
\fmf{derplain}{v6,v1}
\fmf{plain}{v0,v4}
\fmf{plain}{v0,v5}
\fmf{plain}{v0,v6}
\fmf{plain,tension=0.25,right=0.25}{v2,v0,v2}
\fmffixed{(0,0.4h)}{v0,v2}
\fmffixed{(0,0.3h)}{v5,v0}
\fmffixed{(0.4w,0)}{v6,v4}
\fmffreeze
\end{fmfchar*}}
&=
\frac{1}{20\varepsilon^2}-\frac{1}{\varepsilon}\Big(\frac{13}{20}+\frac{4\zeta(3)}{5}\Big)
\\
J_{30}=\raisebox{\eqoff}{%
\begin{fmfchar*}(20,15)
\fmfleft{in}
\fmfright{out}
\fmf{plain}{in,v1}
\fmf{plain}{v4,out}
\fmffixed{(0.8w,0)}{v1,v4}
\fmffixed{(whatever,0)}{v1,in}
\fmffixed{(whatever,0)}{out,v4}
\fmfpoly{phantom}{v1,v2,v3,v4,v5,v6}
\fmffreeze
\fmf{derplainpt}{v2,v1}
\fmf{plain}{v2,v7}
\fmf{derplainpt}{v7,v3}
\fmf{plain}{v3,v4}
\fmf{derplain}{v5,v4}
\fmf{plain}{v5,v6}
\fmf{derplain}{v6,v1}
\fmf{plain}{v0,v2}
\fmf{plain}{v0,v7}
\fmf{plain}{v0,v3}
\fmf{plain}{v0,v5}
\fmf{plain}{v0,v6}
\fmffixed{(0.4w,0)}{v1,v0}
\fmffixed{(whatever,0)}{v2,v7}
\fmffreeze
\end{fmfchar*}}
&=
\frac{1}{\varepsilon}\Big(\frac{9}{10}+\frac{11\zeta(3)}{5}-2\zeta(5)\Big)
\\
J_{31}=\raisebox{\eqoff}{%
\begin{fmfchar*}(20,15)
\fmfleft{in}
\fmfright{out}
\fmf{plain}{in,v1}
\fmf{plain}{v4,out}
\fmffixed{(0.8w,0)}{v1,v4}
\fmffixed{(whatever,0)}{v1,in}
\fmffixed{(whatever,0)}{out,v4}
\fmfpoly{phantom}{v1,v2,v3,v4,v5,v6}
\fmffreeze
\fmf{derplainpt}{v2,v1}
\fmf{plain}{v2,v7}
\fmf{plain}{v7,v3}
\fmf{derplainpt}{v3,v4}
\fmf{derplain}{v5,v4}
\fmf{plain}{v5,v6}
\fmf{derplain}{v6,v1}
\fmf{plain}{v0,v2}
\fmf{plain}{v0,v7}
\fmf{plain}{v0,v3}
\fmf{plain}{v0,v5}
\fmf{plain}{v0,v6}
\fmffixed{(0.4w,0)}{v1,v0}
\fmffixed{(whatever,0)}{v2,v7}
\fmffreeze
\end{fmfchar*}}
&=
-\frac{1}{\varepsilon}\Big(
\frac{9}{10}+2\zeta(3)-7\zeta(7)\Big)
\\
J_{32}=\raisebox{\eqoff}{%
\begin{fmfchar*}(20,15)
\fmfleft{in}
\fmfright{out}
\fmf{plain}{in,v1}
\fmf{plain}{v4,out}
\fmffixed{(0.8w,0)}{v1,v4}
\fmffixed{(whatever,0)}{v1,in}
\fmffixed{(whatever,0)}{out,v4}
\fmfpoly{phantom}{v1,v2,v3,v4,v5,v6}
\fmffreeze
\fmf{derplain}{v1,v2}
\fmf{derplainpt}{v7,v2}
\fmf{plain}{v7,v3}
\fmf{plain}{v3,v4}
\fmf{derplain}{v4,v5}
\fmf{derplainpt}{v6,v5}
\fmf{plain}{v6,v1}
\fmf{plain}{v0,v2}
\fmf{plain}{v0,v7}
\fmf{plain}{v0,v3}
\fmf{plain}{v0,v5}
\fmf{plain}{v0,v6}
\fmffixed{(0.4w,0)}{v1,v0}
\fmffixed{(whatever,0)}{v2,v7}
\fmffreeze
\end{fmfchar*}}
&=
-\frac{1}{\varepsilon}\Big(
\frac{1}{5}+\frac{2\zeta(3)}{5}+2\zeta(5)-\frac{7\zeta(7)}{2}\Big)
\\
J_{33}=\raisebox{\eqoff}{%
\begin{fmfchar*}(20,15)
\fmfleft{in}
\fmfright{out}
\fmf{plain}{in,v1}
\fmf{plain}{v4,out}
\fmffixed{(0.8w,0)}{v1,v4}
\fmffixed{(whatever,0)}{v1,in}
\fmffixed{(whatever,0)}{out,v4}
\fmfpoly{phantom}{v1,v2,v3,v4,v5,v6}
\fmffreeze
\fmf{derplainpt}{v1,v2}
\fmf{derplain}{v7,v2}
\fmf{plain}{v7,v3}
\fmf{plain}{v3,v4}
\fmf{derplain}{v4,v5}
\fmf{derplainpt}{v6,v5}
\fmf{plain}{v6,v1}
\fmf{plain}{v0,v2}
\fmf{plain}{v0,v7}
\fmf{plain}{v0,v3}
\fmf{plain}{v0,v5}
\fmf{plain}{v0,v6}
\fmffixed{(0.4w,0)}{v1,v0}
\fmffixed{(whatever,0)}{v2,v7}
\fmffreeze
\end{fmfchar*}}
&=
-\frac{1}{\varepsilon}\Big(
\frac{3}{10}+\frac{3\zeta(3)}{5}-2\zeta(5)\Big)
\\
J_{34}=\raisebox{\eqoff}{%
\begin{fmfchar*}(20,15)
\fmfleft{in}
\fmfright{out}
\fmf{plain}{in,v1}
\fmf{plain}{v4,out}
\fmffixed{(0.8w,0)}{v1,v4}
\fmffixed{(whatever,0)}{v1,in}
\fmffixed{(whatever,0)}{out,v4}
\fmfpoly{phantom}{v1,v2,v3,v4,v5,v6}
\fmffreeze
\fmf{derplain}{v1,v2}
\fmf{derplain}{v7,v2}
\fmf{plain}{v7,v3}
\fmf{plain}{v3,v4}
\fmf{derplainpt}{v4,v5}
\fmf{derplainpt}{v6,v5}
\fmf{plain}{v6,v1}
\fmf{plain}{v0,v2}
\fmf{plain}{v0,v7}
\fmf{plain}{v0,v3}
\fmf{plain}{v0,v5}
\fmf{plain}{v0,v6}
\fmffixed{(0.4w,0)}{v1,v0}
\fmffixed{(whatever,0)}{v2,v7}
\fmffreeze
\end{fmfchar*}}
&=
\frac{1}{\varepsilon}\Big(
\frac{1}{10}+\frac{\zeta(3)}{5}-2\zeta(5)+\frac{7\zeta(7)}{2}\Big)
\\
& \\
J_{35}^{\mu\nu\rho\sigma}=\raisebox{\eqoff}{%
\begin{fmfchar*}(20,15)
\fmfleft{in}
\fmfright{out}
\fmf{plain}{in,v1}
\fmf{plain}{v4,out}
\fmffixed{(0.8w,0)}{v1,v4}
\fmffixed{(whatever,0)}{v1,in}
\fmffixed{(whatever,0)}{out,v4}
\fmfpoly{phantom}{v1,v2,v3,v4,v5,v6}
\fmffreeze
\fmf{derplain}{v1,v2}
\fmf{derplain}{v7,v2}
\fmf{plain}{v7,v3}
\fmf{plain}{v3,v4}
\fmf{derplain}{v4,v5}
\fmf{derplain}{v6,v5}
\fmf{plain}{v6,v1}
\fmf{plain}{v0,v2}
\fmf{plain}{v0,v7}
\fmf{plain}{v0,v3}
\fmf{plain}{v0,v5}
\fmf{plain}{v0,v6}
\fmffixed{(0.4w,0)}{v1,v0}
\fmffixed{(whatever,0)}{v2,v7}
\fmffreeze
\fmfipath{p[]}
\fmfiset{p1}{vpath(__v1,__v2)}
\fmfiset{p2}{vpath(__v7,__v2)}
\fmfiset{p3}{vpath(__v4,__v5)}
\fmfiset{p4}{vpath(__v5,__v6)}
\fmfipair{w[]}
\fmfiequ{w1}{point length(p1)/2 of p1}
\fmfiequ{w2}{point length(p2)/2 of p2}
\fmfiequ{w3}{point length(p3)/2 of p3}
\fmfiequ{w4}{point length(p4)/2 of p4}
\fmfiv{l=\footnotesize{$\partial_\mu$},l.a=200,l.d=4}{w1}
\fmfiv{l=\footnotesize{$\partial_\nu$},l.a=-90,l.d=5}{w2}
\fmfiv{l=\footnotesize{$\partial_\rho$},l.a=45,l.d=3}{w3}
\fmfiv{l=\footnotesize{$\partial_\sigma$},l.a=90,l.d=4}{w4}
\end{fmfchar*}}
\end{align*}

\end{fmffile}

\clearpage

\footnotesize
\bibliographystyle{JHEP}
\bibliography{references}

\end{document}
